\documentclass[preprint,12pt]{elsarticle}

\usepackage{amssymb,amsmath}
\usepackage{amsthm}
\biboptions{sort&compress}
\newdefinition{rmk}{Remark}
\DeclareMathOperator{\tr}{tr}
\DeclareMathOperator{\diag}{diag}

\newcommand{\subfigimg}[3][,]{%
  \setbox1=\hbox{\includegraphics[#1]{#3}}
  \leavevmode\rlap{\usebox1}
  \rlap{\hspace*{0pt}\raisebox{\dimexpr\ht1-0.5\baselineskip}{\footnotesize{#2}}}
  \phantom{\usebox1}
}

\newcommand{\pderiv}[2]{\dfrac{\partial#1}{\partial#2}}

\renewcommand{\vec}{\boldsymbol}
\newcommand{\sedf}{\Psi}
\newcommand{\psedf}[1]{\sedf_{,#1}}
\newcommand{\point}{{\vartheta}}
\newcommand{\gp}{\mathcal{GP}}
\newcommand{\dk}[1]{k_{,#1}}
\newcommand{\cgp}{\mathcal{F}}
\newcommand{\cmean}{\mathcal{M}}
\newcommand{\ccov}{\mathcal{K}}
\newcommand{\observestate}{\vec{y}}
\newcommand{\observepoints}{\vec{\point}}
\newcommand{\predictstate}{\vec{f}_*}
\newcommand{\predictpoints}{\vec{\point}_*}
\newcommand{\predictmean}{\bar{\predictstate}}

\newcommand{\error}{\epsilon}
\newcommand{\rand}[1]{{#1}^{\mathcal{R}}}
\newcommand{\vecrand}[1]{\rand{\vec{#1}}}
\newcommand{\observe}[1]{\widetilde{#1}}
\newcommand{\sigmap}{\varsigma} 
\newcommand{\sigmapvec}{\boldsymbol{\varsigma}} 
\newcommand{\extract}[1]{\langle{#1}\rangle}
\newcommand{\prob}{\mathbb{P}}
\ifdefined\pdfliteral
    \let\griffPdfliteral\pdfliteral
\else \def\griffPdfliteral#1{} \fi

\newcommand\scriptr[1][2]{\leavevmode\hbox{\kern1pt\vbox to1ex{}\griffPdfliteral{%
    q 1 J .27 0 0 .27 0 0 cm #1 w
    0 2 m
    0 2 8.1 9.7 9.2 13.2 c
    10.4 16.8 8.4 15.4 8 14.7 c
    7.6 14 6.8 12.6 12 13 c
    17 13.5 14.5 7.8 13.7 6 c
    12.8 4.3 10.3 1.2 11.4 .2 c
    12.6 -.7 18.8 3.6 18.8 3.6 c
    18.8 3.6 l S Q
}\kern6pt}}
 
\newcommand{\x}{\vartheta}

\renewcommand{\u}{\vec{f}_\mathcal{I}}
\newcommand{\Z}{\vec{\point}_\mathcal{I}}

\newcommand{\m}{\mathbf{m}}
\renewcommand{\S}{\mathbf{S}}
\renewcommand{\L}{\mathcal{L}}

\newcommand{\Q}{\mathcal{Q}}

\usepackage{cancel}
\usepackage[ruled,vlined, linesnumbered]{algorithm2e}

\newcommand{\N}[1]{\mathcal{N}\left(#1\right)}
\newcommand*\diff{\mathop{}\!\mathrm{d}}

\journal{Journal}

\begin{document}

\begin{frontmatter}



\title{Strain energy density as a Gaussian process and its utilization in stochastic finite element analysis: application to planar soft tissues}


\author[gcec]{Ankush Aggarwal\corref{cor1}}

\affiliation[gcec]{organization={Glasgow Computational Engineering Centre, James Watt School of Engineering},
            addressline={University of Glasgow}, 
            city={Glasgow},
            postcode={G12~8LT}, 
            state={Scotland},
            country={United Kingdom}}

\author[compsci,compscidtu]{Bj\o{}rn Sand Jensen}
\affiliation[compsci]{organization={School of Computing Science},
            addressline={University of Glasgow}, 
            city={Glasgow},
            postcode={G12~8LT}, 
            state={Scotland},
            country={United Kingdom}}
\affiliation[compscidtu]{organization={Department of Applied Mathematics and Computer Science},
            addressline={Technical University of Denmark}, 
            city={Kgs. Lyngby},
            postcode={2800},             
            country={Denmark}}
            
\author[swansea]{Sanjay Pant}

\affiliation[swansea]{organization={Zienkiewicz Centre for Computational Engineering},
            addressline={Swansea University}, 
            city={Swansea},
            postcode={SA1 8EP}, 
            state={Wales},
            country={United Kingdom}}

\author[ou]{Chung-Hao Lee}

\affiliation[ou]{organization={School of Aerospace and Mechanical Engineering},
            addressline={The University of Oklahoma}, 
            city={Norman},
            postcode={73019}, 
            state={OK},
            country={United States of America}}

\cortext[cor1]{Correspondence to ankush.aggarwal@glasgow.ac.uk}

\begin{abstract}
Data-based approaches are promising alternatives to the traditional analytical constitutive models for solid mechanics. Herein, we propose a Gaussian process (GP) based constitutive modeling framework, specifically focusing on planar, hyperelastic and incompressible soft tissues. The strain energy density of soft tissues is modeled as a GP, which can be regressed to experimental stress-strain data obtained from biaxial experiments. Moreover, the GP model can be weakly constrained to be convex. A key advantage of a GP-based model is that, in addition to the mean value, it provides a probability density (i.e.~associated uncertainty) for the strain energy density. To simulate the effect of this uncertainty, a non-intrusive stochastic finite element analysis (SFEA) framework is proposed. The proposed framework is verified against an artificial dataset based on the Gasser--Ogden--Holzapfel model and applied to a real experimental dataset of a porcine aortic valve leaflet tissue. Results show that the proposed framework can be trained with limited experimental data and fits the data better than several existing models. The SFEA framework provides a straightforward way of using the experimental data and quantifying the resulting uncertainty in simulation-based predictions.
\end{abstract}



\end{frontmatter}


\section{Introduction}
\label{sec:intro}

Even with advanced numerical techniques, predictive mechanical modeling of complex materials, such as soft tissues, remains an unresolved challenge. Although the governing equations for solid mechanics (based on equilibrium) are deterministic and have been well established, uncertainty can arise through unknown variabilities in the domain shape, boundary conditions, and/or material properties. These three primary sources of uncertainty have been investigated in the literature. The present work focuses on the material properties that, in the context of solid mechanics, enter through the constitutive model defining the relationship between stresses and strains. Soft tissues are chosen as an application due to their nonlinear behavior and commonly observed variability in their response, which makes predictive modeling particularly challenging.

Soft tissues are usually modeled as hyperelastic, wherein a strain energy density function (SEDF) is defined to represent their stress-strain behavior. Traditionally, analytical forms of SEDF have been proposed based on experimental observations -- both macro- and micro-scopic, and dozens of models can be found in literature \cite{maurel1998}. Recently, we proposed a Bayesian framework to compare the models at describing the experimental data and found that the existing models do not fully capture the observed behavior. Hence, there is still a room for further improvements in constitutive models of soft tissues.

A recent, novel direction in constitutive modeling is using data-driven and machine learning approaches, which forgo analytical forms in favor of numerical or statistical representations. 

\citet{conti2018data} proposed a purely data-driven approach and solved one-dimensional linear elasticity problems. This approach can be thought of as a nearest-neighbor model, and it has been further developed for various problems. \citet{KIRCHDOERFER201681} extended it to one- and two-dimensional linear elasticity problems, and further works have extended the approach to inelasticity \cite{EGGERSMANN201981}, dynamics \cite{https://doi.org/10.1002/nme.5716}, fracture \cite{CARRARA2020113390}, and large strain elasticity \cite{PLATZER2021113756}. Recently, an enhancement in this approach was proposed to deal with outliers \cite{EGGERSMANN2021113499}. One of the issues in this approach is that it is heavily influenced by the outliers. A solution to this issue has been proposed by \citet{HE2021110124}. However, the authors of that study concluded with ``reinstates the importance of having sufficiently rich data coverage''. Such data-driven approaches are being further extended to reduce sensitivity to noise and find lower-dimensional representations \cite{HE2021114034, HE2020112791}.

Another promising approach is the use of a neural network (NN) to model the constitutive behavior \cite{https://doi.org/10.1002/pamm.202100072, tac2021datadriven, https://doi.org/10.1002/cnm.3438, https://doi.org/10.1002/nme.6459, KLEIN2022104703}. Since one key feature of NNs is their flexible architecture that can be adapted to a wide range of problems, several studies have explored varying versions of NNs \cite{https://doi.org/10.1002/pamm.202100072}. For example, \citet{https://doi.org/10.1002/cnm.3438} used strain components as the inputs to the neural network, \citet{KLEIN2022104703} used the deformation gradient, its cofactor and determinant as inputs, and \citet{tac2021datadriven} used strain invariants as inputs. Another difference between different formulations are whether (and how) they enforce convexity of the constitutive model. However, none of these models naturally account for the variability in the responses, commonly observed in soft tissues.

An alternative to neural networks in the machine learning literature is the Gaussian processes (GPs), which have been used to model stochastic systems. One of the attractive features of GPs is that they are naturally Bayesian. GPs also offer flexible regression and have a rigorous mathematical foundation that provides a control over their smoothness. In the field of mechanics, GPs have been primarily used as surrogate models, for example, for reduced-order modeling \cite{GUO2018807}, for metamodeling \cite{WANG2021104532}, for uncertainty propagation \cite{LEE2020112724, zeraatpisheh_bordas_beex_2021, STOWERS2021104340}, and for inverse problems \cite{10.3389/fphys.2018.01002}. However, their use to model constitutive relationships remains uncommon. To the authors' best knowledge, only \citet{frankel2020tensor} proposed using GPs for constitutive modeling of hyperelastic materials. However, their work was limited to isotropic materials and did not enforce any convexity constraints. 

Herein, we propose to treat the constitutive model as a stochastic process, more specifically a Gaussian process that allows us to directly incorporate the experimental data, capture the observed experimental variations in the stress-strain responses, and quantify the uncertainty through a natural Bayesian framework. Moreover, we propose a straightforward way to propagate the uncertainty through a nonlinear elasticity problem via stochastic finite element method. The remainder of this paper is organized as follows. In Section~\ref{sec:methods}, we delineate the development of a GP-based constitutive model and how convexity can be enforced in this framework. In Section~\ref{sec:test}, we verify the formulation based on an artificial dataset. In Section~\ref{sec:result}, we apply the proposed framework to a real experimental dataset of planar soft tissue. In Section~\ref{sec:stochastic}, we present the stochastic finite element analysis framework and the results obtained using the framework for analysing valve leaflet closure under static follower pressure load. Lastly, in Section~\ref{sec:discuss}, we discuss the advantages of this proposed framework and compare it to other approaches in the literature, followed by some concluding remarks.


\section{Methods\label{sec:methods}}
\subsection{Nonlinear elasticity}
\label{nonlinear-problem-section}
Given a domain $\Omega\subset\mathbb{R}^d$, a (static) nonlinear elasticity problem involves finding the deformation mapping, i.e., a map from undeformed (also called reference) to deformed positions $\varphi:\vec{X}\rightarrow\vec{x}$ over the domain $\Omega$, such that it satisfies the mechanical equilibrium \cite{holzapfel-book}
\begin{equation}
    \nabla_{\vec{X}}\cdot \mathbf{P} + \vec{B} = \vec{0},\label{equilibrium}
\end{equation}
where $\mathbf{P}$ is the 1st Piola-Kirchhoff (PK) stress tensor, $\nabla_{\vec{X}}\cdot\mathbf{P}$ denotes the divergence of $\mathbf{P}$ with respect to the undeformed configuration $\vec{X}$, and $\vec{B}$ is the applied body force per unit undeformed volume. 

The equilibrium equation \eqref{equilibrium} is completed with boundary conditions on domain boundary $\partial\Omega$. Assume $x_{i^d}$ denotes the $i^d$-th component of $\vec{x}$, where $i^d\in\left\{1,\dots,d\right\}$. To denote the boundary conditions in the $i^d$-th component, the boundary $\partial\Omega$ is categorized into two types: Dirichlet boundary $\Omega_D^{i^d}$ and Neumann boundary $\Omega_N^{i^d}$, such that  $\partial\Omega_D^{i^d}\cup\partial\Omega_N^{i^d}=\partial\Omega$ and $\partial\Omega_D^{i^d}\cap\partial\Omega_N^{i^d}=\emptyset$ $\forall i^d$. Thus, the boundary conditions can be expressed as 
\begin{subequations}\label{boundary}
\begin{align}
x_{i^d} &= \bar{x}_{i^d} \text{ on } \vec{X}\in\partial\Omega_D^{i^d} \text{ and}\\
P_{i^dj^d}N_{j^d} &= \bar{t}_{i^d} \text{ on } \vec{X}\in\partial\Omega_N^{i^d},
\end{align}
\end{subequations}
where $\bar{x}_{i^d}$ is the prescribed position on the Dirichlet boundary $\partial\Omega_D^{i^d}$, $\bar{t}_{i^d}$ is the prescribed traction on the Neumann boundary $\partial\Omega_N^{i^d}$ with surface normal $\vec{N}$ in the undeformed configuration, and summation is implied on repeated indices. 

Following the standard definitions \cite{bonet1997nonlinear,holzapfel-book,belytschko2013nonlinear}, the deformation gradient is $\mathbf{F}=\nabla_{\vec{X}}\varphi=\partial \vec{x}/\partial \vec{X}$ and the right Cauchy-Green deformation tensor is $\mathbf{C}=\mathbf{F}^{\top}\mathbf{F}$ with three isotropic invariants 
\begin{subequations}\label{invariants}
\begin{align}
{I}_1 &:= \tr \left(\mathbf{C}  \right) ,  \\ 
{I}_2 &:= \frac{1}{2} \left[ \tr^2 \left( \mathbf{C} \right) - \tr \left(\mathbf{C}^2  \right) \right] \text{ and}  \\
J &:= \sqrt{\det \left(\mathbf{C} \right)}.
\end{align}
\end{subequations}
Additional pseudo-invariants have been defined for anisotropic materials. A commonly used invariant for modeling single-fiber anisotropy is \cite{holzapfel2000new}
\begin{equation}
    I_4:=\vec{M}\cdot\mathbf{C}\vec{M},\label{I4-def}
\end{equation}
which is also equal to the square of the stretch along the preferred fiber direction $\vec{M}$.

\subsection{Constitutive models, frame invariance, and poly-convexity}
In order to close the governing system of equations, a relationship between stress and deformation (strain) needs to be defined through a constitutive model. In hyperelasticity, the constitutive model is described using a strain energy density function (SEDF) $\sedf(\mathbf{F})$ from which stresses are derived through differentiation \cite{holzapfel-book}. Specifically, for a compressible material, the first PK stress is $\mathbf{P}=\partial \sedf/\partial \mathbf{F}$ and the Cauchy stress is $\boldsymbol{\sigma}=J^{-1}\mathbf{P}\mathbf{F}^{\top}$. For an incompressible material, a constraint $J=1$ is imposed by adding a Lagrange multiplier term. Thus, for an incompressible material, $\mathbf{P} = \partial \sedf/\partial \mathbf{F} - p \mathbf{F}^{-\top}$ and $\boldsymbol{\sigma} = \mathbf{P}\mathbf{F}^{\top}-p\mathbf{I}$, where $p$ is the hydrostatic pressure acting as the Lagrange multiplier and $\mathbf{I}$ is an identity tensor. 

A constitutive model must satisfy certain properties in order to ensure a unique solution of the elasticity problem. Specifically, a model must be invariant with respect to rigid body rotation, which means that the SEDF is a function of the right Cauchy-Green deformation tensor $\mathbf{C}$. Moreover, a model must be invariant with respect to material symmetry. This implies that, for isotropic materials, $\sedf$ must be a function of the three isotropic invariants of $\mathbf{C}$: $I_1$, $I_2$, and $J$ defined in Eq.~\eqref{invariants}. For anisotropic materials, the list needs to be expanded to include the pseudo-invariants that account for the material directions, such as $I_4$ defined in Eq.~\eqref{I4-def}. Herein, we focus on planar soft tissues, which are nearly incompressible (i.e., $J$ is constrained to be equal to 1) and have a single preferred fiber direction $\vec{M}$. Thus, we restrict our focus to solids where the SEDF $\sedf$ is a function of $I_1$ and $I_4$. While some authors have demonstrated the need to  include other pseudo-invariants while modeling biological tissues \cite{MURPHY201390}, most of the existing hyperelastic models for planar biological tissues are only formulated based on $I_1$ and $I_4$ \cite{maurel1998}. Consequently, the 1st PK stress can be written as
\begin{equation}
    \mathbf{P} = \pderiv{\sedf}{\mathbf{F}} - p\mathbf{F}^{-\top} = 2\psedf{1}\,\mathbf{F} + 2\psedf{4}\,\mathbf{F}\vec{M}\otimes\vec{M} - p\mathbf{F}^{-\top},
\end{equation}
where a short-hand notation $(\cdot)_{,i}:=\partial (\cdot)/\partial I_i$ is used for partial derivatives with respect to the invariants. Similar notation is adopted for higher order derivatives, such as $(\cdot)_{,ij}:=\partial^2(\cdot)/\partial I_i \partial I_j$ etc. 

Another important property that the SEDF $\sedf$ must satisfy is convexity \cite{SCHRODER20054352, https://doi.org/10.1002/pamm.200510099,schroder2010anisotropie}, which ensures ellipticity of the governing equations that in turn guarantees the existence and uniqueness of a solution to the problem of nonlinear elasticity. In our case, this means that the second derivatives of $\sedf$ with respect to $I_1$ and $I_4$ are always positive, i.e.
\begin{subequations}\label{convexity-conditions}
\begin{align}
    \psedf{11} &\ge 0\text{ and} \label{conv1}\\
    \psedf{44} &\ge 0, \label{conv2}
\end{align}
as well as the determinant of the Hessian is positive, i.e.,
\begin{equation}
    \psedf{11}\psedf{44}-\psedf{14}^2 \ge 0.
\end{equation}
\end{subequations}
\begin{rmk}
Several definitions of convexity have been proposed in literature, with full convexity \eqref{convexity-conditions} being the strongest condition and a rank-one convexity being the fundamental requirement \cite{dacorogna2008polyconvex}. For constitutive models dependent only on $I_1$ and $I_4$, some of the convexity definitions become equivalent. Thus, for simplicity, we seek to enforce convexity \emph{weakly} through Eqs.~\eqref{conv1} and \eqref{conv2} only. That is, the positivity of determinant is not enforced. However, if desired, it is possible to enforce the full convexity using the proposed framework.
\end{rmk}

We note that for the incompressible case, $I_1\ge3$ always, while $I_4>0$. Lastly, when the material is undeformed, i.e., $\mathbf{F}=\mathbf{I}$, $I_1=3$ and $I_4=1$. As a short-hand notation, a point in the $I_1$-$I_4$ space is denoted as $\point:=(I_1,I_4)$, and therefore, we write the SEDF as a function of $\point$, i.e., $\sedf(\point)$. To denote a set/vector of points in the $I_1$-$I_4$ space, $\observepoints$ is used.


\subsection{Probability notation}
A random scalar variable is denoted as $\rand{u}$ and its realization is denoted as $u\in\mathbb{R}$. To denote a higher dimensional random variable, $\vecrand{u}$ is used, and its realization is denoted as $\vec{u}\in\mathbb{R}^n$. The probability density function (PDF) of random variable $\vecrand{u}$ is denoted as $\prob_{\vecrand{u}}(\vec{u})$; the PDF indicates that the probability of $\vecrand{u}$ realizing a value in the neighborhood of $\vec{u}$ is given by $\prob_{\vecrand{u}}(\vec{u})\mathrm{d}U$, where $\mathrm{d}U$ is the volume of the infinitesimal neighborhood around $\vec{u}$ in $\mathbb{R}^n$. Moreover, the probability given some information (or data) $\mathcal{I}$ is written as $\prob_{\vecrand{u}}(\vec{u} \mid \mathcal{I})$. Two commonly used measures of a random variable are its mean (or expected) valued vector of length $n$
\begin{equation}
    \mathbb{E}({\vecrand{u}}) := \int_{\mathbb{R}^n} \vec{u}\,\prob_{\vecrand{u}}(\vec{u})\mathrm{d}U,
\end{equation}
and a positive semi-definite covariance matrix of dimension $n\times n$
\begin{equation}
    \mathbb{V}({\vecrand{u}}) := \int_{\mathbb{R}^n} \left[\vec{u} - \mathbb{E}({\vecrand{u}}) \right]\otimes\left[\vec{u} - \mathbb{E}({\vecrand{u}}) \right] \,\prob_{\vecrand{u}}(\vec{u})\mathrm{d}U.
\end{equation}

A Gaussian, also called normal, probability distribution is fully described in terms of the mean vector and covariance matrix. Specifically, for a normally distributed random variable $\vecrand{u}$ with mean $\vec{\mu}=\mathbb{E}({\vecrand{u}})$ and co-variance matrix $\mathbf{\Sigma}=\mathbb{V}({\vecrand{u}})$, its probability density function is given by
\begin{equation}
    \prob_{\vecrand{u}}(\vec{u}\mid\vec{\mu},\mathbf{\Sigma}) = \dfrac{1}{\sqrt{(2\pi)^n\det(\mathbf{\Sigma})}} \exp\left[ -\dfrac{1}{2} [\vec{u}-\vec{\mu}]\cdot\mathbf{\Sigma}^{-1}[\vec{u}-\vec{\mu}] \right]. \label{pdf-normal}
\end{equation}
The following short-hand is used to denote a normally distributed random variable: 
\begin{equation}
    \vecrand{u} \sim \mathcal{N}\left( \vec{\mu}, \mathbf{\Sigma} \right).
\end{equation}
A slightly abusive short-hand to denote the PDF \eqref{pdf-normal} evaluated at a general point $\vec{u}$ is also adopted, i.e.,
\begin{equation}
    \prob_{\vecrand{u}}(\vec{u} \mid \vec{\mu}, \mathbf{\Sigma}) = \mathcal{N}(\vec{u} \mid \vec{\mu}, \mathbf{\Sigma}).
\end{equation}
A normally distributed scalar variable with a zero mean and unit variance follows a PDF known as the standard normal distribution function, and it is denoted as
\begin{equation}
    \phi(u) := \dfrac{\exp\left[{-u^2/2}\right]}{\sqrt{2\pi}}.\label{standard-normal}
\end{equation}
Lastly, the cumulative distribution function (CDF) of the standard normal distribution function \eqref{standard-normal} is denoted as
\begin{equation}
    \Phi(u) := \int\limits_{-\infty}^u \phi(t)\,\mathrm{d}t.\label{CDF}
\end{equation}
We note that the above function \eqref{CDF} maps real numbers to a finite set, $\Phi:(-\infty,\infty)\rightarrow[0,1]$. To simplify the notation of probability, it is common to skip the subscript when writing the PDF, i.e., $\prob(\vec{u})$ is used instead of $\prob_{\vecrand{u}}(\vec{u})$. Thus, from here on, we will follow this slightly abusive but simpler notation.

\subsection{Bayes' theorem}
For two continuous random (scalar) variables $\rand{u}$ and $\rand{v}$, let the joint prior probability density function be denoted by $\prob(u,v)$. Further, the prior marginal probability densities of $\rand{u}$ and $\rand{v}$ are denoted as $\prob(u)$ and $\prob(v)$, respectively. The posterior probability density of $\rand{u}$ given $\rand{v}=v$ (known as the conditional probability) is given by the Bayes' theorem:
\begin{equation*}
    \overbrace{\prob(u \mid v)}^{\text{Posterior}} = \frac{\overbrace{\prob(v \mid u)}^{\text{Likelihood}} \overbrace{\prob(u)}^{\text{Prior}}}{\underbrace{\prob(v)}_{\text{Normalization term}}},
\end{equation*}
where $\prob(v\mid u)$ is the likelihood term. The denominator on the right-hand side is also the normalization term, i.e.,:
\begin{equation*}
    \prob(v) = \int \prob(u, v) \, \mathrm{d}u = \int \prob(v \mid u) \prob(u) \, \mathrm{d}u.
\end{equation*}

\begin{figure}[h!]
    \centering
    \includegraphics{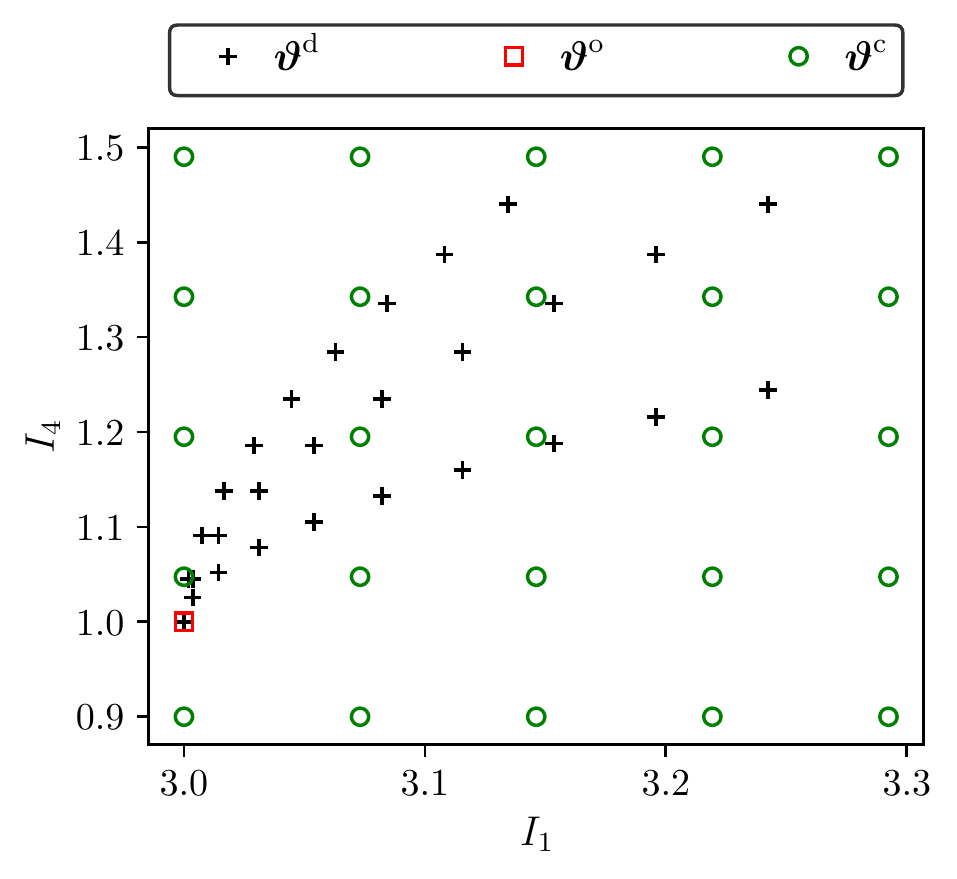}
    \caption{Three types of inputs are used to train the GP model: stress-strain measurements (related to derivatives of SEDF) at $\observepoints^{\text{d}}$, reference value of SEDF at $\observepoints^{\text{o}}$, and convexity constraints (related to second derivatives of SEDF) at $\observepoints^{\text{c}}$.}
    \label{point-defs}
\end{figure}

In the present work, we model SEDF $\sedf$ as \emph{a random process}. In order to find the PDF of $\sedf$, we use Bayes' theorem to incorporate three types of data/information, and write the posterior probability density of $\sedf$ as:
\begin{equation}
\prob(\sedf \mid \observestate,\observepoints)
= 
\dfrac{
\overbrace{\prob(\observestate^{\text{o}}\mid \observepoints^{\text{o}}, \sedf) 
\prob(\observestate^{\text{d}}\mid \observepoints^{\text{d}}, \sedf)
\prob(\mathbf{c} \mid \observepoints^{\text{c}}, \sedf)}^{\text{Three likelihood terms}}
\prob(\sedf)
}
{\prob(\observestate^{\text{o}}, \observestate^{\text{d}}, \mathbf{c} \mid \observepoints^{\text{o}}, \observepoints^{\text{d}},\observepoints^{\text{c}})},
\label{bayes2}
\end{equation}
where $\observestate^{\text{o}}$ are observations related to the original function $\sedf$ at $\observepoints^{\text{o}}$, $\observestate^{\text{d}}$ are observations related to the derivatives of $\sedf$ at $\observepoints^{\text{d}}$, $\mathbf{c}$ are constraints related to second derivatives of $\sedf$ at $\observepoints^{\text{c}}$,  $\observestate=\observestate^{\text{o}\,}\cup\,\observestate^{\text{d}}\,\cup\,\mathbf{c}$, and $\observepoints=\observepoints^{\text{o}}\,\cup\,\observepoints^{\text{d}}\,\cup\,\observepoints^{\text{c}}$. The three types of data/information are summarized in Table~\ref{summary-table} and depicted in Fig.~\ref{point-defs}. Next we describe the three terms one by one, starting with the observations related to the derivatives that come from experiments.

\begin{table}[h!]
\centering
\caption{A summary of the observations (or constraints) at locations in the $\point$-space and the associated likelihood functions required to predict the posterior PDF of $\sedf$ \label{summary-table}}
\small{
\begin{tabular}{|c|c|c|c|}
\hline
\textbf{Type} & \textbf{Locations (counter)} & \textbf{Observations} & \begin{tabular}[c]{@{}c@{}}\textbf{Likelihood,} \\ \textbf{Hyperparameters}\end{tabular} \\ \hline\hline
\begin{tabular}[c]{@{}c@{}}Experimental \\ measurements\end{tabular} & $\observepoints^{\text{d}}$ ($i^{\text{d}}=1,\dots,N^{\text{d}}$)  & $\observestate^\text{d}$ & Eq.~\eqref{observe-llh}, $e_x^2$, $e_y^2$        \\ \hline
\begin{tabular}[c]{@{}c@{}}Reference \\ point\end{tabular}           & $\observepoints^{\text{o}}$ ($i^{\text{o}}=1$)                     & $\observestate^\text{o}$ & Eq.~\eqref{W-llh}, $e_0^2$                       \\ \hline
\begin{tabular}[c]{@{}c@{}}Convexity \\ constraints\end{tabular}     & $\observepoints^{\text{c}}$  ($i^{\text{c}}=1,\dots,N^{\text{c}}$) & $\mathbf{c}$             & Eq.~\eqref{probit-llh}, $\nu$ \\ \hline
\end{tabular}
}
\end{table}
\subsection{Experimental observations}
\begin{figure}[h!]
\centering
\includegraphics[width=\textwidth]{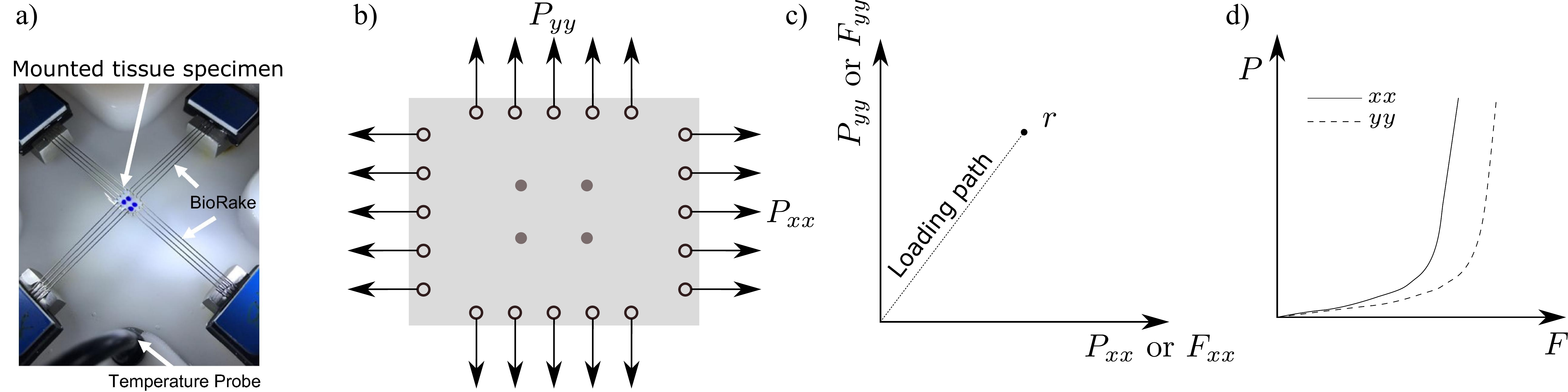}
\caption{Biaxial stretching is a commonly used ex-vivo experiment for thin planar soft tissues: a) a biaxial stretcher using BioRakes to mount the tissue specimen; b) a schematic of the tissue sample under biaxial stresses along two axes aligned with $x-$ and $y-$axes and fiducial markers to measure strain; c) one protocol is defined as a loading path in the stress or deformation space; d) the resulting stress-deformation curves along the two directions. \label{exp}}
\end{figure}

Constitutive models are empirical relationships that are based on experimental observations. For planar soft tissues, a common experiment is biaxial stretching, where a rectangular sample of the tissue is stretched in two orthogonal directions, with the fiber direction commonly aligned with one of the two directions (Fig.~\ref{exp}). If Cartesian coordinates are aligned with the sample edges, the applied forces $\observe{f}_x$ and $\observe{f}_y$ in $x$- and $y$-directions are converted into averaged components of 1$^{\text{st}}$ PK stresses $\observe{P}_{xx}=\observe{f}_x/L_yt$ and $\observe{P}_{yy}=\observe{f}_y/L_xt$, where $L_x$ and $L_y$ are the dimensions of the rectangular sample and $t$ is its thickness in the undeformed configuration. Here, the notation $\observe{\cdot}$ is used to denote quantities that are experimentally observed and therefore may contain observation noise. 

In biaxial stretching experiments, the deformation is tracked via fiducial markers on the sample and the applied deformation gradient $\mathbf{F}$ is derived using bilinear finite element shape functions \cite{billiar2000biaxial}. Generally, the shear components are difficult to control with the BioRake tissue mounting and therefore neglected. Thus, based on incompressibility, we have
\begin{equation}
    \mathbf{F} = \diag\left[ \lambda_x, \lambda_y, \frac{1}{\lambda_x\lambda_y}\right],
\end{equation}
where $\lambda_x$ and $\lambda_y$ are the ratios of sample dimensions in deformed and undeformed configurations (i.e., stretches) along the $x$- and $y$-directions, respectively. Generally, a straight line in the deformation or stress space is followed along which multiple points are recorded, which gives us the stress-stretch curves. One loading path (denoted using the symbol $r=F_{xx}/F_{yy}$ or $r=P_{xx}/P_{yy}$) is known as one protocol, and multiple protocols are combined to create a set of the experimental observations $\left\{\lambda_{x}^{i^\text{d}}, \lambda_{y}^{i^\text{d}}, \observe{P}_{xx}^{i^\text{d}}, \observe{P}_{yy}^{i^\text{d}} \right\}_{{i^\text{d}}=1}^{N^\text{d}}$, where $N^\text{d}$ is the total number of observations. 

In the experiments, it is a common practice to align the fiber direction with the $x-$axis, i.e., $\vec{M}=\left[1,0,0 \right]$. Thus, the observed stretches can be transformed into deformation invariants as 
\begin{subequations}\label{invariants-exp}
\begin{align}
    I_1^{i^\text{d}} &= \left(\lambda_{x}^{i^\text{d}}\right)^2 + \left(\lambda_{y}^{i^\text{d}}\right)^2 + \left(\frac{1}{\lambda_{x}^{i^\text{d}}\lambda_y^{i^\text{d}}}\right)^2 \\
    I_4^{i^\text{d}} &= \left(\lambda_{x}^{i^\text{d}}\right)^2.
\end{align}
\end{subequations}
Furthermore, since the out-of-plane stress is zero, i.e. $P_{zz}=0$, it is  used to determine the hydrostatic pressure $p=2\psedf1/\left(\lambda_{x}^{i^\text{d}}\lambda_y^{i^\text{d}}\right)^2$ \cite{FAN20142043, KIENDL2015280}. Subsequently the model stresses can be written in terms of the derivatives of $\sedf$ as:
\begin{subequations}\label{model-stresses}
\begin{align}
{P}_{xx}^{i^\text{d}} &= 2\psedf1(I_1^{i^\text{d}},I_4^{i^\text{d}}) \left[ \lambda_x^{i^\text{d}}  - \frac{1}{\left(\lambda_{x}^{i^\text{d}}\right)^3\left(\lambda_{y}^{i^\text{d}}\right)^2} \right] + 
2\psedf4(I_1^{i^\text{d}},I_4^{i^\text{d}}) \lambda_x^i\\
{P}_{yy}^{i^\text{d}} &= 2\psedf1(I_1^{i^\text{d}},I_4^{i^\text{d}}) \left[ \lambda_y^{i^\text{d}}  - \frac{1}{\left(\lambda_{x}^{i^\text{d}}\right)^2\left(\lambda_{y}^{i^\text{d}}\right)^3} \right].
\end{align}
\end{subequations}
These model stresses differ from the observed stresses by observation errors,
\begin{subequations}\label{stresses-relate}
\begin{align}
    \observe{P}_{xx}^{i^\text{d}} &= P_{xx}^{i^\text{d}} + \error_x \\
    \observe{P}_{yy}^{i^\text{d}} &= P_{yy}^{i^\text{d}} + \error_y ,
\end{align}
\end{subequations}
where $\error_x$ and $\error_y$ are experimental noises in the two measurements, and are assumed to be independent, uniform (i.e., same for all $i^{\text{d}}$), and zero-mean Gaussian. That is, $\error_x \sim \mathcal{N}(0,e_x^2)$ and $\error_y \sim \mathcal{N}(0,e_y^2)$, with $e_x^2$ and $e_y^2$ being two hyperparameters to be determined. Thus, the observation points and observations are denoted as $\observepoints^\text{d}=\left\{(I_1^{i^\text{d}},I_4^{i^\text{d}}) \right\}_{{i^\text{d}}=1}^{N^\text{d}}$ and $\observestate^\text{d}=\observestate^\text{d}_x \cup \observestate^\text{d}_y = \left\{\observe{P}_{xx}^{i^\text{d}} \right\}_{{i^\text{d}}=1}^{N^\text{d}} \cup \left\{\observe{P}_{yy}^{i^\text{d}} \right\}_{{i^\text{d}}=1}^{N^\text{d}}$, respectively, where superscript $^\text{d}$ denotes the observations related to derivatives. In other words, for a given model $\sedf(\point)$, the likelihood of the observed stresses is given by:
\begin{equation}
    \prob(\observestate^\text{d} \mid \observepoints^\text{d}, \sedf) = \prod\limits_{{i^\text{d}}=1}^{N^\text{d}} \mathcal{N} (\observe{P}_{xx}^{i^\text{d}} -P_{xx}^{i^\text{d}}\mid 0, e_{x}^2 ) \, \mathcal{N} (\observe{P}_{yy}^{i^\text{d}} -P_{yy}^{i^\text{d}}\mid 0, e_{y}^2 ),\label{observe-llh}
\end{equation}
where $P_{xx}^{i^\text{d}}$ and $P_{yy}^{i^\text{d}}$ are derived from $\sedf$ using Eq.~\eqref{model-stresses}.

\subsection{Reference point}
There are no direct observations on $\sedf$, thus making it arbitrary to an additive constant. However, customarily, $\sedf$ is set to be null at the reference configuration. That is, at $\observepoints^\text{o}=\left\{(3,1) \right\}$, $\observestate^{\text{o}}=\left\{0\right\}$, where the superscript $^\text{o}$ indicates observations related to the original function. Note that there is no error associated with this observation. In other words, the likelihood function is a Dirac delta function, which, in practice, is implemented by using a normal distribution with a fixed small variance $e_0^2=10^{-5}$:
\begin{equation}
    \prob(\observestate^\text{o} \mid \observepoints^\text{o}, \sedf) = \mathcal{N} (\sedf(\observepoints^\text{o})\mid 0, e_{0}^2 ).\label{W-llh}
\end{equation}

\subsection{Convexity constraints}
Using only the experimental observations, the resulting SEDF can have negative second derivatives, thus violating the convexity requirement \eqref{convexity-conditions}. To resolve this issue, we propose a technique based on the monotonic GPs developed by \citet{riihimaki2010gaussian}. In their work, \citeauthor{riihimaki2010gaussian} enforced monotonicity at a finite number of locations by constraining the first derivatives to be positive through a likelihood function based on the CDF \eqref{CDF}. 

\begin{figure}[h!]
\centering
\includegraphics[width=0.5\textwidth]{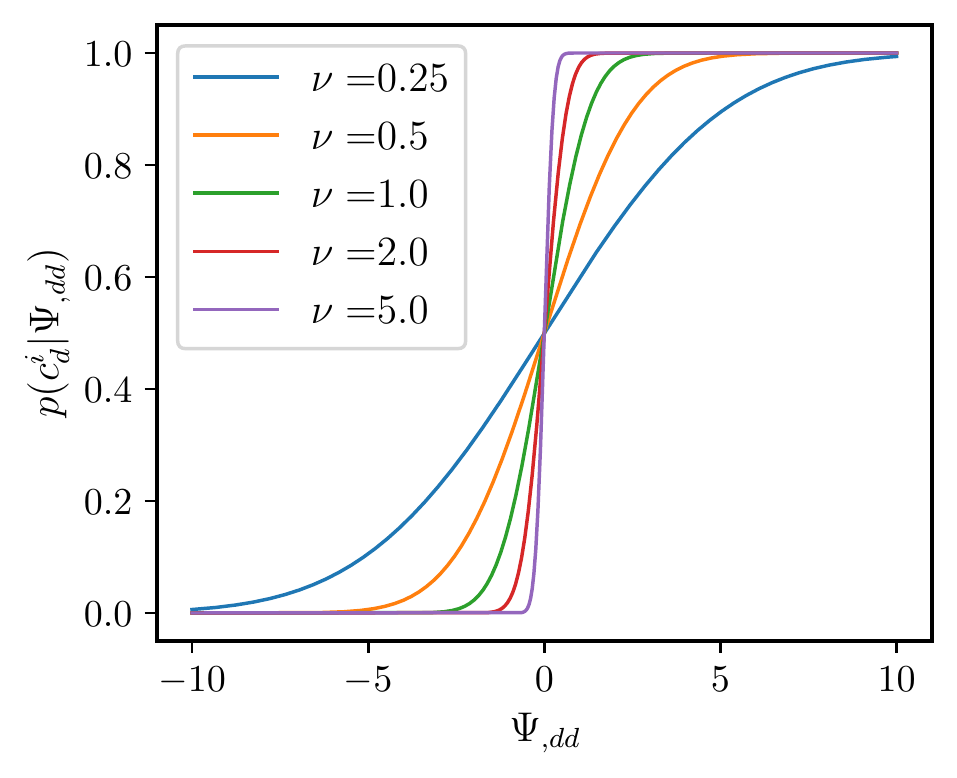}
\caption{Likelihood function based on the CDF \eqref{CDF} used to enforce convexity constraints \eqref{constraint-eq}.}\label{fig:probit}
\end{figure}

Equivalently, herein we enforce the convexity constraints at a finite number of locations in the $\point$-space, denoted as $\point^{(i^{\text{c}})}$, $i^{\text{c}}=1,\dots, N^{c}$. Thus, the constraints $c^{i^{\text{c}}}_d$ are denoted as:
\begin{equation}
    c^{i^{\text{c}}}_d: \psedf{dd}(\point^{i^{\text{c}}}) \ge 0, \;\; d=1,4.\label{constraint-eq}
\end{equation}
From a probabilistic perspective, the above constraints can be viewed as follows. The likelihood of negative second derivatives is zero, while the likelihood of positive second derivatives is non-zero but uniform (i.e., all positive second-derivatives are equally likely). To approximate such a likelihood function, following \cite{riihimaki2010gaussian}, the CDF \eqref{CDF} is adopted, i.e.,
\begin{equation}
    \prob\left(c^{i^{\text{c}}}_d \mid \point^{i^{\text{c}}},\sedf \right) \propto \Phi \left( \nu \psedf{dd}(\point^{i^{\text{c}}}) \right),\; i^{\text{c}}=1,\dots,N^{\text{c}},\,d=1,4. \label{probit-func} 
\end{equation}
 Although the above likelihood function tolerates small violations of the constraints for finite values of $\nu$, it approaches the desired step function (Eq.~\ref{constraint-eq}) when $\nu\rightarrow\infty$ (Fig.~\ref{fig:probit}). The location of constraints, $\point^{i^{\text{c}}}$, are chosen to be equally spaced in a rectangular subspace  $(I_1,I_4)\in[3,I_1^{\text{max}}]\otimes[I_4^{\text{min}},I_4^{\text{max}}]$, where $I_1^{\text{max}}$, $I_4^{\text{min}}$ and $I_4^{\text{max}}$ are chosen based on the target range of predictive deformation. Thus, the likelihood of all constraints combined $\mathbf{c}$ is written as
\begin{equation}
    \prob\left(\mathbf{c} \mid \observepoints^{\text{c}},\sedf \right) = \frac{1}{Z} \prod\limits_{i^{\text{c}}=1}^{N^{\text{c}}} \Phi \left( \nu \psedf{11}(\point^{i^{\text{c}}}) \right)\Phi \left( \nu \psedf{44}(\point^{i^{\text{c}}}) \right), \label{probit-llh} 
\end{equation}
where $Z$ is a normalisation factor.

\subsection{Gaussian processes}
Now that the three likelihood terms in \eqref{bayes2} have been defined, a prior probability distribution of SEDF, $\prob(\sedf)$, is required. In this study, we propose to model the SEDF prior as a Gaussian process (GP), which can be viewed as a generalization of a multivariate normal probability distribution to functions. More specifically, the strain energy density is a GP dependent on $\point$ and its prior distribution is denoted as \cite{books/lib/RasmussenW06}
\begin{equation}
    \rand{\sedf}(\point) \sim \gp\left(m(\point),k(\point,\point')\right),\label{gp-prior}
\end{equation}
where the prior mean $m(\point)$ and covariance $k(\point,\point')$ functions are defined as
\begin{subequations}\label{prior-mean-cov}
\begin{align}
    m(\point) &= \mathbb{E}\left(\rand{\sedf}(\point) \right) \text{ and}\\
    k(\point,\point') &= \text{Cov}\left(\rand{\sedf}(\point),\rand{\sedf}(\point') \right) = \mathbb{E}\left(\left(\rand{\sedf}(\point)-m(\point)\right)\left(\rand{\sedf}(\point')-m(\point')\right) \right). \label{kernel-func}
\end{align}
\end{subequations}
While a zero-mean prior is common in the literature, a linear function (i.e., a linear hyperelastic model) is used here:
\begin{equation}
    m(\point) = \alpha(I_1-3) + \beta(I_4-1).
\end{equation}
On the other hand, various options have been proposed for the covariance function \cite{books/lib/RasmussenW06}; the most commonly used squared exponential covariance function (also called the radial basis kernel) is adopted here
\begin{equation}
    k(\point,\point') = \sigma_f^2\exp\left(-\frac{1}{2\scriptr^2}\|\point-\point'\|^2\right).\label{kernel}
\end{equation}
The above kernel choice gives a stationary and infinitely differentiable GP \cite{books/lib/RasmussenW06}. $\alpha$, $\beta$, $\sigma_f$ and $\scriptr$ used in the definitions above are the hyperparameters of the GP prior. 

Since differentiation is a linear operator, the derivatives of a GP are also GPs as long as the kernel is differentiable \cite{books/lib/RasmussenW06}. More specifically, the mean of the derivative is equal to the derivative of the mean. Therefore, the  mean of the combined vector of $\rand{\sedf}$ and its derivatives is a vector function given by
\begin{equation}
    \cmean (\point) := \mathbb{E}\begin{bmatrix}
    \rand{\sedf} \\
    \rand{\psedf1}\\
    \rand{\psedf4}\\
    \rand{\psedf{11}}\\
    \rand{\psedf{44}}\
    \end{bmatrix}
    =
    \begin{bmatrix}
    \alpha(I_1-3)+\beta(I_4-1)\\
    \alpha\\
    \beta\\
    0\\
    0
    \end{bmatrix},\label{combined-mean}
\end{equation}
and its full covariance matrix can be written as:
\begin{equation}
    \ccov(\point,\point') := \text{Cov}\begin{bmatrix}
    \rand{\sedf},\rand{\sedf'} & \rand{\sedf},\rand{\psedf1'} & \rand{\sedf},\rand{\psedf4'} & \rand{\sedf},\rand{\psedf{11}'} & \rand{\sedf},\rand{\psedf{44}'} \\
    \rand{\psedf1},\rand{\sedf'} & \rand{\psedf1},\rand{\psedf1'} & \rand{\psedf1},\rand{\psedf4'} & \rand{\psedf1},\rand{\psedf{11}'} & \rand{\psedf1},\rand{\psedf{44}'} \\
    \rand{\psedf4},\rand{\sedf'} & \rand{\psedf4},\rand{\psedf1'} & \rand{\psedf4},\rand{\psedf4'} & \rand{\psedf4},\rand{\psedf{11}'} & \rand{\psedf4},\rand{\psedf{44}'} \\
    \rand{\psedf{11}},\rand{\sedf'} & \rand{\psedf{11}},\rand{\psedf1'} & \rand{\psedf{11}},\rand{\psedf4'} & \rand{\psedf{11}},\rand{\psedf{11}'} & \rand{\psedf{11}},\rand{\psedf{44}'} \\
    \rand{\psedf{44}},\rand{\sedf'} & \rand{\psedf{44}},\rand{\psedf1'} & \rand{\psedf{44}},\rand{\psedf4'} & \rand{\psedf{44}},\rand{\psedf{44}'} & \rand{\psedf{44}},\rand{\psedf{44}'} \\
    \end{bmatrix} \label{combined-cov}
\end{equation}
Here, $\sedf'$ is the SEDF evaluated at $\point'$. Similar to the mean, the covariance between the function and its derivatives can be derived by differentiating the covariance function (Eq.~\ref{kernel-func}). Using the symmetry property of the squared exponential kernel function (Eq.~\ref{kernel}) about its two arguments, it is easy to see that 
\begin{equation}
    \ccov(\point,\point') = \begin{bmatrix}
     k & \dk1    & \dk4 & \dk{11} & \dk{44} \\
       & \dk{11} & \dk{14} & \dk{111} & \dk{144} \\
       &         & \dk{44} & \dk{411} & \dk{444} \\
       &         &         & \dk{1111} & \dk{1144} \\
     \text{Sym.}  &         &         &           & \dk{4444} \\
     \end{bmatrix},\label{combined-cov2}
\end{equation}
where the short-hand notation for partial derivatives has been extended to the kernel function and all terms are evaluated at $(\point,\point')$. The combined state of $\sedf$ and its derivatives is denoted as $\cgp$, and described as the following joint GP:
\begin{equation}
    \rand{\cgp} := \left[\rand{\sedf},\rand{\psedf1},\rand{\psedf4},\rand{\psedf{11}},\rand{\psedf{44}}\right] \sim \gp (\cmean(\point),\ccov(\point,\point')).\label{cgp}
\end{equation}
Based on this prior, the distribution of $\rand{\sedf}$ and its derivatives at all observation points $\observepoints$, is denoted as $\vecrand{f}$:
\begin{equation}
    \vecrand{f} \sim \mathcal{N}\left(\cmean(\observepoints),\ccov(\observepoints,\observepoints)\right).\label{prior}
\end{equation}
The distribution of $\rand{\sedf}$ and its derivatives at desired prediction point (or a set of points, in general) $\predictpoints$ is denoted as $\rand{\predictstate}$, and follows the following joint prior distribution with $\vecrand{f}$:
\begin{equation}
    \begin{bmatrix}
    \vecrand{f}\\
    \rand{\predictstate}
    \end{bmatrix} \sim \mathcal{N} \left( 
    \begin{bmatrix}
    \cmean(\observepoints)\\
    \cmean(\predictpoints)
    \end{bmatrix},
    \begin{bmatrix}
    \ccov(\observepoints,\observepoints) & \ccov(\observepoints,\predictpoints)\\
    \ccov(\predictpoints,\observepoints) & 
    \ccov(\predictpoints,\predictpoints)
    \end{bmatrix}
    \right).\label{joint-distribution}
\end{equation}
Since the set $\observepoints$ has $(N^{\text{c}} + N^{\text{d}} + 1)$ points, the length of $\vecrand{f}$, which includes the SEDF, its two first derivatives and two second derivatives, is $5\times(N^{\text{c}} + N^{\text{d}} + 1)$. However, not all the terms in $\vecrand{f}$ are used in the three likelihood terms of Eq.~\eqref{bayes2}. For example, only the first derivatives are needed at $\observepoints^{\text{d}}$. Therefore, we extract the relevant parts of the mean vector and covariant matrix and denote them by $\extract{\cdot}^\text{o}$ for original function at $\observepoints^{\text{o}}$ only, $\extract{\cdot}^\text{d}$ for derivatives  at $\observepoints^{\text{d}}$ only, $\extract{\cdot}^\text{c}$ for second derivatives at $\observepoints^{\text{c}}$ only, and $\extract{\cdot}$ for all the above three. Thus,
\begin{align}
    \extract{\vecrand{f}} &\sim \mathcal{N} \left( \extract{\cmean(\observepoints)}, \extract{\ccov(\observepoints,\observepoints)} \right) \text{ and} \label{prior2}\\
    \begin{bmatrix}
    \extract{\vecrand{f}}\\
    \rand{\predictstate}
    \end{bmatrix} &\sim \mathcal{N} \left( 
    \begin{bmatrix}
    \extract{\cmean(\observepoints)}\\
    \cmean(\predictpoints)
    \end{bmatrix},
    \begin{bmatrix}
    \extract{\ccov(\observepoints,\observepoints)} & \extract{\ccov(\observepoints,\predictpoints)}\\
    \extract{\ccov(\predictpoints,\observepoints)} & 
    \ccov(\predictpoints,\predictpoints)
    \end{bmatrix}
    \right).\label{joint-distribution2}
\end{align}
From, Eq.~\eqref{joint-distribution2}, we get an expression for $\prob(\predictstate\mid\extract{\vec{f}})$
\begin{subequations}\label{fstart-givenf}
\begin{equation}
    \rand{\predictstate} \mid \extract{\vec{f}} \sim \mathcal{N} \left(\predictmean,\text{Cov}(\predictstate) \right),
\end{equation}
where
\begin{align}
    \predictmean &:= \extract{\ccov(\predictpoints,\observepoints)} \left[ \extract{\ccov(\observepoints,\observepoints)} \right]^{-1} \extract{\vec{f}} \text{ and} \nonumber\\
    \text{Cov}(\predictstate) &:= \ccov(\predictpoints,\predictpoints) - \extract{\ccov(\predictpoints,\observepoints)}\left[ \ccov(\observepoints,\observepoints) \right]^{-1} \extract{\ccov(\observepoints,\predictpoints)}.
\end{align}
\end{subequations}
Using Eq.~\eqref{prior2} we write the Bayes' theorem \eqref{bayes2} more explicitly as
\begin{equation}
\prob(\extract{\vec{f}} \mid \observestate,\observepoints)
= 
\dfrac{
{\prob(\observestate^{\text{o}}\mid \observepoints^{\text{o}}, \extract{\vec{f}}^{\text{o}}) 
\prob(\observestate^{\text{d}}\mid \observepoints^{\text{d}}, \extract{\vec{f}}^{\text{d}})
\prob(\mathbf{c} \mid \observepoints^{\text{c}}, \extract{\vec{f}}^{\text{c}})}
\prob(\extract{\vec{f}})
}
{\prob(\observestate^{\text{o}}, \observestate^{\text{d}}, \mathbf{c} \mid \observepoints^{\text{o}}, \observepoints^{\text{d}},\observepoints^{\text{c}})},
\label{bayes3}
\end{equation}
and, finally, we arrive at the probability distribution of predictions based on all the observations
\begin{equation}
    \prob(\predictstate\mid \observestate,\observepoints,\observepoints_*) = \int\limits_{\extract{\vec{f}}} \underbrace{\prob(\predictstate\mid\extract{\vec{f}})}_{\text{Eq. }\eqref{fstart-givenf}} \underbrace{\prob(\extract{\vec{f}} \mid \observestate,\observepoints)}_{\text{Eq. }\eqref{bayes3}} \diff\extract{\vec{f}}. \label{final-bayes}
\end{equation}
Out of the three likelihood terms, two are Gaussian, while the likelihood for constraints is non-Gaussian. In the absence of  constraints, a closed-form solution is available for Eq.~\eqref{bayes3} (called the exact GP, see \ref{sec:appendix}). However, the presence of non-Gaussian constraints requires an alternative approach that we describe next.

\subsection{Approximate GP}
When using non-Gaussian likelihood, such as the likelihood in Eq.~\eqref{probit-func}, a closed-form solution for the predicted mean and covariance is no longer possible. 
Markov Chain Monte Carlo (MCMC) is a commonly used approach to sample the posterior probability distribution in such cases. However, the latent variables of a Gaussian process are highly correlated, making convergence of MCMC extremely challenging to achieve using standard MCMC methods (see e.g. \cite{titsias_rattray_lawrence_2011}). An alternative and by now popular approach based on variational inference is adopted, which poses the problem in terms of an optimization problem to find an approximation to the posterior probability distribution \cite{Titsias2019, Hensman2013}.

To simplify the notation, we first rewrite Eq.~\eqref{bayes3} using short-hand notation
\begin{equation}
    \prob(\extract{\vec{f}}  \mid \observestate,\observepoints)
= \frac{\mathrm{L} \prob(\extract{\vec{f}})}{\mathrm{D}},\label{bayes5}
\end{equation}
where $\mathrm{L}:=\prob(\observestate^{\text{o}}\mid \observepoints^{\text{o}}, \extract{\vec{f}}^{\text{o}}) 
\prob(\observestate^{\text{d}}\mid \observepoints^{\text{d}}, \extract{\vec{f}}^{\text{d}})
\prob(\mathbf{c} \mid \observepoints^{\text{c}}, \extract{\vec{f}}^{\text{c}})$ is the combination of all likelihood terms and $\mathrm{D}:=\prob(\observestate^{\text{o}}, \observestate^{\text{d}}, \mathbf{c} \mid \observepoints^{\text{o}}, \observepoints^{\text{d}},\observepoints^{\text{c}})$ is denominator, also called the evidence. 
The approach of variational inference aims to find an approximation for the posterior, $\Q(\extract{\vec{f}}) \approx \prob(\extract{\vec{f}} \mid \observestate,\observepoints)$. The difference between the two probability distributions is quantified in terms of the Kullback--Leibler (KL) Divergence, denoted as
\begin{equation}
    {\mathrm{KL}} \left[ \Q(\extract{\vec{f}}) \;\|\; \prob(\extract{\vec{f}} \mid \observestate,\observepoints) \right] := \int\limits_{\extract{\vec{f}}} \Q(\extract{\vec{f}}) \log\left( \frac{\Q(\extract{\vec{f}})}{\prob(\extract{\vec{f}} \mid \observestate,\observepoints)} \right) \,\diff\extract{\vec{f}}.
\end{equation}
Our aim is to find $\Q$ that minimizes its KL divergence from the true posterior. When expanded using Eq.~\eqref{bayes5}, we get
\begin{equation}
    {\mathrm{KL}} \left[ \Q(\extract{\vec{f}}) \;\|\; \prob(\extract{\vec{f}} \mid \observestate,\observepoints) \right] = \log(\mathrm{D}) - \L.\label{Klequal}
\end{equation}
where
\begin{equation}
    \L \triangleq \int\limits_{\extract{\vec{f}}} \Q(\extract{\vec{f}}) \log\left( \frac{\mathrm{L}\prob(\extract{\vec{f}})}{\Q(\extract{\vec{f}})} \right)\,\diff\extract{\vec{f}} \label{ELBO}
\end{equation}
is defined as the loss function. Since $\mathrm{D}$ does not depend on $\extract{\vec{f}}$ (i.e., a constant) and the KL divergence is non-negative, Eq.~\eqref{Klequal} implies that $\log(\mathrm{D}) \ge \L$. Thus, the loss function \eqref{ELBO} is called the evidence lower bound (ELBO), and if one maximizes the ELBO, the KL divergence is minimized. 

Next, the approximate distribution $\Q$ needs to be parameterized such that the optimization of the ELBO is computationally tractable. To achieve that, the statistical model is augmented with a set of $M$ inducing points $(\u, \Z)$,  where the vector $\Z$ contains the locations of the inducing points in the original input space and $\u$ is governed by the GP prior, i.e., 
\begin{equation}
    \prob(\u \mid \Z) =  \mathcal{N}\left(\u \mid \cmean(\Z),\ccov(\Z,\Z)\right).
\end{equation}
The approximate distribution is assumed to have a form 
\begin{equation}
    {\Q := \prob(\extract{\vec{f}} \mid \observestate, \u, \Z)\mathbb{Q}(\u \mid \Z)}.
\end{equation}
The specific augmentation decouples the variables and leads to a general expression for the ELBO in the sparse variational GP case
\begin{align}
\L &= \mathbb{E}_\Q\left[\log \mathrm{L}\right] - {\mathrm{KL}}[\mathbb{Q}(\u \mid \Z) \;\|\; \prob(\u \mid \Z)].
\label{eq:original_ELBO}
\end{align} 
The inference problem is now reduced to determine the set of parameters determining $\Q$ through numerical optimization.

To support the setting where the likelihood $\mathrm{L}$ is not Gaussian, we follow \cite{Hensman2013} and parameterize $\mathbb{Q}(\u \mid \Z)$ as a Gaussian distribution with free mean and covariance parameters $(\m, \S)$, in which case the expectation over conditionally independent data points can be estimated by Monte Carlo sampling. Therefore, the final loss function, to be maximized, is given by 
\begin{align*}
\L &= \sum_{i=1}^M \mathbb{E}_{q(f_i; \m, \S)} \left[\log \mathrm{L}_i\right] - {\mathrm{KL}}[\mathbb{Q}(\u \mid \Z; \m, \S) \;\|\; \prob(\u \mid \Z)], \nonumber
\\
q(f_i; \m, \S) &\sim \N{\beta_i\m, \ccov(\x_i, \x_i) - \beta_i(\ccov(\Z,\Z) - \S)\beta_i^T}, 
\\
\beta_i &= \ccov(\x_i, \Z)\ccov(\Z,\Z)^{-1}. \end{align*}
This loss function can be optimized using standard mini-batch stochastic gradient methods similar to \cite{Hensman2013} as long as we can evaluate the three likelihood functions present in Eq.~\eqref{bayes3}. The set of optimized parameters, $\m$, $\S$, $\Z$ and kernel hyper-parameters, ensures that the predictive distribution $ \prob(\predictstate\mid \observestate,\observepoints,\observepoints_*)$ can be computed.

\subsection{Optimization algorithm and parameters}
Based on the GP prior, the posterior distribution depends on the following hyperparameters: $\alpha$ and $\beta$ define the mean function, $\sigma_f$ and $\scriptr$ define the covariance function, $e_x^2$ and $e_y^2$ are the errors in the derivatives-based likelihood, and $e_0^2$ is the error in the reference point likelihood. The constraint likelihood depends on the hyperparameter $\nu$. In addition, GP also depends on the location of the inducing points. Two of the hyperparameters, $e_0^2=10^{-5}$ and $\nu=10^4$ are kept fixed, while the rest of the hyperparameters need to be determined. Regression of the GP (also called training) is an iterative process of finding values of hyperparameters that optimize the loss function. 

\begin{figure}
    \centering
    \subfigimg[width=0.49\textwidth]{a)}{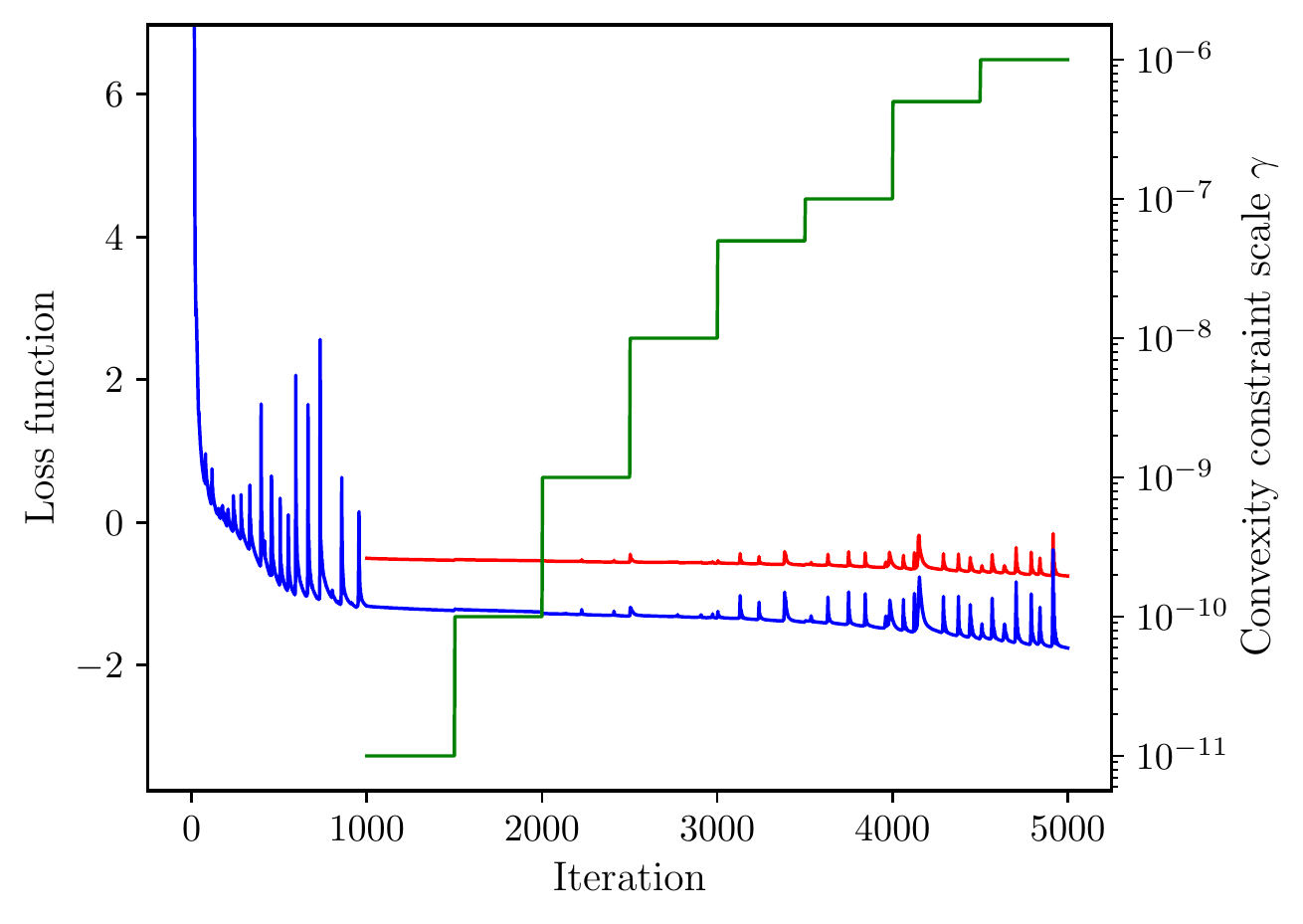}
    \subfigimg[width=0.49\textwidth]{b)}{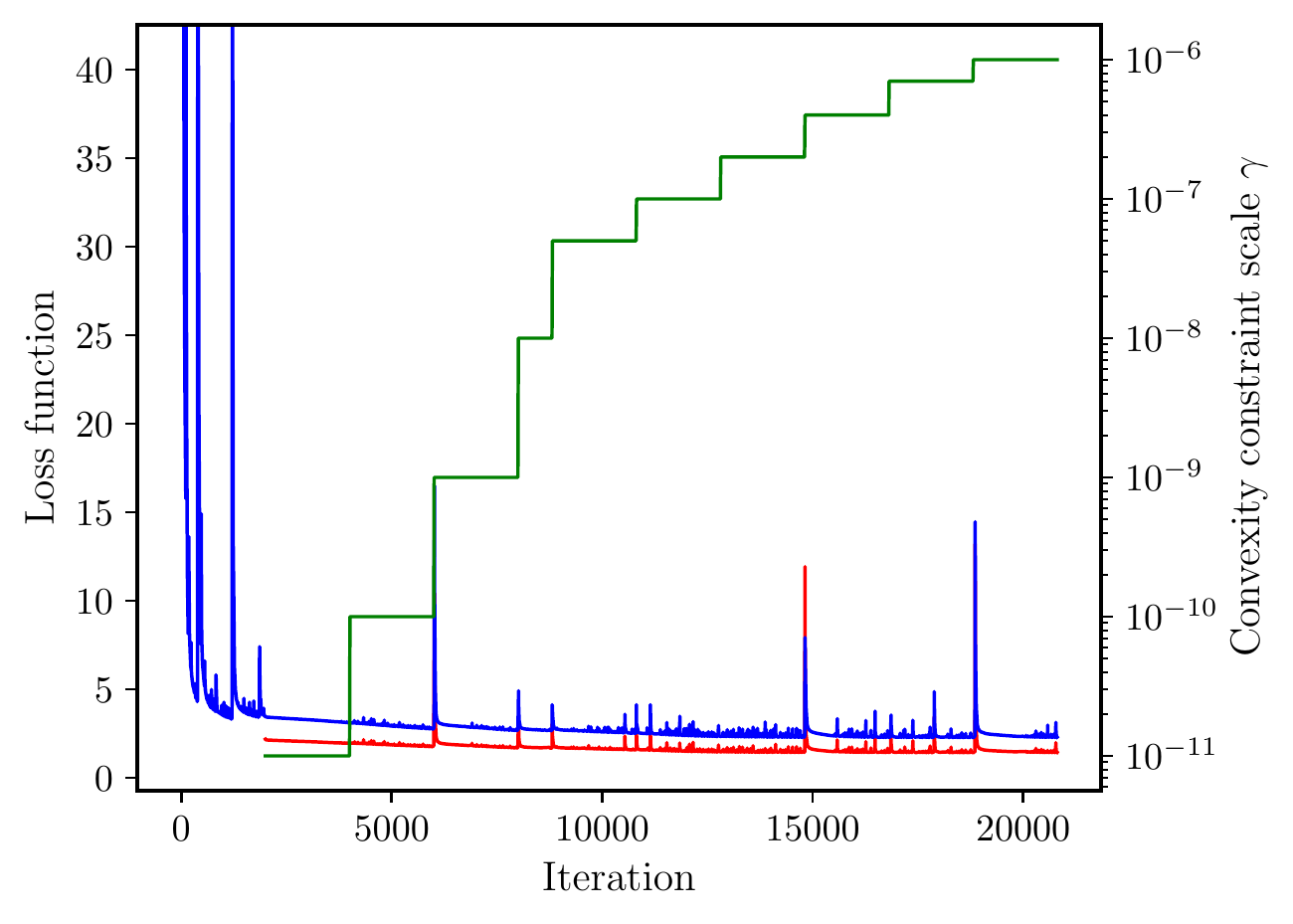} \\
    \includegraphics[width=0.95\textwidth]{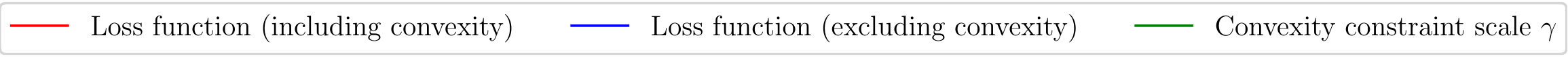}
    \caption{Iteration plots for two GPs trained using the proposed framework: a) GOH with $\ell=8$ and convexity constraints and b) real experimental data for AV leaflet tissue.}
    \label{fig:convergence}
\end{figure}

The optimization is performed as follows. Negative of the loss function is minimized iteratively via a stochastic gradient method \cite{Hensman2013}, where the step size is scaled by a parameter called the ``learning rate''. First, 1000 iterations are performed without the constraints and with a learning rate of 0.05. Then, the convexity constraint log-likelihood terms multiplied by a scaling parameter $\gamma$, which is gradually increased in steps. It is initially set at a value of $10^{-11}$, increasing to a final value of $10^{-6}$. At each value of $\gamma$, 500 iterations are performed with a learning rate of 0.01 (Fig.~\ref{fig:convergence}).

\section{Verification test\label{sec:test}}
In this section, the proposed framework is verified by training a GP based on artificial biaxial stretching data for a known hyperelastic model (referred to as the ground truth) and comparing the results. Next, the methodology for data creation is described, followed by the details of GP training procedure and the results.

\subsection{Artificial data creation}
\begin{figure}[h!]
    \centering
    \includegraphics[width=\textwidth]{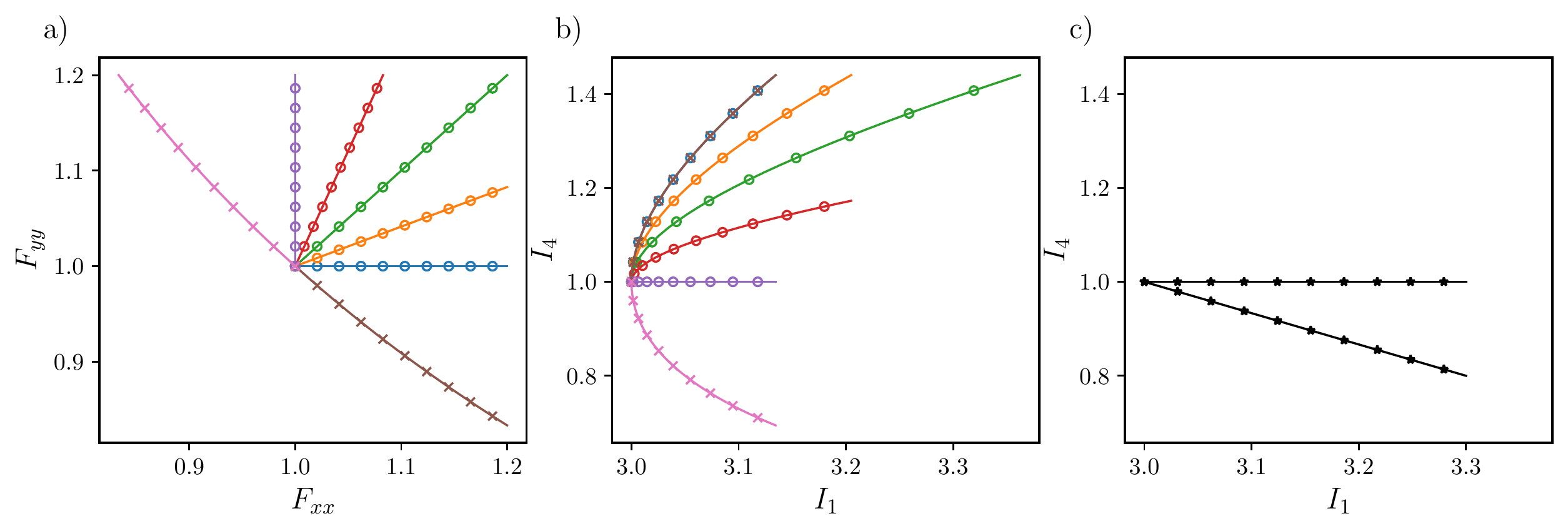}\\
    \includegraphics[width=\textwidth]{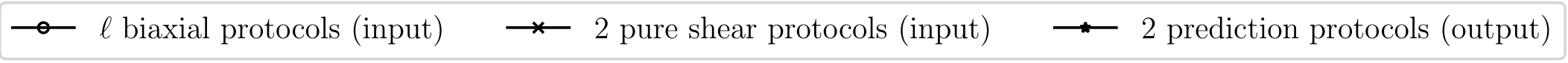}
    \caption{Training and prediction protocols used for verification of the proposed GP model.}
    \label{protocol-fig}
\end{figure}
A widely used model for soft tissues, the Gasser--Ogden--Holzapfel (GOH) model \cite{gasser2006hyperelastic}
\begin{equation}
     \sedf^{\text{true}} = \dfrac{\mu}{2}(I_1-3) + \frac{k_1}{2k_2}\left[\exp(k_2(\kappa I_1+(1-3\kappa)I_4-1)^2)-1\right] \label{goh-model}
\end{equation}
with parameters $\mu=5$ kPa, $k_1=4$ kPa, $k_2=10$, $\kappa=0.1$ and $\vec{M}=[1,0,0]$ is used to create artificial biaxial stretch observations. $\ell$ straight lines in the deformation space (Fig.~\ref{protocol-fig}a), equi-spaced between 0 and $\pi/2$, are used as inputs. These represent biaxial stretch protocols. In addition to the $\ell$ protocols, two pure shear protocols
\begin{align*}
    \mathbf{F} &= \diag\left[\lambda,1/\lambda,1 \right] \text{ and}\\
    \mathbf{F} &= \diag\left[1/\lambda,\lambda,1 \right], 
\end{align*}
with $\lambda>1$, are also used. It is typical for soft tissues to experience up to 20\% stretches under physiological conditions. Thus, for all of the protocols, a maximum tensile stretch of 1.2 is used, and the corresponding $I_1$ and $I_4$ ranges are used for training and testing. To emulate the experimental error, a normally-distributed random noise with mean 0 and variance 0.02 is added to the resulting stresses $\observe{P}_{xx}$ and $\observe{P}_{yy}$. Thus, data from a total of $(\ell+2)$ protocols is used to fit a constitutive model. The effect of the number of protocols on the predictive capability of the proposed framework is studied by using $\ell=3$ and $\ell=8$, as well as by removing the pure shear protocols. The effect of noise is studied by increasing the variance of added error to 0.2. The locations of the experimental observations, the straight lines in the deformation space (Fig.~\ref{protocol-fig}a), map to nonlinear paths in the $\point$ space (Fig.~\ref{protocol-fig}b). The extent of these observations is approximately $I_1\in \left[3,3.36\right]$ and $I_4\in\left[0.7,1.44 \right]$. To enforce convexity, this region is padded with a 0.1 on each side and uniformly spaced points are used. Specifically, $20\times20$ uniformly spaced points in $I_1\in \left[2.9,3.46\right]$ and $I_4\in\left[0.6,1.54 \right]$ are used to enforce convexity.

In order to verify the proposed GP model $\rand{\sedf}$ against the ground truth $\sedf^{\text{true}}$, the mean of the posterior SEDF ${\sedf}$ and its first derivatives ${\psedf1}$ and ${\psedf4}$ are plotted at $50\times50$ points in the $\point$ space. To quantify the difference between the ground truth and the fitted GP, the following error is is defined:
\begin{equation}
    \text{Error}_{\sedf}(\point) = \frac{|\sedf(\point) - \sedf^{\text{true}}(\point)|}{\max\limits_{\point\in\predictpoints}(\sedf^{\text{true}}(\point))}\times 100.
\end{equation}
Similar error definitions are used for the first derivatives. In addition to the mean, the GP framework also provides us with covariance of $\sedf$ and its derivatives, from which the standard deviation at each point in $\point$-space is calculated and plotted. Lastly, to check the the convexity, the mean second derivatives are plotted at these points. 

In order to test the predictive capability of the proposed GP model outside the training range (i.e., extrapolation), two protocols different from the training protocols are used (Fig.~\ref{protocol-fig}c). First prediction protocol follows the same path as one of the training protocols (i.e., $I_4$ fixed), but extends to a larger stretch of 1.31. The second prediction protocol follows a straight path in the $\point$-space from $(3,1)$ to $(3.3,0.8)$ which requires a shear strain (i.e., an off-diagonal term in $\mathbf{F}$). As a result, the second prediction protocol also generates shear stress, a situation which is not used for training the GP. Lastly, one of the key advantages of GP models is that they provide a distribution rather than point estimates. To use the distribution information, the mean and standard deviation of predicted stresses using the GP model are computed and compared with the ground truth GOH model \eqref{goh-model}.

\subsection{Results}
\begin{figure}[h!]
\centering
\includegraphics[width=\textwidth]{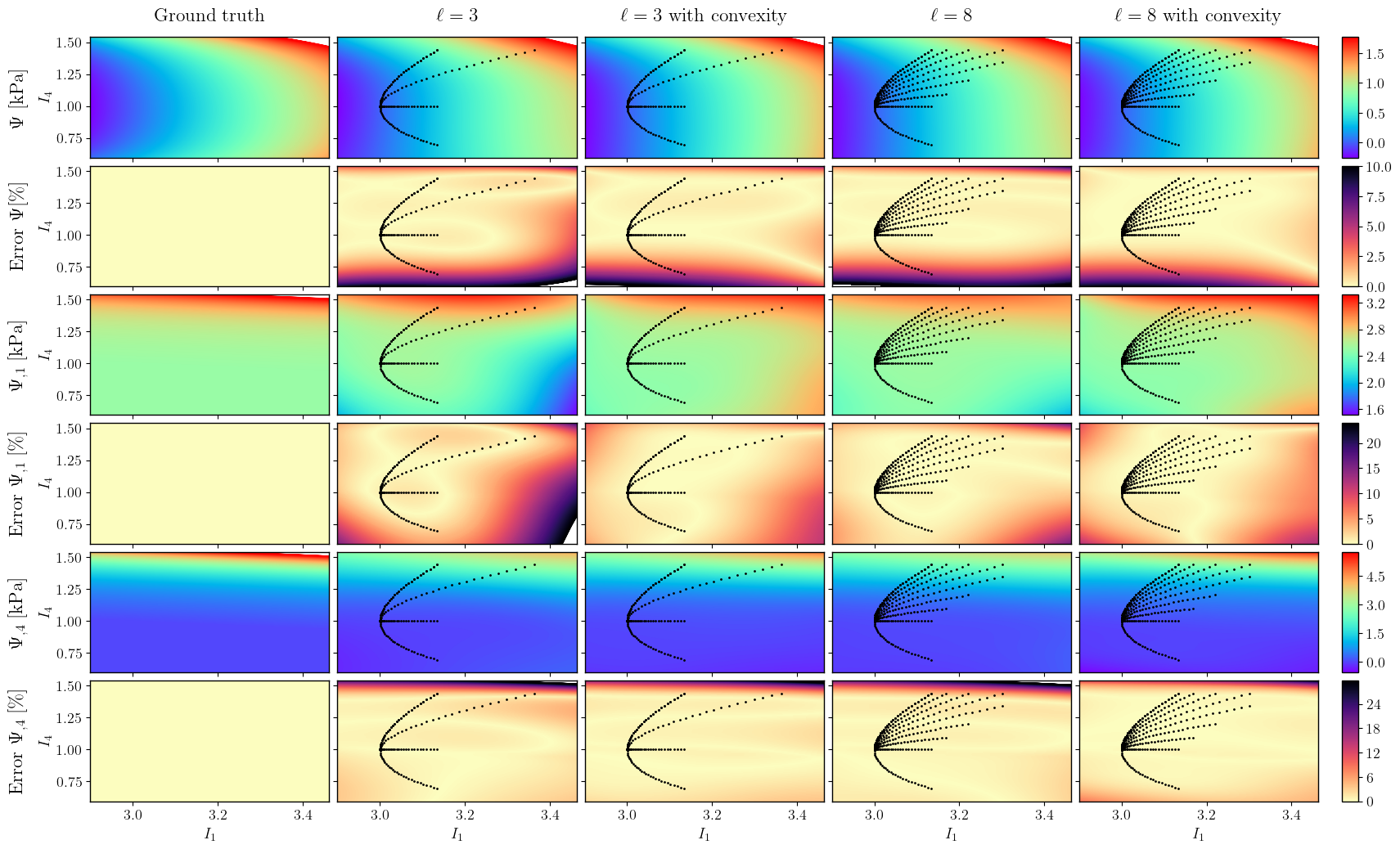}
\caption{Verification of the proposed GP-based model was performed against a ground truth of GOH model. The resulting mean SEDF and its derivatives for different numbers of protocols ($\ell$), both with and without convexity constraints are plotted. Also the corresponding errors with respect to the ground truth are plotted.}
\label{GOH-sedf}
\end{figure}

The results using $\ell=3$ and $\ell=8$, with and without convexity constraints are shown in Fig.~\ref{GOH-sedf}. The error in $\sedf$ for all cases is reasonably small ($<12$\%), especially near the training points (denoted as dots). The accuracy of the derivatives of $\sedf$ is more important, since the derivatives are directly related to the stresses. Again, the errors in derivatives near the training points are small. However, the error farther from the training points reduces when using convexity. Moreover, when using higher $\ell$, the errors also decrease slightly. Overall, the difference between the accuracy using $\ell=3$ and $\ell=8$ is not significant. Thus, the proposed GP framework works well with a small number of protocols ($\ell\gtrsim3$). In practice, it is common to use between 3 and 10 protocols, thus a limited amount of experimental data is needed to train the proposed GP model. To fully understand the effect of training points and noise on the results, two additional settings are compared in \ref{additional-verification}.

\begin{figure}[h!]
\centering
\includegraphics[width=\textwidth]{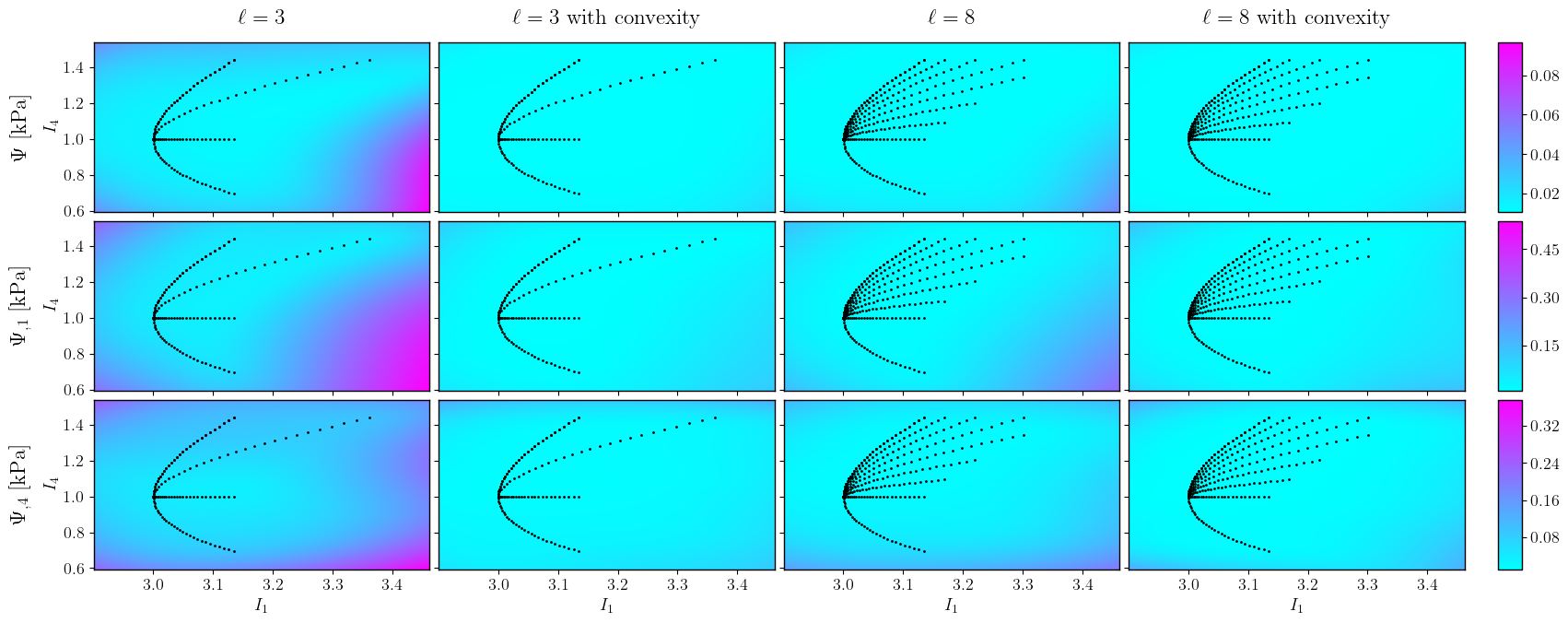}
\caption{Standard deviation in SEDF and its derivatives for different numbers of protocols ($\ell$), both with and without convexity constraints are plotted.}
\label{GOH-std}
\end{figure}

To quantify the resulting uncertainty in SEDF, the standard deviations of the fitted model are plotted in Fig.~\ref{GOH-std}. Interestingly, the standard deviation is reduced substantially when using convexity and when using a higher number of protocols. Thus, the confidence in the GP results is increased when we enforce convexity and as we increase the amount of experimental information.

\begin{figure}[h!]
\centering
\includegraphics[width=\textwidth]{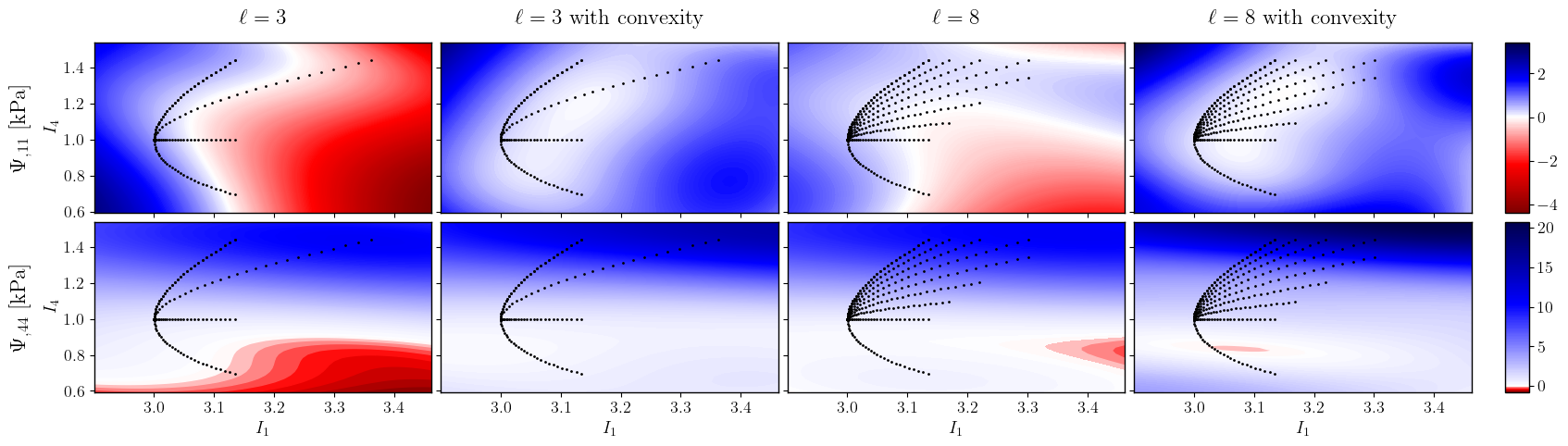}
\caption{Mean second derivatives of the resulting SEDF for different numbers of protocols ($\ell$), both with and without convexity constraints are plotted.}
\label{GOH-second-deriv}
\end{figure}

Lastly, as mentioned previously, convexity is an important requirement for $\sedf$ to satisfy. To verify convexity, the second derivatives are plotted in Fig.~\ref{GOH-second-deriv}. Clearly, when the convexity constraints are not enforced, the second derivatives attain large negative values, especially away from the training points. However, including the convexity constraints resolves the issue, and the resulting second derivatives of $\sedf$ are positive in the chosen range of $I_1$ and $I_4$.

\begin{figure}[h!]
\centering
\includegraphics[width=\textwidth]{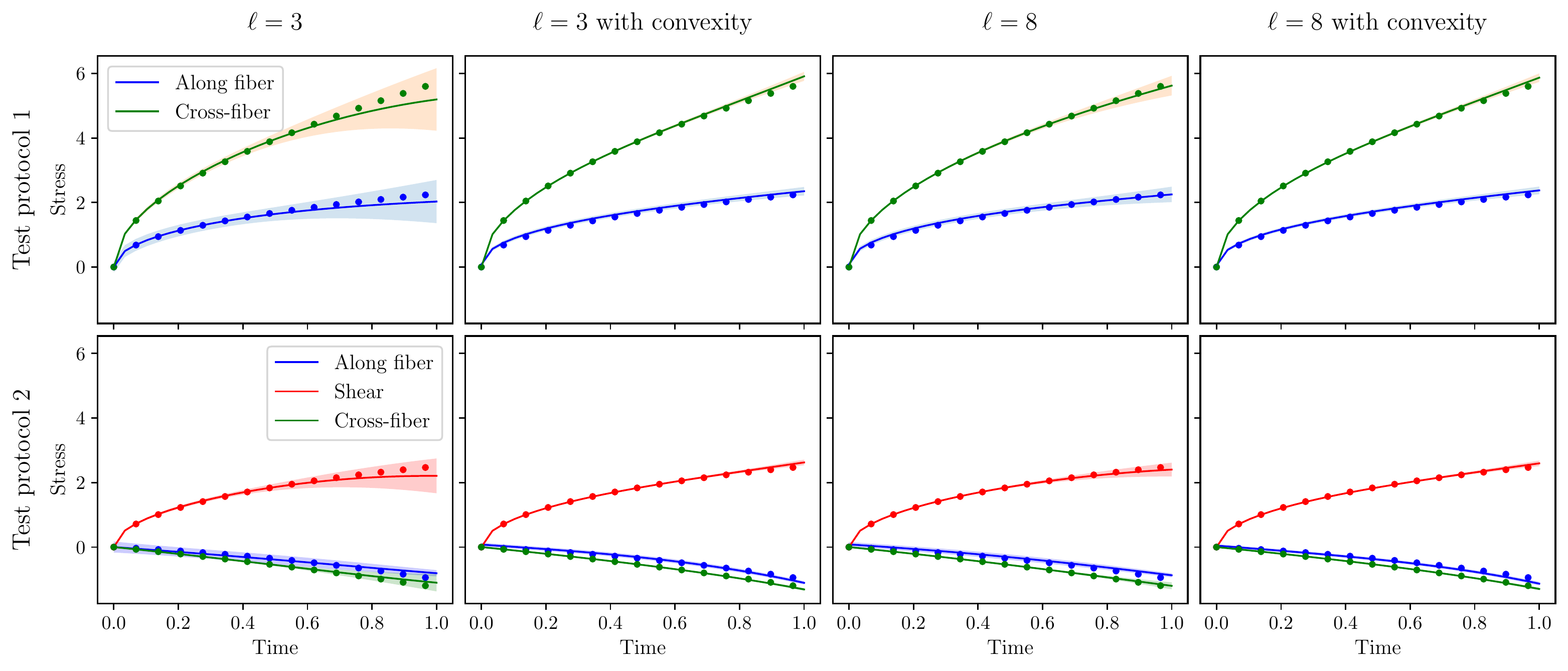}
\caption{Results for the two testing protocols with four settings, where the points denote the ground truth, lines are GP predictions, and the shaded areas indicate two standard deviations of the GP. Test protocol 1 represents a biaxial stretch (i.e., no shear), while the test protocol 2 includes a shear deformation and stress.}
\label{GOH-predict}
\end{figure}

The results of the \emph{prediction} protocol are shown in Fig.~\ref{GOH-predict}. The mean response of the GP model (lines) matches very well with the ground truth (points). The only significant deviation is in the $\ell=3$ case without enforcing convexity. This is a remarkable result considering that in these prediction protocols we are also testing the extrapolation capability of the GP. The shaded areas denote two standard deviations of the GP model, which are get smaller as we increase the number of training protocols and include convexity. Thus, the proposed GP framework also provides high confidence in the results. 

\section{Application to an experimental biaxial testing dataset\label{sec:result}}

\subsection{Experimental methods}
The experimental data used in this study is the same as that reported in a previous study \cite{HUDSON2022104907}, and its experimental procedure is summarized next. A porcine heart was obtained from a USDA-approved abattoir (Chickasha Meat Company, Chickasha, OK). The heart was dissected, and the aortic valve (AV) tissue was extracted from the aorta. The excised tissue was then briefly stored at $-20^\circ$C prior to mechanics testing within 6--12 hours. Prior to biaxial testing, the excised AV specimen was thawed in an in-house phosphate-buffered saline (PBS) solution at room temperature. Once thawed, a square region of the tissue was dissected and thickness measurements were made using a non-contact laser displacement sensor (Keyence IL-030, Itaska, IL) at three different locations to determine the average tissue thickness. 

For biaxial testing, the tissue specimen was mounted to a commercial biaxial testing system (BioTester, CellScale, Canada, 1.5~N load cells) via BioRake tines, resulting in an effective testing region of $6.5\times6.5$ mm. During mounting, the tissue's circumferential and radial directions were aligned with the $x$- and $y$-directions of the biaxial testing system, respectively. Because the tissue's fiber orientation was aligned with the biaxial testing direction in the experimental setting, the off-diagonal terms in the deformation tensor $\mathbf{F}$ were small, and therefore any shear deformation was neglected. 

For testing, four glass beads (with a diameter of $300-500$ {\textmu}m) were placed on the center region of the specimen to serve as fiducial markers for quantifying the in-plane strains. The specimen was submerged in a 32$^\circ$C PBS bath during the testing. The force readings from the load cells and CCD camera images capturing the bead positions were recorded at 15 Hz throughout the test. 
A preconditioning protocol, consisting of six loading/unloading cycles at a target first PK peak stress of $P = 240$ kPa, was first applied to restore the tissue to its in-vivo functional configuration. The preconditioning protocols were followed by seven testing protocols. 

\begin{table}[ht]
\caption{List of seven chosen invariant-based hyperelastic models from literature.}
\label{model-list}
\small{
\begin{tabular}{ll}
\hline
\bf{Model} & \bf{Strain energy density function} \\
\hline

GOH & $\sedf = \frac{\mu}{2}(I_1-3) + \frac{k_1}{2k_2}\left[\exp(k_2(\kappa I_1+(1-3\kappa)I_4-1)^2)-1\right]$  \\[3pt]

HGO & $\sedf =  \frac{\mu}{2}(I_1-3) + \frac{k_1}{2k_2}\left[\exp(k_2(I_4-1)^2)-1\right]$ \\[3pt]

HGO2 & $\sedf = \frac{k_1}{k_2}\left[\exp(k_2(I_1-3))-1\right] + \frac{k_3}{2k_4}\left[\exp(k_4(I_4-1)^2)-1\right]$ \\[3pt]

Holzapfel & $\sedf = \frac{\mu}{2}(I_1-3) +  \frac{k_1}{2k_2}\left[ \exp(k_2(\kappa(I_1-3)^2+(1-\kappa)(I_4-1)^2))-1\right]$ \\[3pt]

HY & $\sedf = \frac{k_1}{k_2}\left[\exp(k_2(I_1-3))-1\right] +  \frac{k_3}{k_4}\left[\exp(k_4(\sqrt{I_4}-1)^2)-1\right]$ \\[3pt]

LS & $\sedf = \frac{\mu}{2}(I_1-3) + \frac{k_1}{2}\left[\kappa\exp(k_2(I_1-3)^2) + (1-\kappa)\exp(k_3(I_4-1)^2)-1\right]$ \\[3pt]

MN & $\sedf = \frac{\mu}{2}(I_1-3) + {k_1}\left[\exp(k_2(I_1-3)^2+k_3(\sqrt{I_4}-1)^4)-1\right]$ \\[3pt]

\hline
\end{tabular}
}
\end{table}

For using the above data in our GP framework, the measured deformations are converted into invariants $I_1$ and $I_4$ (Eq.~\ref{invariants-exp}), and the measured stresses from the seven protocols are used to train the GP. To enforce convexity, similar to the verification case in the last section, the range of $I_1$ and $I_4$ is padded with 0.1 in all directions, and convexity is enforced on uniformly spaced $20\times20$ points. In this case there is no ground truth to compare with. Therefore, just the fit to the input experimental data is quantified by the $L_2$ norm of the difference between the modeled mean and experimental stresses, i.e.,
\begin{equation}
    L_2 = \sqrt{\sum\limits_{i^{\text{d}}=1}^{N^{\text{d}}} \left\{ \left[\observe{P}^{i^{\text{d}}}_{xx} - P^{i^{\text{d}}}_{xx} \right]^2 + \left[\observe{P}^{i^{\text{d}}}_{yy} - P^{i^{\text{d}}}_{yy} \right]^2\right\}}.
\end{equation}
Also, the goodness of fit is quantified in terms of the coefficient of determination
\begin{equation}
    R^2 = 1 - \frac{\sum\limits_{i^{\text{d}}=1}^{N^{\text{d}}} \left\{ \left[\observe{P}^{i^{\text{d}}}_{xx} - P^{i^{\text{d}}}_{xx} \right]^2 + \left[\observe{P}^{i^{\text{d}}}_{yy} - P^{i^{\text{d}}}_{yy} \right]^2\right\}}{\sum\limits_{i^{\text{d}}=1}^{N^{\text{d}}} \left\{ \left[\observe{P}^{i^{\text{d}}}_{xx} - \bar{P} \right]^2 + \left[\observe{P}^{i^{\text{d}}}_{yy} - \bar{P} \right]^2\right\}},
\end{equation}
where $\bar{P}=\frac{1}{2N^{\text{d}}} \sum\limits_{i^{\text{d}}=1}^{N^{\text{d}}} \left[ \observe{P}^{i^{\text{d}}}_{xx} + \observe{P}^{i^{\text{d}}}_{yy} \right] $ is the mean observed stress. To compare against some of the existing models in the literature, the following seven invariant-based models were chosen: (i) the Lee--Sacks (LS) model for the mitral valve leaflet tissue \cite{LEE20142055}; (ii) the May--Newman (MN) model with another form proposed for the mitral valve tissue \cite{may1998constitutive}; (iii and iv) two variants of a model proposed by Holzapfel, Gasser, and Ogden for arterial tissue with an additive split of isotropic and anisotropic components \cite{holzapfel2000new} (HGO with linear isotropic term and HGO2 with an exponential isotropic term); (v) Holzapfel model proposed for coronary arteries \cite{doi:10.1152/ajpheart.00934.2004}; (vi) another model proposed by Gasser, Ogden and Holzapfel (GOH) for coronary arteries \cite{doi:10.1098/rsif.2005.0073}, and (vii) Humphrey--Yin (HY) model developed for myocardium \cite{HUMPHREY1987563}. These models and their corresponding SEDF are summarized in Table~\ref{model-list}.

\subsection{Results}
\begin{figure}[h!]
\centering
\subfigimg[width=\textwidth]{a)}{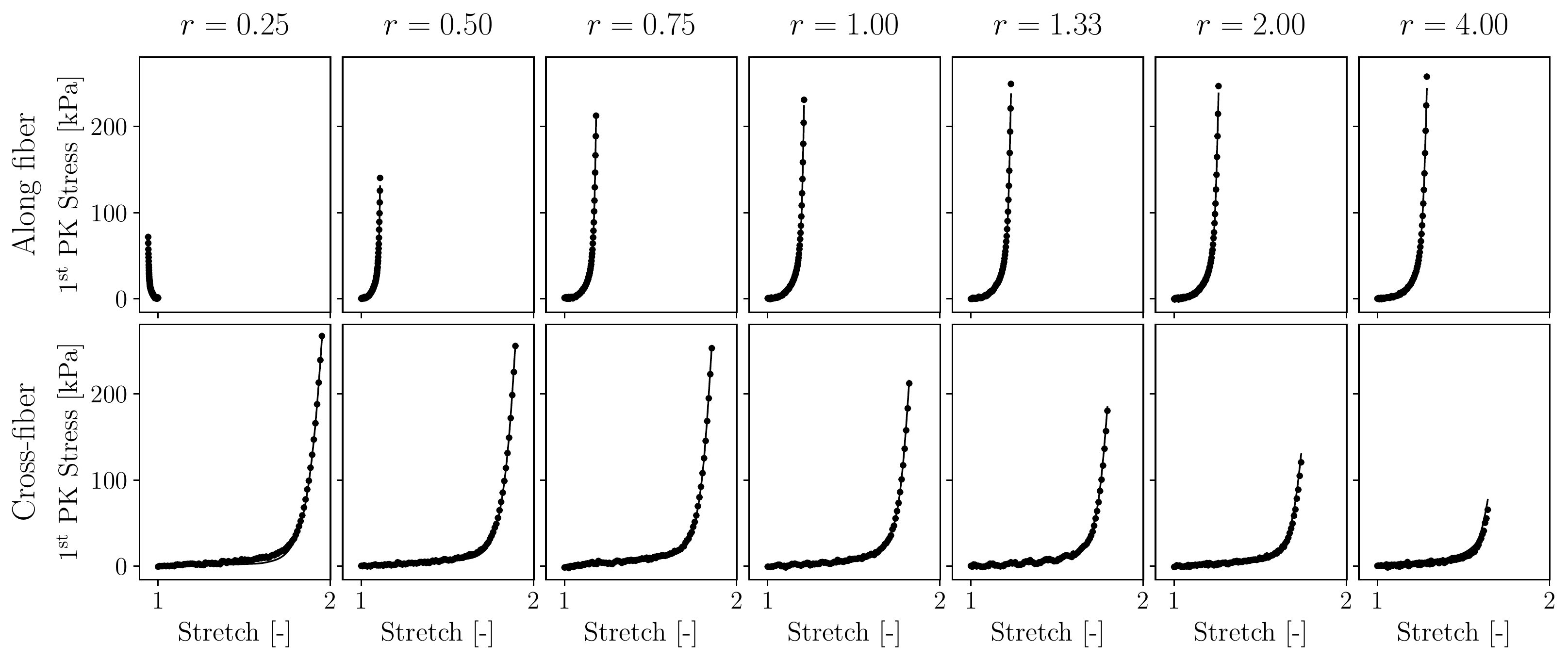}\\
\subfigimg[width=0.49\textwidth]{b)}{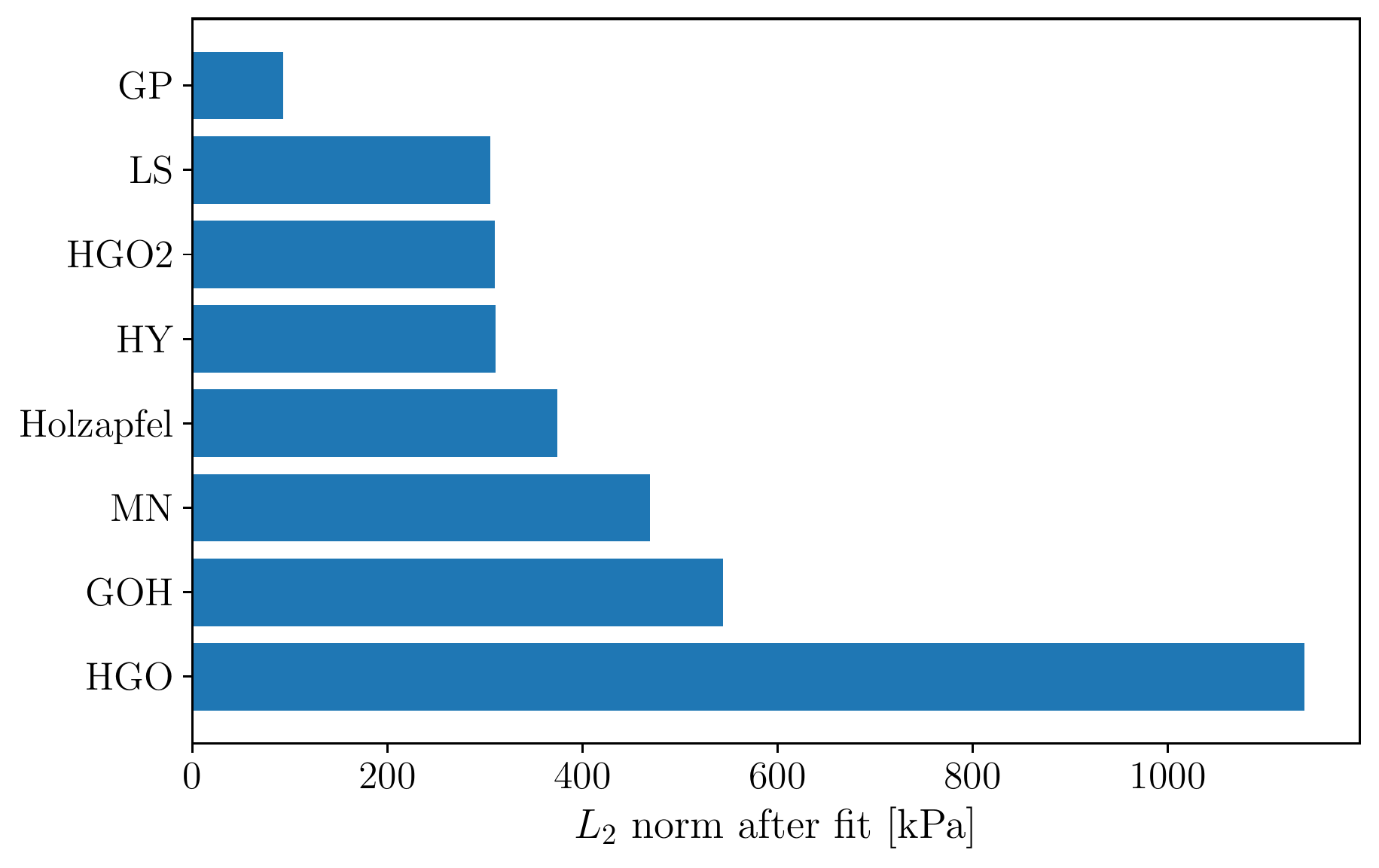}
\subfigimg[width=0.49\textwidth]{c)}{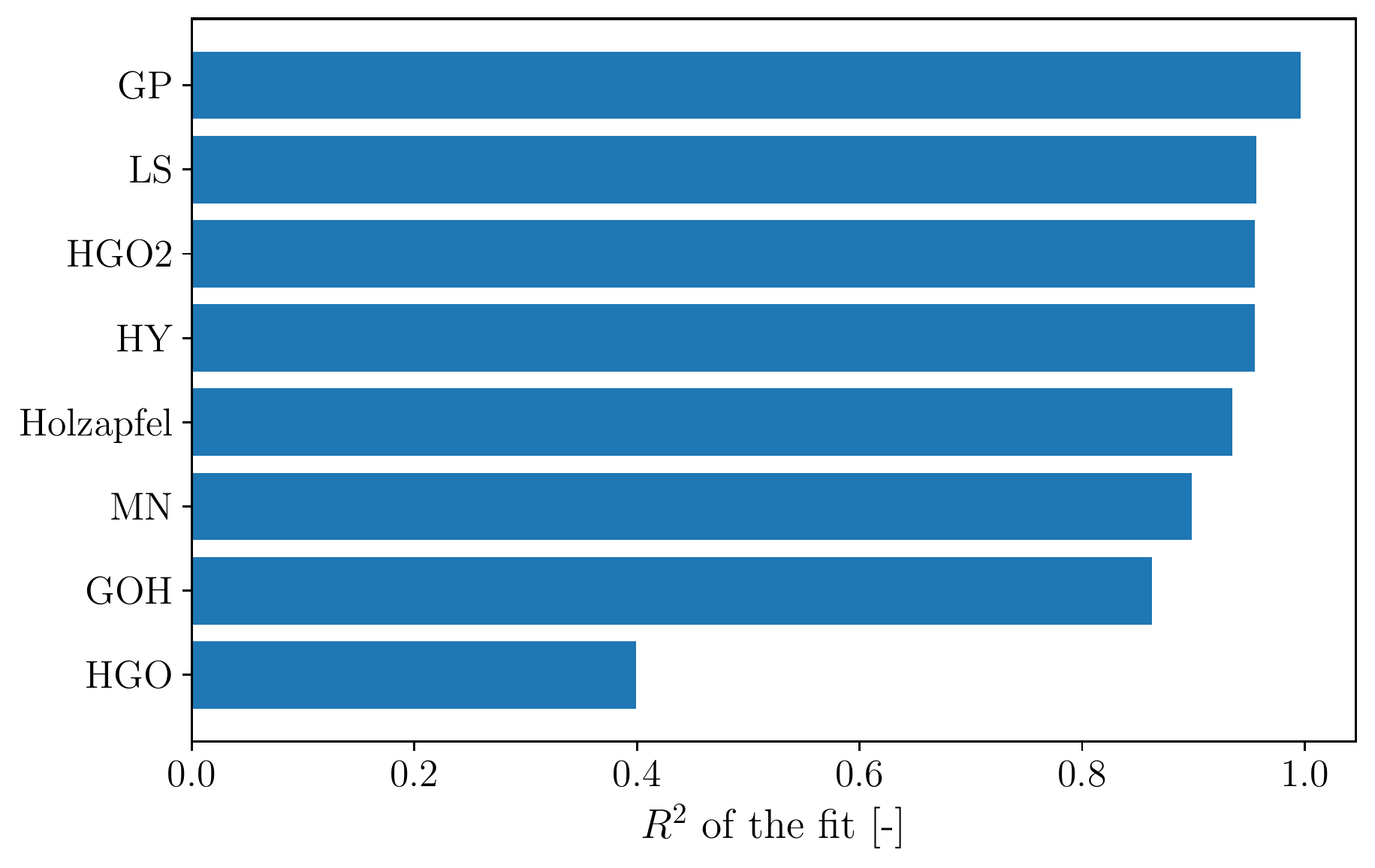}
\caption{Results of GP fitted to experimental dataset of AV leaflet tissue: a) comparison of the trained GP (lines) versus the original data (points); b) the $L_2$ norm of the GP fit compared with that of seven models from the literature; c) the $R^2$ coefficient of the GP fit compared with that of seven models from the literature.\label{exp-data-results}}
\end{figure}

The trained GP fits extremely well to the experimental data (Fig.~\ref{exp-data-results}a). Compared with existing invariant-based models in the literature, the GP model has the least fitting $L_2$ norm (Fig.~\ref{exp-data-results}b) and highest (and almost perfect 1) coefficient of determination (Fig.~\ref{exp-data-results}c). 

\section{Stochastic finite element analysis\label{sec:stochastic}}
One of the advantages of a Gaussian process over the traditional models, and even other data-driven models, is that it is naturally Bayesian and provides a distribution of the SEDF $\rand{\sedf}$. Thus, in addition to the mean values, one also obtains the variation in the SEDF and resulting stress-strain behavior. In this section, a framework is proposed to use this distribution to carry out a stochastic finite element analysis (SFEA) in a non-intrusive manner. That is, the aim is to use existing finite element solvers with the new GP-based constitutive model to find the distribution of the finite element analysis (FEA) results, such as displacements and stresses. 

In nonlinear FEA, even if the distributions of inputs are Gaussian, the distributions of the outputs are, in general, not necessarily Gaussian. Thus, finding the exact distribution of each FEA result becomes an extremely high-dimensional problem, where each scalar variable in the original FEA (such as displacement along one of the axes at one node at one load/time) becomes a function in the probability space. In its full generality, the problem of finding these distributions is prohibitively expensive. Thus, to simplify the SFEA, we focus on quantifying the expected value (i.e., the mean) and standard deviation of the FEA results. 

\subsection{Proposed methodology}
A practical challenges for SFEA is how to use a GP as a constitutive model in a traditional finite element solver. The most straightforward approach to quantify the effect of uncertainty in $\rand{\sedf}$ on FEA results is by generating samples of the \emph{function} $\sedf(I_1,I_4)$ and performing the FEA using each of the samples. This approach is also called propagating the samples through FEA or a Monte Carlo simulation. However, there is no functional form of the posterior distribution of GP. In general, a GP can only be sampled at a finite number of locations in the $I_1$-$I_4$ space. One could directly use the mean of a GP for an FEA since every time a GP mean is evaluated at the same point, it will be the same value (and same for the derivatives of the GP). However, for a random (not mean) sample from the GP distribution, because of its randomness, every time it is evaluated, the value (and derivatives) will be different, which cannot be used in FEA. To resolve this issue, we propose to use tensor product splines as intermediary functions, as detailed below.

\begin{figure}[h!]
\centering
\subfigimg[width=0.45\textwidth]{a)}{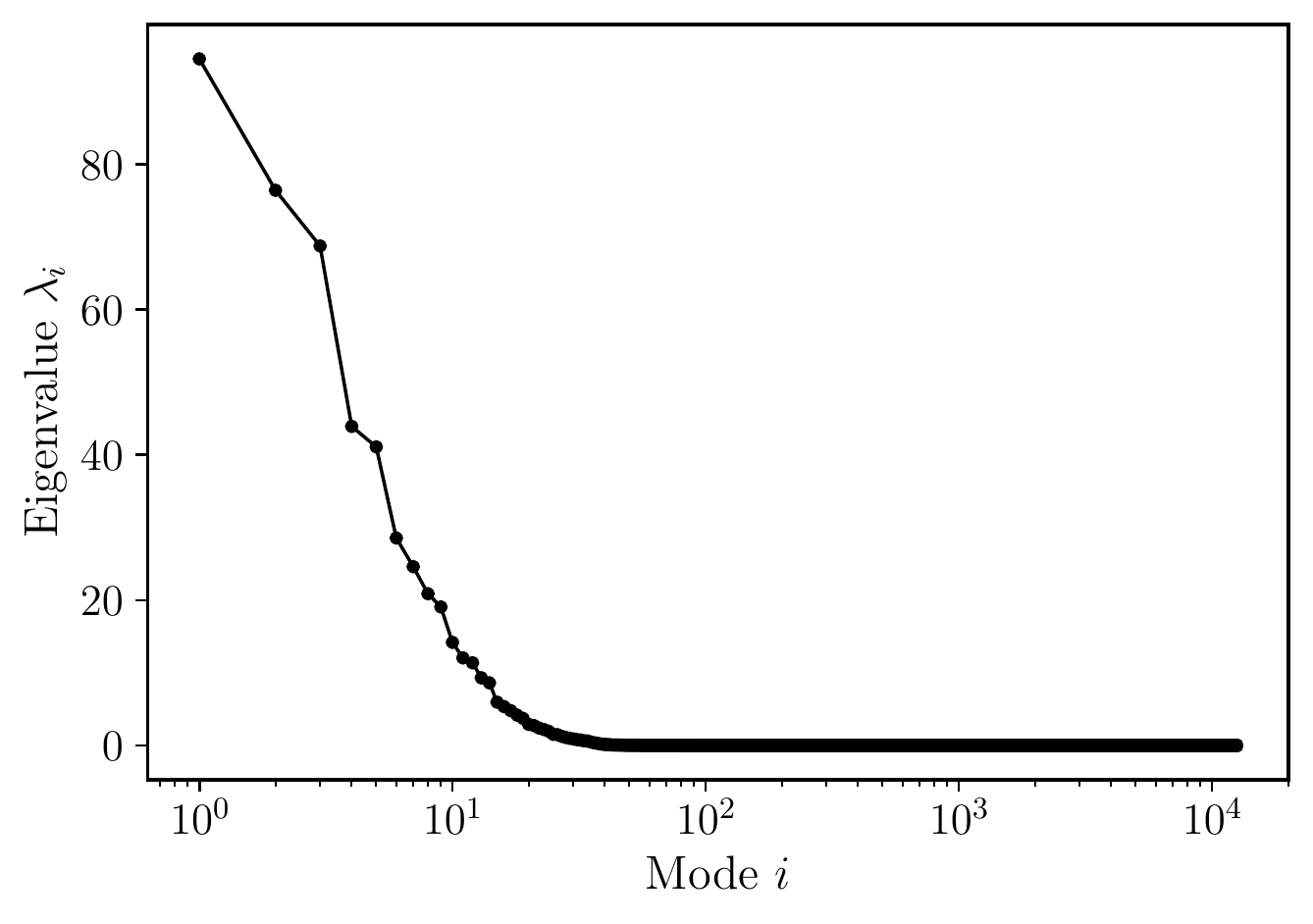}
\subfigimg[width=0.45\textwidth]{b)}{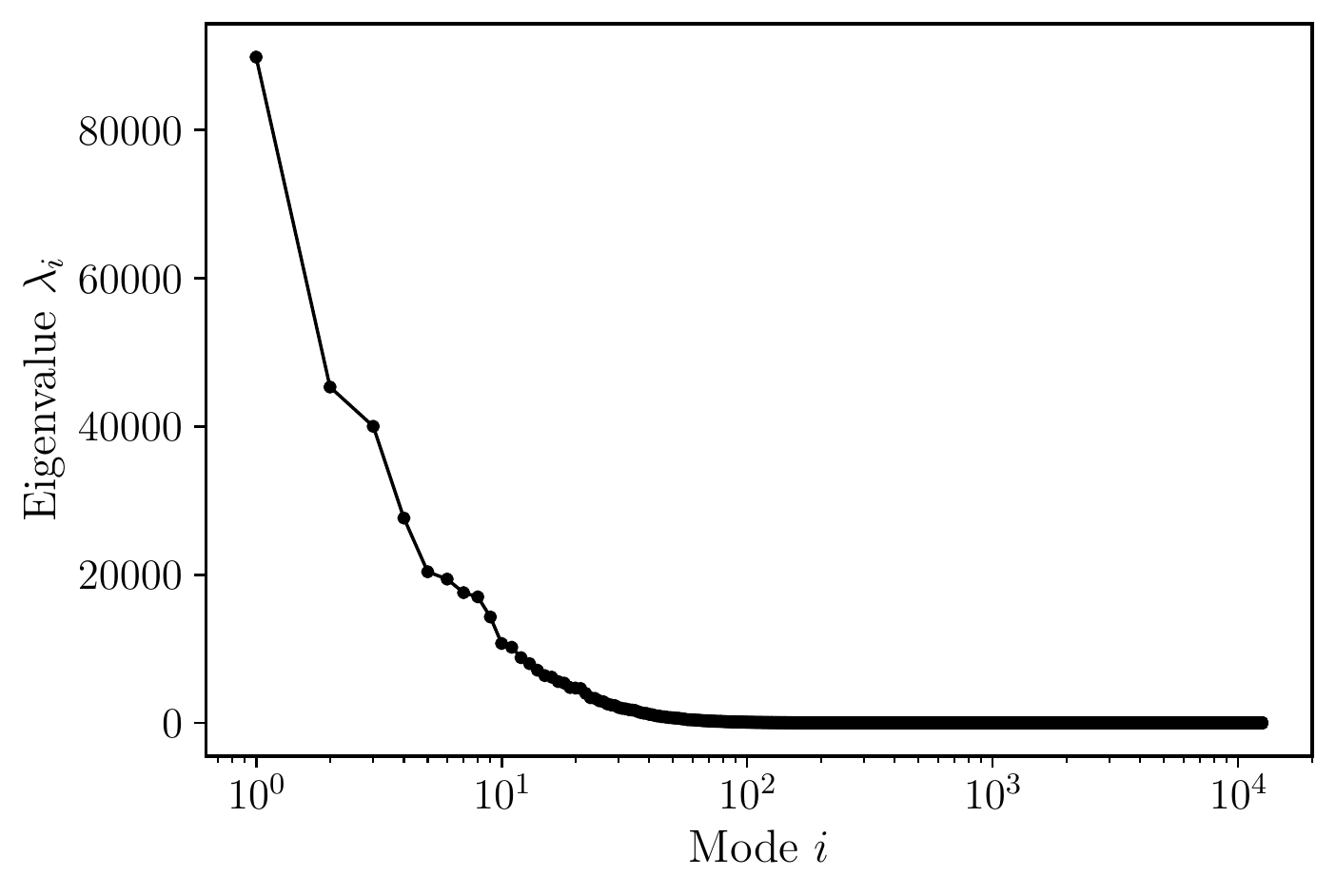}
\caption{The eigenvalues of the covariance matrix of two GPs trained on a) GOH model (with $\ell=8$ and convexity) and b) real experimental data of AV leaflet show a spectrum much smaller than the size of the matrix}\label{fig:spectrum}
\end{figure}

Once the posterior of the GP $\prob(\predictstate \mid \observestate,\observepoints)$ has been obtained, it is sampled at $\predictpoints$ --- a fine grid of $N$ points in the $\point$-space. We denote the vector of SEDF $\sedf$ evaluated at $\predictpoints$ as $\vecrand{W}$, and thus obtain its distribution:
\begin{equation}
    \vecrand{W} \sim \mathcal{N} \left(\bar{\vec{W}},\mathbf{\Sigma}\right),
\end{equation}
where $\bar{\vec{W}}$ is the mean SEDF at $N$ points (i.e., a vector of length $N$) and $\mathbf{\Sigma}$ is the covariance of SEDF at those $N$ points (i.e., a matrix of size $N\times N$). Next, we perform an eigenvalue decomposition of the covariance matrix $\mathbf{\Sigma}$ and sort its eigenvalues in the decreasing order. Thus, the covariance matrix can be written as:
\begin{equation}
    \mathbf{\Sigma} = \sum\limits_{i=1}^N \lambda_i \vec{E}_i\otimes\vec{E}_i,
\end{equation}
where $\vec{E}_i$ is the $i$-th eigenvector of $\mathbf{\Sigma}$ with $\lambda_i$ the corresponding eigenvalue. Because of the high correlation present in a GP, the spectrum (i.e., the number of non-zero eigenvalues) of $\mathbf{\Sigma}$ is expected to be much smaller than $N$ (Fig.~\ref{fig:spectrum}). Therefore, we approximate the eigenvalue decomposition as
\begin{equation}
    \mathbf{\Sigma} \approx \sum\limits_{i=1}^{m\ll N} \lambda_i \vec{E}_i\otimes\vec{E}_i,
\end{equation}
where the approximation keeps only the $m$ largest eigenvalues and it is desired to have $m\ll N$. The decision on how many eigenvalues to keep is based on how much variation we would like to capture. A common way is to keep $m$ dominant modes such that
\begin{equation}
    1- \frac{\sum\limits_{i=1}^m \lambda_i^2}{\sum\limits_{j=1}^N \lambda_j^2}<\text{TOL},
\end{equation}
where $\text{TOL}$ is a tolerance defining the error in the approximation. For most practical purposes, a tolerance of 0.05 is reasonable so that 95\% of the variation is captured. Based on this approximation, the distribution of sampled SEDF can now be written as
\begin{equation}
    \vecrand{W} \approx \bar{\vec{W}} + \sum\limits_{i=1}^m \rand{\sigmap_i} \vec{E}_i,
\end{equation}
where $\rand{\sigmap_i} \sim \mathcal{N}(0,\lambda_i)$ is the normally distributed coefficients along $i$-th eigenvector.

Once this approximation has been made, a two-dimensional (i.e., the dimension of the $\point$-space) tensor product spline is interpolated through the mean $\bar{\vec{W}}$ and each eigenvector $\vec{E}_i$. The interpolated splines are represented as $\bar{\mathcal{S}}(\point)$ and $\mathcal{S}_i(\point)$, respectively. Interpolation property implies that $\bar{\mathcal{S}}(\point^j)={\bar{W}}_j$ and $\mathcal{S}_i(\point^j)=E_{ij}$ $\forall\,\point^j\in\predictpoints$ (see \ref{spline-appendix} for more details on splines). Thus, a spline-based \emph{functional} approximation of the SEDF $\rand{\sedf}$ can now be written as
\begin{equation}
    \rand{\sedf}(\point) \approx \rand{\sedf}_{\mathcal{S}}(\point) = \bar{\mathcal{S}}(\point) + \sum\limits_{i=1}^m \rand{\sigmap_i} \mathcal{S}_i(\point).
\end{equation}
The advantage of this approach is that it provides a straightforward reduced dimensional representation of the stochastic $\rand{\sedf}$ in terms of $m$ independent normally distributed scalar variables $\rand{\sigmap_i}$, with a realization $\sigmapvec\in \mathbb{R}^m$. 

Next, this lower-dimensional representation allows us to use any of the existing stochastic methods to propagate the distributions of $\sigmap_i$ through a finite element model. Here, we employ a sigma-point technique, which constructs $2m+1$ points in the $\sigmapvec$-space \cite{uhlmann1995dynamic}. Each of these points are used in the finite element simulation, and the results are weighted to obtain the mean and covariance of finite element results. That is, if the finite element result of each simulation is $\vec{R}_k$, for $k=1,\dots,2m+1$, its variation can be represented as a normal distribution:
\begin{equation}
    \vecrand{R} \sim \mathcal{N}(\bar{\vec{R}},\text{Cov}(\vec{R})),
\end{equation}
where mean is $\bar{\vec{R}} = \sum_k w_k \vec{R}_k$ and covariance is $\text{Cov}(\vec{R}) = \sum_k v_k (\vec{R}_k-\bar{\vec{R}}) \otimes (\vec{R}_k-\bar{\vec{R}})$. There are a variety of methods for determining the $\sigmap$ point locations and the corresponding weights $w_k$ and $v_k$. Due to the independence of $\sigmap_i$, the following simple choice is adopted here \cite{van2004sigma}:
\begin{subequations}\label{sigma-point-defs}
\begin{align}
    \left\{\nu_k \right\} &= \left\{0\right\} \cup \left\{\pm\sqrt{m\lambda_j} \right\}_{j=1}^m \\
    \displaystyle w_k &= \begin{cases}
			0 & \text{if $k=1$}\\
            \frac{1}{2m} & \text{otherwise}
		 \end{cases} \\
	\displaystyle v_k &= \begin{cases}
			2 & \text{if $k=1$}\\
            \frac{1}{2m} & \text{otherwise}
		 \end{cases}.	 
\end{align}
\end{subequations}

\subsection{Test case}
\begin{figure}[h!]
\centering
\includegraphics[width=0.5\textwidth]{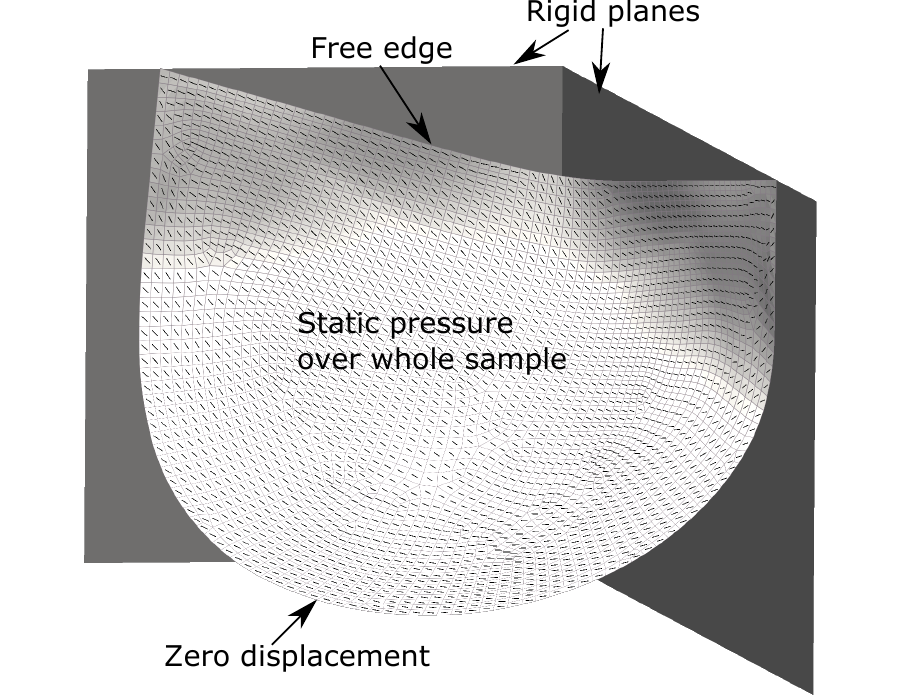}
\caption{Simulation setup of a bioprosthetic valve leaflet closure under static pressure head with the contact with other leaflets being modeled as a contact with two rigid planes symmetrically arranged. The fibers are oriented approximately circumferentially and are depicted with black lines.}\label{fig:leaflet-setup}
\end{figure}

A semilunar shaped single tissue representing a bioprosthetic valve leaflet with an area of 2.3~cm$^2$ is simulated using Reissner--Mindlin thin shell elements in FEBio \cite{maas2012febio}. The sample geometry is meshed using 2789 quadrilateral elements with a constant thickness of 0.38~mm \citep{betsch19964}. Displacement is interpolated using bilinear shape functions and stresses are integrated through the thickness using three-point Gauss quadrature rule to obtain bending moments. The fibers are aligned approximately in the circumferential direction. Contact of the sample with other leaflets is modeled using two idealized rigid planes placed at $\pm60^\circ$ with respect to the sample's plane of symmetry (Fig.~\ref{fig:leaflet-setup}).

A uniform normal follower pressure load is applied on the tissue with a maximum value of ${P}_0=80$ Pa. The contact is solved using augmented Lagrange method as the pressure is linearly increased. A spline-based constitutive model is implemented as a plugin in FEBio, which allows us to use the interpolated spline $\sedf_{\mathcal{S}}$ as an input to the simulations. The incompressibility condition is relaxed by adding a volumetric term to the SEDF and using isochoric version of invariants. That is, $I_1$ is replaced with $\tilde{I}_1=J^{-2/3}I_1$ and $I_4$ is replaced with $\tilde{I}_4=J^{-1/3}I_4)$. At each load step, static equilibrium equations \eqref{equilibrium} are solved using the BFGS solver to obtain the deformed shape of the tissue sample \citep{aggarwal_inverse_2015}. 

Two SFEA simulations are performed, and the mean and standard deviation of the displacement and von-Mises stresses are calculated for each case at four pressure values: $P=0$, $P=P_0/3$, $P=2P_0/3$, and $P=P_0$. The first simulation is performed using the GP trained on the artificial dataset from the GOH model (from Section~3, $\ell=8$ with convexity). The ground truth $\sedf^{\text{true}}$ allows us to compare our SFEA results with the standard FEA using the GOH model. The second simulation is performed using the GP trained on the real experimental dataset of AV leaflet tissue (from Section~4). 

\subsection{Results}
\begin{figure}[h!]
\centering
\subfigimg[width=\textwidth]{a)}{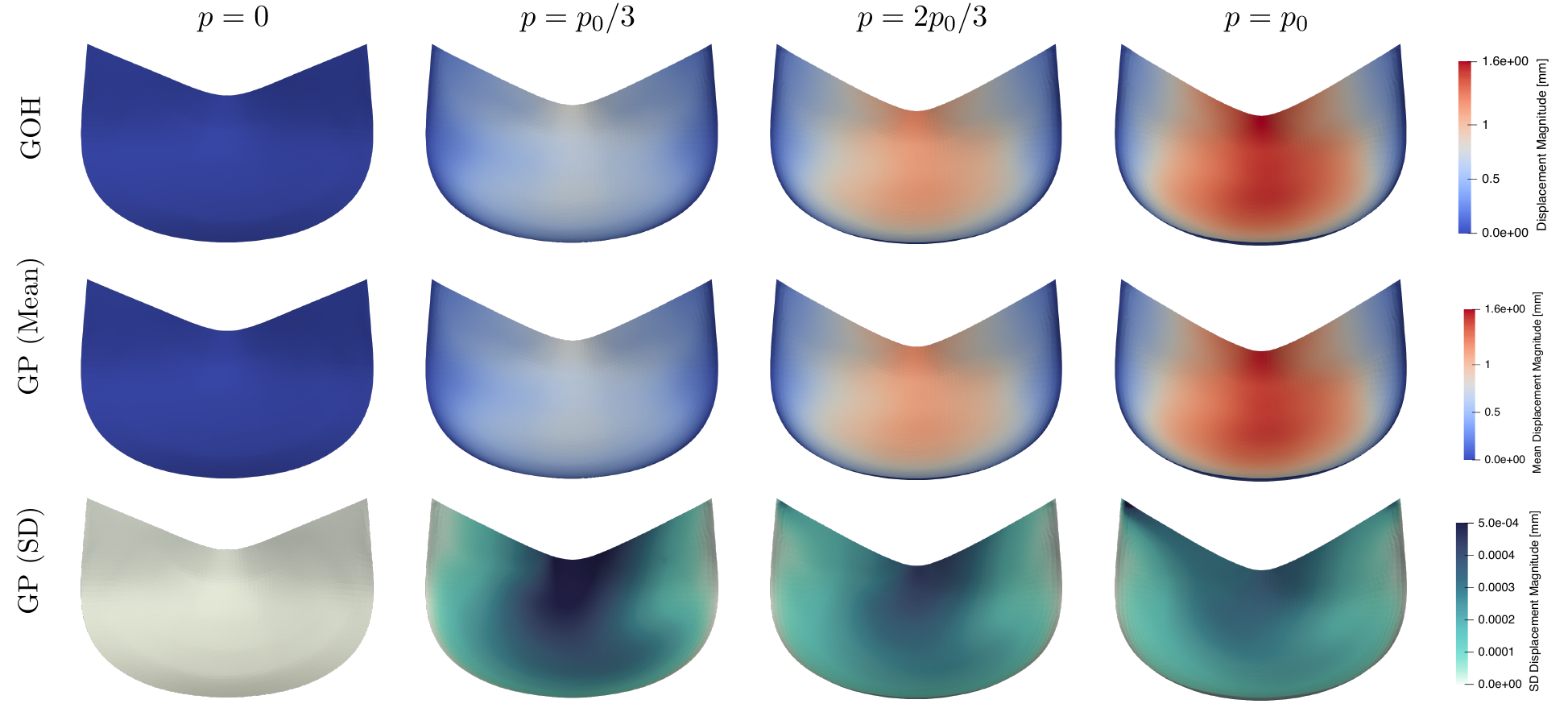} \\
\subfigimg[width=\textwidth]{b)}{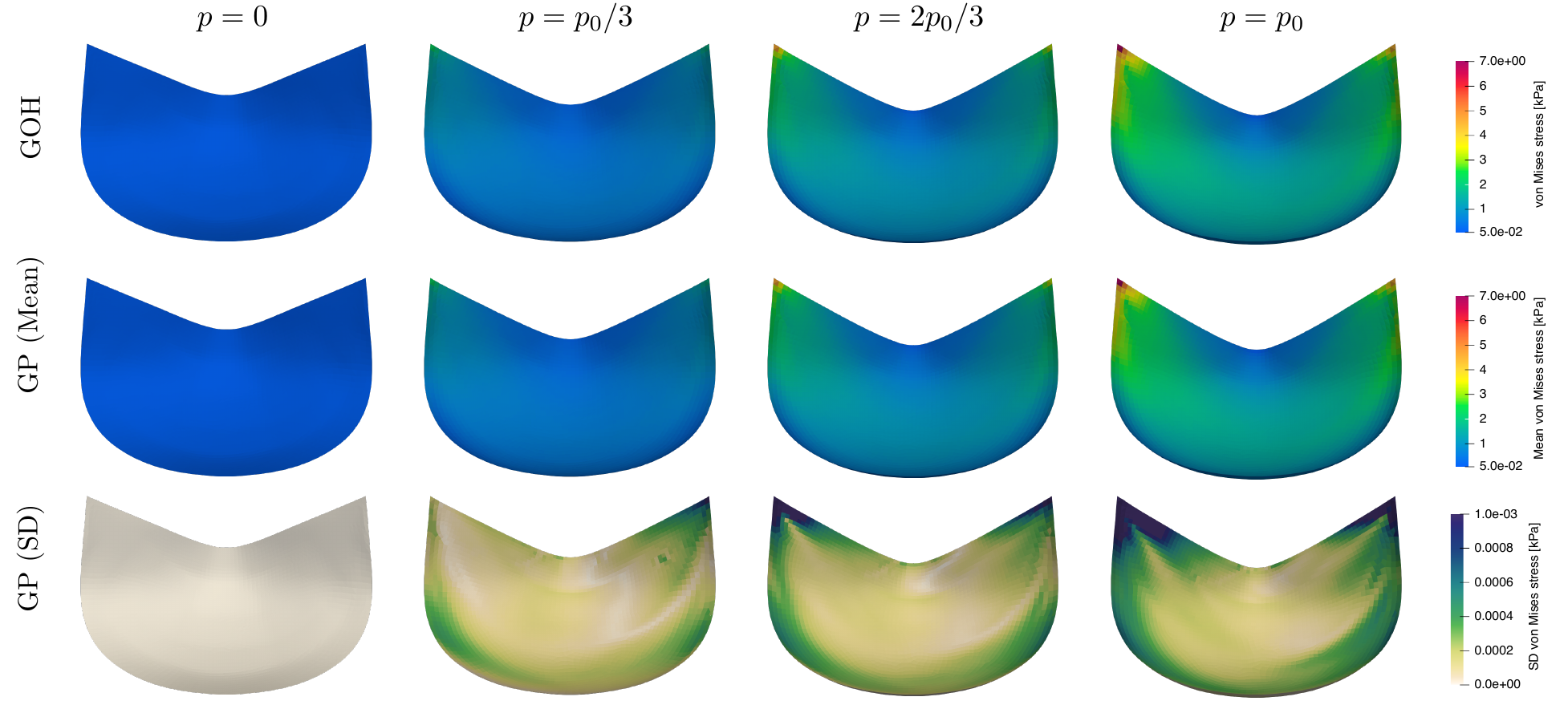}
\caption{Stochastic FE simulation result using the proposed GP-based framework (bottom two rows in a and b) compared with the standard FE simulation using GOH model (top rows in a and b) shows good agreement with the mean. The standard deviation values are small almost everywhere, except at the commissures, especially for stresses.}\label{fig:leaflet-goh-results}
\end{figure}

The resulting displacement magnitudes and von-Mises stresses are shown in Fig.~\ref{fig:leaflet-goh-results}. The mean values from GP match well with the ground truth GOH model. Also the standard deviations are small almost everywhere, except at the commissures. The commissures are known to have high stress concentrations, which is also what we observe in our results. In addition to high stresses, the results also show high standard deviation in those areas, indicating that, based on the input data, our confidence on those stress concentration values is low. Importantly, the spline-based model does not incur any significant additional computational cost. The solution time of the GP-based approach is found to be only 10\% higher, which is due to the need of slightly smaller load steps for convergence.

\begin{figure}[h!]
\centering
\includegraphics[width=\textwidth]{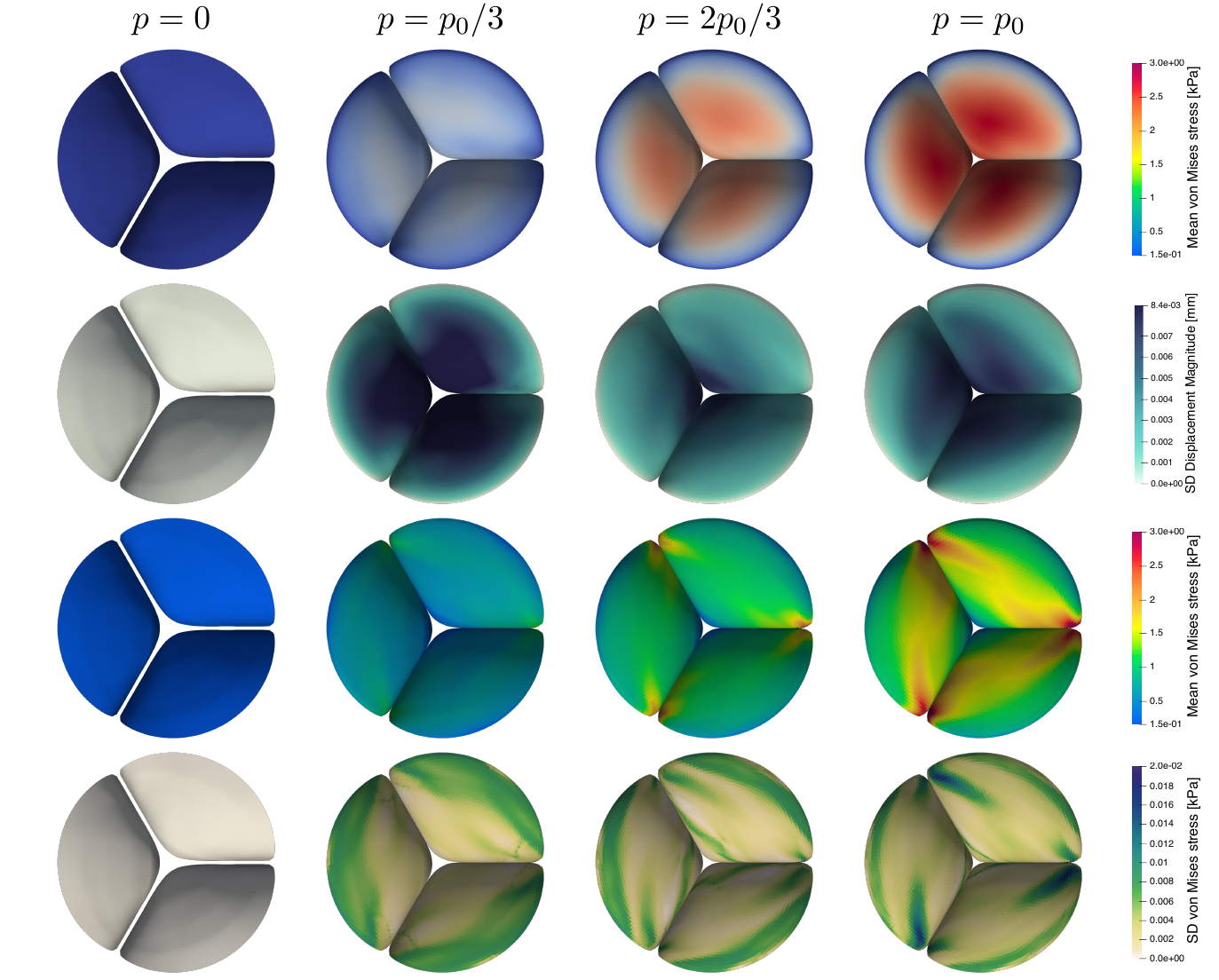} \\
\caption{Stochastic FE simulation result using the proposed GP-based framework using the real experimental dataset for AV leaflet. Results show higher standard deviations than the GOH-based model, thus capturing the effect of experimental uncertainty.}\label{fig:leaflet-data-results}
\end{figure}

The resulting displacement and von-Mises stress -- both mean and standard deviation (SD) -- using GP trained on the experimental data for AV leaflet tissue are shown in Fig.~\ref{fig:leaflet-data-results}. The AV tissue is highly nonlinear, with very small stiffness at the reference configuration (called the toe region) that increases rapidly at high stretches. This results in more interesting displacement and stress patterns. The displacement magnitude is higher, and the variation in displacement is largely in the belly region of the leaflet. The variation in stresses is also much higher compared to the GOH model, which is expected when using real experimental data. Von-mises stress is highest at the commisures, where we also find large variations. Moreover, the variation in stresses is also high at the fixed edge. 

\section{Discussion\label{sec:discuss}}
In spite of the decades of important developments, constitutive modeling of materials remains an active area of research. Traditionally, these models have been developed based on continuum mechanics/thermodynamic requirements and understanding of a material's microstructure. With the advances in data science and technology, there is a drive to use experimental data to inform constitutive model development. One approach is to choose models based on the data; since there is a large number of analytical constitutive models available in literature, it becomes a challenge to select an appropriate model. We have addressed such a problem of model selection using a Bayesian framework \cite{https://doi.org/10.48550/arxiv.2209.13038}. Another approach to discover (new) analytical forms of the constitutive models from data, so that the traditional computational setup (such as finite element analysis) can be preserved \cite{HAUSEUX201886, HAUSEUX2017917, JOSHI2022115225, THAKOLKARAN2022105076, DESHPANDE2022115307}. 

An alternative and attractive approach is to forgo analytical constitutive models in favor of data-based ones. There have been several recent efforts in this direction, both using purely data-based approach (which can also be thought of as a nearest-neighbor model) \cite{conti2018data,KIRCHDOERFER201681,EGGERSMANN201981,https://doi.org/10.1002/nme.5716,CARRARA2020113390,PLATZER2021113756,EGGERSMANN2021113499,HE2021110124,HE2021114034, HE2020112791} and using neural network models \cite{https://doi.org/10.1002/pamm.202100072, tac2021datadriven, https://doi.org/10.1002/cnm.3438, https://doi.org/10.1002/nme.6459, KLEIN2022104703}. The aim of this work is to propose a constitutive model based on Gaussian processes, which are naturally Bayesian. Given the strengths of the Bayesian framework, interested readers may benefit from the tutorials on their application in mechanics \cite{rappel2020tutorial}. 

\subsection{Advantages and features of the proposed GP model}

A key feature of the Bayesian approach is in addition to the mean response, it also provides a distribution that can be used to quantify the uncertainty and establish confidence in the results. The proposed GP framework makes three main improvements upon the work by \citet{frankel2020tensor}: 1) enforcement of convexity constraints, 2) extension to anisotropic material, and 3) application to real experimental data of soft biological tissues. The new features have been implemented in a github fork of GPytorch and are openly available. 

The primary role of convexity constraint is to ensure that the resulting model can be readily used in solving boundary value problems. This is because, without convexity constraints, the second derivatives could take large negative values, especially in the region of extrapolation (Fig.~\ref{GOH-second-deriv}), which will result in a non-elliptic problem and, possibly, a non-unique solution. Any effect of convexity constraint on the accuracy is expected to be restricted to the areas of extrapolation where there is no experimental data to guide the GP model.

The focus here was on planar soft tissues, which are hyperelastic, incompressible, and show single-fiber anisotropy. As a result, their strain energy density is a function of only two strain invariants, $I_1$ and $I_4$. Comparison of the proposed framework with an establish GOH model shows good agreement (Fig.~\ref{GOH-sedf}), even when extrapolated outside the training range (Fig.~\ref{GOH-predict}). The GP-based model is straightforward to use for real experimental data combining multiple protocols, and shows better fit to biaxial data from an aortic valve leaflet compared to several of the establish soft tissue hyperelastic models (also based on $I_1$ and $I_4$, Fig.~\ref{exp-data-results}). 

The presented framework work adds to the several other advances being made in the field of data-based constitutive modeling. Some of the other data-driven approaches \cite{conti2018data,KIRCHDOERFER201681,EGGERSMANN201981,https://doi.org/10.1002/nme.5716,CARRARA2020113390,PLATZER2021113756,EGGERSMANN2021113499,HE2021110124,HE2021114034, HE2020112791} rely on having a much larger number of measurements such that the entire deformation space is filled. However, this may require an inordinately large number of experiments to be performed. This limitation was also highlighted by \citet{HE2021110124}. In comparison, the proposed framework works with the number of protocols typically used in practice, and also allows us to quantify the confidence in the results. 

To fully utilize the distribution of the strain energy density provided by the GP framework, a non-intrusive stochastic finite element framework has been proposed. An intermediate spline-based interpolation has been used to take the GP predictions and use them in a finite element solver. The spline-based constitutive model has been implemented as a plug-in for FEBio \cite{maas2012febio}. The results of the traditional FE model using the GOH model have been found to be well comparable with those using SFEM with the fitted GP (Fig.~\ref{fig:leaflet-goh-results}). Also, the formulation allowed us to simulate the leaflet closure directly using real experimental data, without assuming any functional form of the strain energy density, while quantifying the uncertainty in addition to the mean results (Fig.~\ref{fig:leaflet-data-results}).

Based on the results, we believe that the GP-based framework is a strong contender for data-based constitutive modeling. Its strength lies in its naturally Bayesian setup and a rigorous mathematical foundation it is built upon. The proposed framework can be thought of as the first step towards exploring the full potential of GP-based constitutive modeling. Such future developments are briefly discussed next.

\subsection{Future work}

A natural extension will be made to hyperelastic solids where the strain energy depends on other strain invariants, such as $I_2$, $I_5$ and $J$. If a GP cannot fit a given experimental dataset, that might indicate that the considered list of invariants is insufficient and therefore needs to be expanded or modified. Given that the proposed framework does not incur additional computational cost during FEA, one could use GP-based framework as a surrogate for hierarchical meso-scale and multi-scale modeling. More generally, the GP-based framework could be used with reduced-order models, especially since GP allows one to quantify the uncertainty associated with model reduction.

The choice of kernel should be comprehensively explored. For example, in the squared exponential kernel function, one could use an anisotropic length scale, that might provide more flexibility. There are also other kernel functions proposed in the literature, such as periodic kernel could be used to model solids exhibiting hysteresis, fatigue, and other similar inelastic effects. Newer, more sophisticated approaches for monotonic GP with better theoretical properties have been proposed \cite{https://doi.org/10.48550/arxiv.1905.12930, andersen2018non}, and these could be explored for enforcing convexity constraints instead.

In addition, the proposed framework could be implemented within FEniCS, which has been used in recent open-source codes for performing uncertainty quantification \cite{HAUSEUX201886}, design of experiments \cite{SUTULA2020103999} and parameter estimation \cite{ELOUNEG2021106620}. However, implementing a spline-based material model in the unified field language (used in FEniCS) was found to be not straightforward, specifically the required conditional statements.

An important uncertainty in soft tissues is their reference configuration. While we considered the input stretches (and therefore the strain invariants) as fixed in this work, one could allow them to vary during the model fitting, thereby naturally determining pre-stretches as part of the GP training. We also considered a uniform experimental error in the observed stresses. However, depending on the exact experimental setup, the error could be non-uniform (such as proportional to the applied force magnitude). Alternatively, one could constrain the range of error hyperparameters ($e_x$ and $e_y$) based on knowledge of the experimental setup, such as the least count of load cells in biaxial testing setup.  

Lastly, Bayesian optimization \cite{clyde2001experimental} and design of experiments \cite{10.1214/15-BA977} could also be leveraged as the next step. For example, if the resulting model shows large variations for certain deformations, one could go back and design experiments to specifically measure stresses at those deformations and feed them back into the GP training, thereby reducing the uncertainty. The ability to make these choices provides a large flexibility in configuring the GP-based framework to one's needs. The proposed framework addresses the uncertainty in material behavior. Combining it with other uncertainties, such as geometry and loading, remains a challenge that requires advanced computational approaches to deal with multiple uncertainties \cite{DSOUZA2021538} and techniques to achieve feasible computation times \cite{DESHPANDE2022115307, HAUSEUX2017917, 10.1371/journal.pone.0189994}.

In conclusion, the proposed GP-based constitutive model development is a promising research direction. Their use in the context of soft tissues is particularly appealing given the ongoing research in constitutive model developments for different soft tissues. Moreover, the option of using GPs to carry out stochastic finite element analysis provides an important computational tool that could be used to improve our understanding and predictive capability of soft tissue mechanics.

\section*{Acknowledgments}
We thank Dr David Kamensky for useful discussions. 


\section*{Conflicts of interest}
The authors have no relevant financial or non-financial interests to disclose.

\clearpage
\appendix

\section{Exact GP}
\label{sec:appendix}
Dropping the convexity constraints (i.e., if $\observepoints^{\text{c}}=\emptyset$), the Bayes' theorem \eqref{bayes3} reduces to:
\begin{equation}
\prob(\extract{\vec{f}} \mid \observestate,\observepoints)
= 
\dfrac{
{\prob(\observestate^{\text{o}}\mid \observepoints^{\text{o}}, \extract{\vec{f}}^{\text{o}}) 
\prob(\observestate^{\text{d}}\mid \observepoints^{\text{d}}, \extract{\vec{f}}^{\text{d}})
}
\prob(\extract{\vec{f}})
}
{\prob(\observestate^{\text{o}}, \observestate^{\text{d}} \mid \observepoints^{\text{o}}, \observepoints^{\text{d}})},
\label{bayes4}
\end{equation}
Using Eqs.~\eqref{stresses-relate}, \eqref{W-llh} and \eqref{joint-distribution2}, it is easy to see that the joint distribution of $\observestate$ and $\predictstate$ is also Gaussian. Specifically,
\begin{equation}
    \begin{bmatrix}
    \rand{\observestate}\\
    \rand{\predictstate}
    \end{bmatrix} \sim \mathcal{N} \left( 
    \begin{bmatrix}
    \extract{\cmean(\observepoints)}\\
    \cmean(\predictpoints)
    \end{bmatrix},
    \begin{bmatrix}
    \extract{\ccov(\observepoints,\observepoints)} + \mathbf{\Lambda} & \extract{\ccov(\observepoints,\predictpoints)}\\
    \extract{\ccov(\predictpoints,\observepoints)} & 
    \ccov(\predictpoints,\predictpoints)
    \end{bmatrix}
    \right),
\end{equation}
where $\mathbf{\Lambda}$ is a diagonal matrix with appropriate entries for the noise variance, i.e., $e_0^2$, $e_x^2$, or $e_y^2$. From the above, the following closed-form solution of the posterior distribution can be derived \cite{books/lib/RasmussenW06}:
\begin{equation}
    \rand{\predictstate} \mid \observepoints,\observestate,\predictpoints \sim \mathcal{N} \left(\predictmean,\text{Cov}(\predictstate) \right),
\end{equation}
where
\begin{align}
    \predictmean &:= \extract{\ccov(\predictpoints,\observepoints)} \left[ \extract{\ccov(\observepoints,\observepoints)} + \mathbf{\Lambda} \right]^{-1} \observestate \text{ and} \nonumber\\
    \text{Cov}(\predictstate) &:= \ccov(\predictpoints,\predictpoints) - \extract{\ccov(\predictpoints,\observepoints)}\left[ \ccov(\observepoints,\observepoints) + \mathbf{\Lambda} \right]^{-1} \extract{\ccov(\observepoints,\predictpoints)}.\label{analyticalGP}
\end{align}

\section{Effect of training points and noise on accuracy\label{additional-verification}}
\begin{figure}
    \centering
    \includegraphics[width=\textwidth]{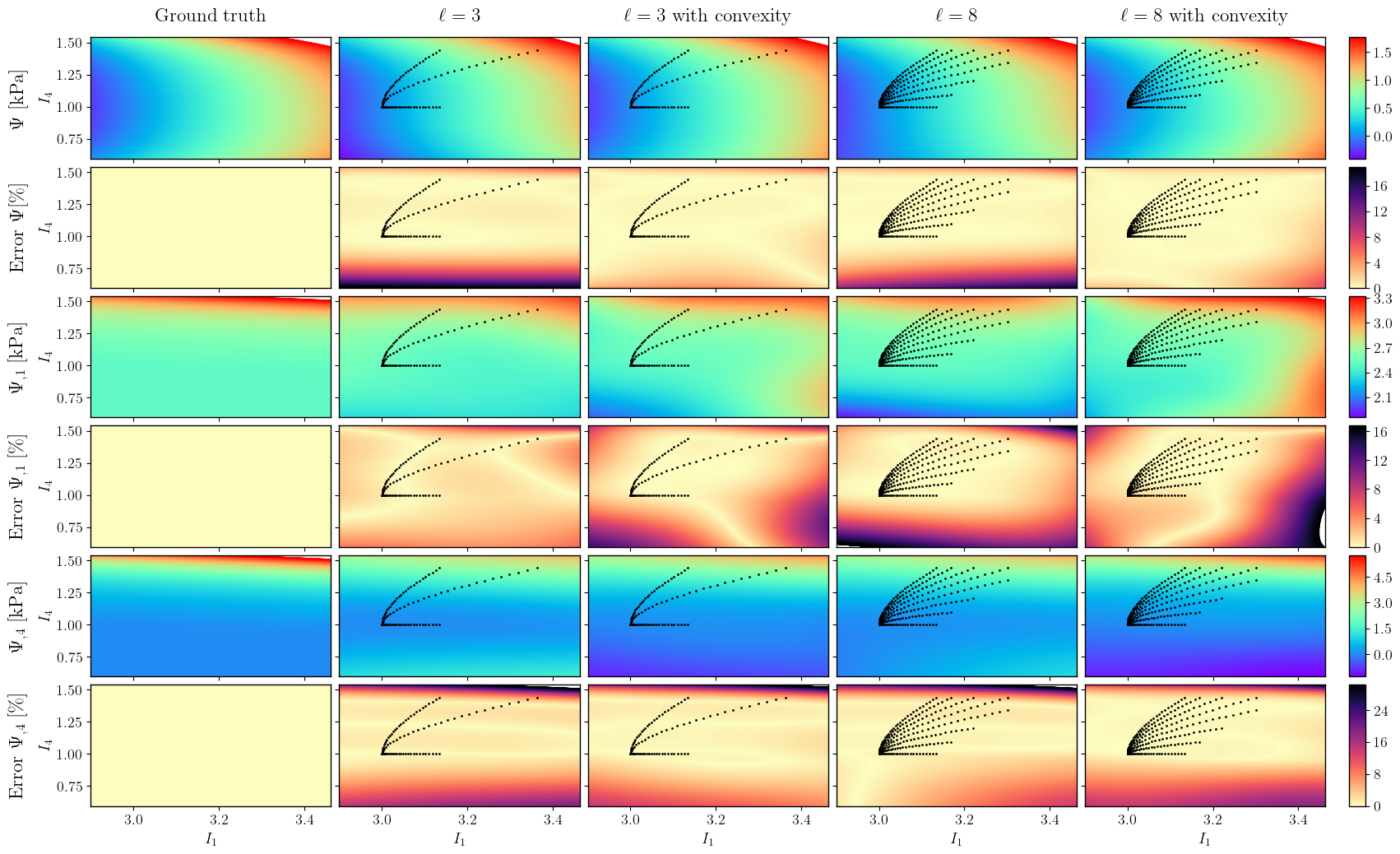}
    \caption{Results when pure shear protocols are not used in training show higher errors in the $I_4<1$ area}
    \label{fig:no-shear}
\end{figure}

\begin{figure}
    \centering
    \includegraphics[width=\textwidth]{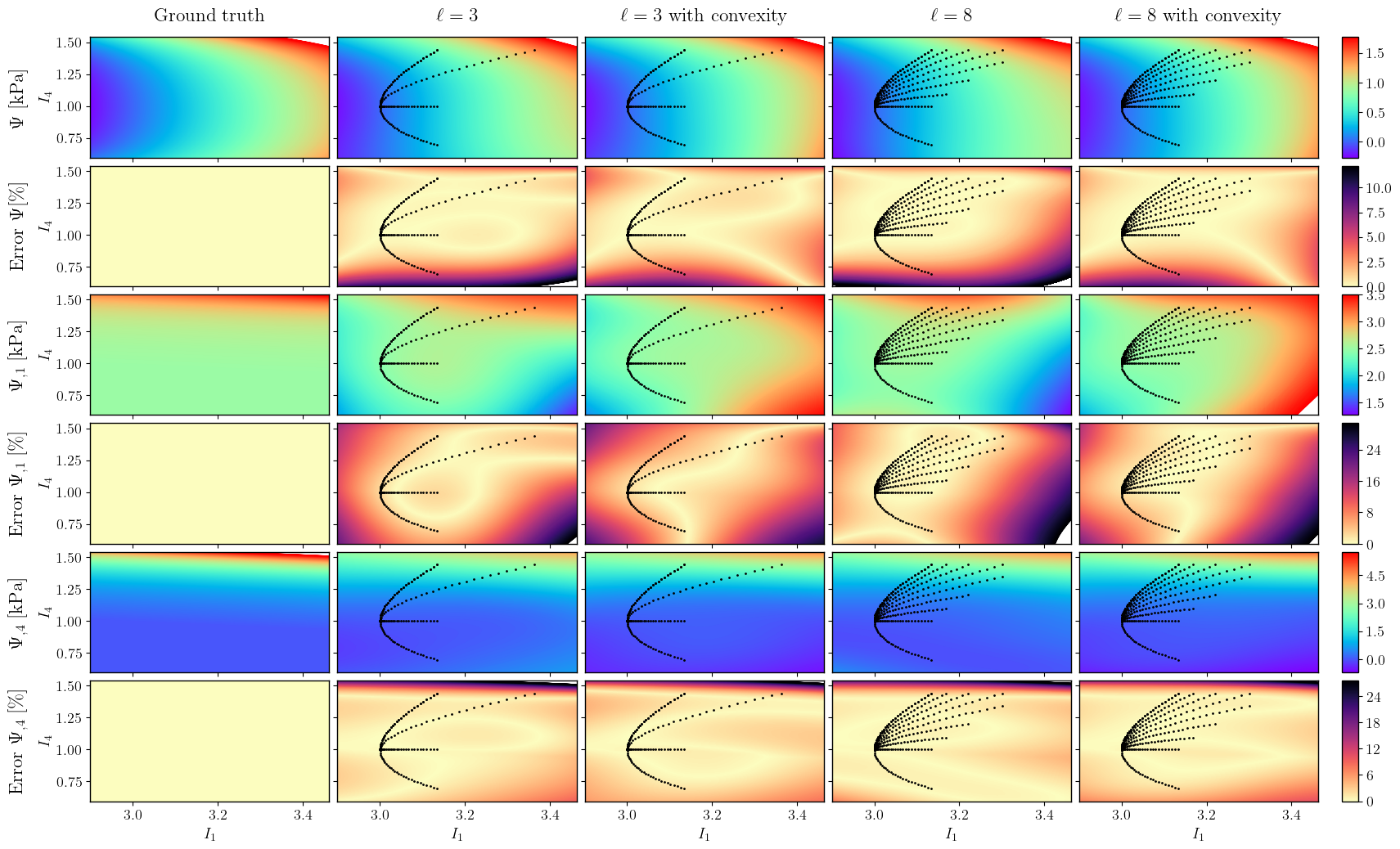}
    \caption{Results when training data includes higher noise (variance of 0.2) show an improvement in accuracy when using convexity conditions}
    \label{fig:higher-noise}
\end{figure}

To further understand the effect of training protocols, results without using the pure shear are presented (Fig.~\ref{fig:no-shear}), which shows higher error in the $I_4<1$ region. Thus, it is more important that the training data covers the deformation space, rather than simply having higher number of protocols/data points. Moreover, to understand the effect of noise in the data, results with higher noise (with a variance of 0.2) are also presented (Fig.~\ref{fig:higher-noise}). These results demonstrate the robustness of the framework with respect to noise and how number of protocols and convexity conditions affect the accuracy.

\section{Spline interpolation\label{spline-appendix}}
Given $n$ data points $(x_i,y_i)$, for $i=1,\dots,n$, a one-dimensional spline is interpolated as follows. A spline function is defined in terms of three quantities: order $q$, a knot vector of non-decreasing values $u_i$, and control points $c_j$. Based on these quantities, the spline function is defined in terms of shape functions $N^q_j$ and (unknown) control points $c_j$ as:
\begin{equation}
    \mathcal{S}^q(x):=\sum\limits_{j=1}^m N^q_j(x)c_j.
\end{equation}
$N^q_j$ are calculated recursively, starting with zero-th order
\begin{equation}
    N^0_j(x) = \begin{cases}
    1 & \text{if }u_j \le x < u_{j+1} \\
    0 & \text{otherwise}
    \end{cases}
\end{equation}
and calculating higher-order functions with the following recursive relation:
\begin{equation}
    N^q_j(x) = \frac{x-u_j}{u_{j+q}-u_j}N^{p-1}_j(x) + \frac{u_{j+q+1}-x}{u_{j+q+1}-u_{j+1}}N^{p-1}_{j+1}(x).
\end{equation}
To interpolate to the data $(x_i,y_i)$, the control points $c_j$ are obtained by solving the following $n$ equations:
\begin{equation}
    \sum\limits_{j=1}^m N^q_j(x_i)c_j = y_i.
\end{equation}
In two dimensions, a tensor product spline is defined using $m\times n$ control points 
\begin{equation}
    \mathcal{S}^q(x,y)=\sum\limits_{j=1}^m \sum\limits_{k=1}^n N_j^q(x)N_k^q(y) c_{jk},
\end{equation}
and the interpolation is done equivalently to find the control points. More details on the spline construction and their properties, such as their derivatives, can be found in \cite{bartels1995introduction}.

\clearpage
 \bibliographystyle{elsarticle-num-names} 
 \bibliography{refs}

\begin{thebibliography}{68}
\expandafter\ifx\csname natexlab\endcsname\relax\def\natexlab#1{#1}\fi
\providecommand{\url}[1]{\texttt{#1}}
\providecommand{\href}[2]{#2}
\providecommand{\path}[1]{#1}
\providecommand{\DOIprefix}{doi:}
\providecommand{\ArXivprefix}{arXiv:}
\providecommand{\URLprefix}{URL: }
\providecommand{\Pubmedprefix}{pmid:}
\providecommand{\doi}[1]{\href{http://dx.doi.org/#1}{\path{#1}}}
\providecommand{\Pubmed}[1]{\href{pmid:#1}{\path{#1}}}
\providecommand{\bibinfo}[2]{#2}
\ifx\xfnm\relax \def\xfnm[#1]{\unskip,\space#1}\fi
\bibitem[{Maurel et~al.(1998)Maurel, Thalmann, Wu, and Thalmann}]{maurel1998}
\bibinfo{author}{W.~Maurel}, \bibinfo{author}{D.~Thalmann},
  \bibinfo{author}{Y.~Wu}, \bibinfo{author}{N.~M. Thalmann},
  \bibinfo{title}{Constitutive Modeling}, \bibinfo{publisher}{Springer Berlin
  Heidelberg}, \bibinfo{address}{Berlin, Heidelberg}, \bibinfo{year}{1998}, pp.
  \bibinfo{pages}{79--120}.
\bibitem[{Conti et~al.(2018)Conti, M{\"u}ller, and Ortiz}]{conti2018data}
\bibinfo{author}{S.~Conti}, \bibinfo{author}{S.~M{\"u}ller},
  \bibinfo{author}{M.~Ortiz},
\newblock \bibinfo{title}{Data-driven problems in elasticity},
\newblock \bibinfo{journal}{Archive for Rational Mechanics and Analysis}
  \bibinfo{volume}{229} (\bibinfo{year}{2018}) \bibinfo{pages}{79--123}.
  \DOIprefix\doi{10.1007/s00205-017-1214-0}.
\bibitem[{Kirchdoerfer and Ortiz(2016)}]{KIRCHDOERFER201681}
\bibinfo{author}{T.~Kirchdoerfer}, \bibinfo{author}{M.~Ortiz},
\newblock \bibinfo{title}{Data-driven computational mechanics},
\newblock \bibinfo{journal}{Computer Methods in Applied Mechanics and
  Engineering} \bibinfo{volume}{304} (\bibinfo{year}{2016})
  \bibinfo{pages}{81--101}. \DOIprefix\doi{10.1016/j.cma.2016.02.001}.
\bibitem[{Eggersmann et~al.(2019)Eggersmann, Kirchdoerfer, Reese, Stainier, and
  Ortiz}]{EGGERSMANN201981}
\bibinfo{author}{R.~Eggersmann}, \bibinfo{author}{T.~Kirchdoerfer},
  \bibinfo{author}{S.~Reese}, \bibinfo{author}{L.~Stainier},
  \bibinfo{author}{M.~Ortiz},
\newblock \bibinfo{title}{Model-free data-driven inelasticity},
\newblock \bibinfo{journal}{Computer Methods in Applied Mechanics and
  Engineering} \bibinfo{volume}{350} (\bibinfo{year}{2019})
  \bibinfo{pages}{81--99}. \DOIprefix\doi{10.1016/j.cma.2019.02.016}.
\bibitem[{Kirchdoerfer and Ortiz(2018)}]{https://doi.org/10.1002/nme.5716}
\bibinfo{author}{T.~Kirchdoerfer}, \bibinfo{author}{M.~Ortiz},
\newblock \bibinfo{title}{Data-driven computing in dynamics},
\newblock \bibinfo{journal}{International Journal for Numerical Methods in
  Engineering} \bibinfo{volume}{113} (\bibinfo{year}{2018})
  \bibinfo{pages}{1697--1710}. \DOIprefix\doi{10.1002/nme.5716}.
\bibitem[{Carrara et~al.(2020)Carrara, {De Lorenzis}, Stainier, and
  Ortiz}]{CARRARA2020113390}
\bibinfo{author}{P.~Carrara}, \bibinfo{author}{L.~{De Lorenzis}},
  \bibinfo{author}{L.~Stainier}, \bibinfo{author}{M.~Ortiz},
\newblock \bibinfo{title}{Data-driven fracture mechanics},
\newblock \bibinfo{journal}{Computer Methods in Applied Mechanics and
  Engineering} \bibinfo{volume}{372} (\bibinfo{year}{2020})
  \bibinfo{pages}{113390}. \DOIprefix\doi{10.1016/j.cma.2020.113390}.
\bibitem[{Platzer et~al.(2021)Platzer, Leygue, Stainier, and
  Ortiz}]{PLATZER2021113756}
\bibinfo{author}{A.~Platzer}, \bibinfo{author}{A.~Leygue},
  \bibinfo{author}{L.~Stainier}, \bibinfo{author}{M.~Ortiz},
\newblock \bibinfo{title}{Finite element solver for data-driven finite strain
  elasticity},
\newblock \bibinfo{journal}{Computer Methods in Applied Mechanics and
  Engineering} \bibinfo{volume}{379} (\bibinfo{year}{2021})
  \bibinfo{pages}{113756}. \DOIprefix\doi{10.1016/j.cma.2021.113756}.
\bibitem[{Eggersmann et~al.(2021)Eggersmann, Stainier, Ortiz, and
  Reese}]{EGGERSMANN2021113499}
\bibinfo{author}{R.~Eggersmann}, \bibinfo{author}{L.~Stainier},
  \bibinfo{author}{M.~Ortiz}, \bibinfo{author}{S.~Reese},
\newblock \bibinfo{title}{Model-free data-driven computational mechanics
  enhanced by tensor voting},
\newblock \bibinfo{journal}{Computer Methods in Applied Mechanics and
  Engineering} \bibinfo{volume}{373} (\bibinfo{year}{2021})
  \bibinfo{pages}{113499}. \DOIprefix\doi{10.1016/j.cma.2020.113499}.
\bibitem[{He et~al.(2021{\natexlab{a}})He, Laurence, Lee, and
  Chen}]{HE2021110124}
\bibinfo{author}{Q.~He}, \bibinfo{author}{D.~W. Laurence},
  \bibinfo{author}{C.-H. Lee}, \bibinfo{author}{J.-S. Chen},
\newblock \bibinfo{title}{Manifold learning based data-driven modeling for soft
  biological tissues},
\newblock \bibinfo{journal}{Journal of Biomechanics} \bibinfo{volume}{117}
  (\bibinfo{year}{2021}{\natexlab{a}}) \bibinfo{pages}{110124}.
  \DOIprefix\doi{10.1016/j.jbiomech.2020.110124}.
\bibitem[{He et~al.(2021{\natexlab{b}})He, He, and Chen}]{HE2021114034}
\bibinfo{author}{X.~He}, \bibinfo{author}{Q.~He}, \bibinfo{author}{J.-S. Chen},
\newblock \bibinfo{title}{Deep autoencoders for physics-constrained data-driven
  nonlinear materials modeling},
\newblock \bibinfo{journal}{Computer Methods in Applied Mechanics and
  Engineering} \bibinfo{volume}{385} (\bibinfo{year}{2021}{\natexlab{b}})
  \bibinfo{pages}{114034}. \DOIprefix\doi{10.1016/j.cma.2021.114034}.
\bibitem[{He and Chen(2020)}]{HE2020112791}
\bibinfo{author}{Q.~He}, \bibinfo{author}{J.-S. Chen},
\newblock \bibinfo{title}{A physics-constrained data-driven approach based on
  locally convex reconstruction for noisy database},
\newblock \bibinfo{journal}{Computer Methods in Applied Mechanics and
  Engineering} \bibinfo{volume}{363} (\bibinfo{year}{2020})
  \bibinfo{pages}{112791}. \DOIprefix\doi{10.1016/j.cma.2019.112791}.
\bibitem[{Hillg{\"a}rtner et~al.(2021)Hillg{\"a}rtner, Linka, Abdolazizi,
  Aydin, Itskov, and Cyron}]{https://doi.org/10.1002/pamm.202100072}
\bibinfo{author}{M.~Hillg{\"a}rtner}, \bibinfo{author}{K.~Linka},
  \bibinfo{author}{K.~P. Abdolazizi}, \bibinfo{author}{R.~C. Aydin},
  \bibinfo{author}{M.~Itskov}, \bibinfo{author}{C.~J. Cyron},
\newblock \bibinfo{title}{Constitutive artificial neural networks: a general
  anisotropic constitutive modeling framework utilizing machine learning},
\newblock \bibinfo{journal}{PAMM} \bibinfo{volume}{21} (\bibinfo{year}{2021})
  \bibinfo{pages}{e202100072}. \DOIprefix\doi{10.1002/pamm.202100072}.
\bibitem[{Tac et~al.(2022)Tac, Sree, Rausch, and Tepole}]{tac2021datadriven}
\bibinfo{author}{V.~Tac}, \bibinfo{author}{V.~D. Sree}, \bibinfo{author}{M.~K.
  Rausch}, \bibinfo{author}{A.~B. Tepole}, \bibinfo{title}{Data-driven modeling
  of the mechanical behavior of anisotropic soft biological tissue},
  \bibinfo{year}{2022}. \DOIprefix\doi{10.1007/s00366-022-01733-3}.
\bibitem[{Zhang et~al.(2021)Zhang, Rossini, Kamensky, Bui-Thanh, and
  Sacks}]{https://doi.org/10.1002/cnm.3438}
\bibinfo{author}{W.~Zhang}, \bibinfo{author}{G.~Rossini},
  \bibinfo{author}{D.~Kamensky}, \bibinfo{author}{T.~Bui-Thanh},
  \bibinfo{author}{M.~S. Sacks},
\newblock \bibinfo{title}{Isogeometric finite element-based simulation of the
  aortic heart valve: Integration of neural network structural material model
  and structural tensor fiber architecture representations},
\newblock \bibinfo{journal}{International Journal for Numerical Methods in
  Biomedical Engineering} \bibinfo{volume}{37} (\bibinfo{year}{2021})
  \bibinfo{pages}{e3438}. \DOIprefix\doi{10.1002/cnm.3438}.
\bibitem[{Chung et~al.(2021)Chung, Im, and
  Cho}]{https://doi.org/10.1002/nme.6459}
\bibinfo{author}{I.~Chung}, \bibinfo{author}{S.~Im}, \bibinfo{author}{M.~Cho},
\newblock \bibinfo{title}{A neural network constitutive model for
  hyperelasticity based on molecular dynamics simulations},
\newblock \bibinfo{journal}{International Journal for Numerical Methods in
  Engineering} \bibinfo{volume}{122} (\bibinfo{year}{2021})
  \bibinfo{pages}{5--24}. \DOIprefix\doi{10.1002/nme.6459}.
\bibitem[{Klein et~al.(2022)Klein, Fern{\'a}ndez, Martin, Neff, and
  Weeger}]{KLEIN2022104703}
\bibinfo{author}{D.~K. Klein}, \bibinfo{author}{M.~Fern{\'a}ndez},
  \bibinfo{author}{R.~J. Martin}, \bibinfo{author}{P.~Neff},
  \bibinfo{author}{O.~Weeger},
\newblock \bibinfo{title}{Polyconvex anisotropic hyperelasticity with neural
  networks},
\newblock \bibinfo{journal}{Journal of the Mechanics and Physics of Solids}
  \bibinfo{volume}{159} (\bibinfo{year}{2022}) \bibinfo{pages}{104703}.
  \DOIprefix\doi{10.1016/j.jmps.2021.104703}.
\bibitem[{Guo and Hesthaven(2018)}]{GUO2018807}
\bibinfo{author}{M.~Guo}, \bibinfo{author}{J.~S. Hesthaven},
\newblock \bibinfo{title}{Reduced order modeling for nonlinear structural
  analysis using gaussian process regression},
\newblock \bibinfo{journal}{Computer Methods in Applied Mechanics and
  Engineering} \bibinfo{volume}{341} (\bibinfo{year}{2018})
  \bibinfo{pages}{807--826}. \DOIprefix\doi{10.1016/j.cma.2018.07.017}.
\bibitem[{Wang et~al.(2021)Wang, Li, Cui, Hui, Yeo, and
  Zehnder}]{WANG2021104532}
\bibinfo{author}{J.~Wang}, \bibinfo{author}{T.~Li}, \bibinfo{author}{F.~Cui},
  \bibinfo{author}{C.-Y. Hui}, \bibinfo{author}{J.~Yeo}, \bibinfo{author}{A.~T.
  Zehnder},
\newblock \bibinfo{title}{Metamodeling of constitutive model using gaussian
  process machine learning},
\newblock \bibinfo{journal}{Journal of the Mechanics and Physics of Solids}
  \bibinfo{volume}{154} (\bibinfo{year}{2021}) \bibinfo{pages}{104532}.
  \DOIprefix\doi{10.1016/j.jmps.2021.104532}.
\bibitem[{Lee et~al.(2020)Lee, Bilionis, and Tepole}]{LEE2020112724}
\bibinfo{author}{T.~Lee}, \bibinfo{author}{I.~Bilionis}, \bibinfo{author}{A.~B.
  Tepole},
\newblock \bibinfo{title}{Propagation of uncertainty in the mechanical and
  biological response of growing tissues using multi-fidelity gaussian process
  regression},
\newblock \bibinfo{journal}{Computer Methods in Applied Mechanics and
  Engineering} \bibinfo{volume}{359} (\bibinfo{year}{2020})
  \bibinfo{pages}{112724}. \DOIprefix\doi{10.1016/j.cma.2019.112724}.
\bibitem[{Zeraatpisheh et~al.(2021)Zeraatpisheh, Bordas, and
  Beex}]{zeraatpisheh_bordas_beex_2021}
\bibinfo{author}{M.~Zeraatpisheh}, \bibinfo{author}{S.~P. Bordas},
  \bibinfo{author}{L.~A. Beex},
\newblock \bibinfo{title}{Bayesian model uncertainty quantification for
  hyperelastic soft tissue models},
\newblock \bibinfo{journal}{Data-Centric Engineering} \bibinfo{volume}{2}
  (\bibinfo{year}{2021}) \bibinfo{pages}{e9}.
  \DOIprefix\doi{10.1017/dce.2021.9}.
\bibitem[{Stowers et~al.(2021)Stowers, Lee, Bilionis, Gosain, and
  Tepole}]{STOWERS2021104340}
\bibinfo{author}{C.~Stowers}, \bibinfo{author}{T.~Lee},
  \bibinfo{author}{I.~Bilionis}, \bibinfo{author}{A.~K. Gosain},
  \bibinfo{author}{A.~B. Tepole},
\newblock \bibinfo{title}{Improving reconstructive surgery design using
  gaussian process surrogates to capture material behavior uncertainty},
\newblock \bibinfo{journal}{Journal of the Mechanical Behavior of Biomedical
  Materials} \bibinfo{volume}{118} (\bibinfo{year}{2021})
  \bibinfo{pages}{104340}. \DOIprefix\doi{10.1016/j.jmbbm.2021.104340}.
\bibitem[{Di~Achille et~al.(2018)Di~Achille, Harouni, Khamzin, Solovyova, Rice,
  and Gurev}]{10.3389/fphys.2018.01002}
\bibinfo{author}{P.~Di~Achille}, \bibinfo{author}{A.~Harouni},
  \bibinfo{author}{S.~Khamzin}, \bibinfo{author}{O.~Solovyova},
  \bibinfo{author}{J.~J. Rice}, \bibinfo{author}{V.~Gurev},
\newblock \bibinfo{title}{Gaussian process regressions for inverse problems and
  parameter searches in models of ventricular mechanics},
\newblock \bibinfo{journal}{Frontiers in Physiology} \bibinfo{volume}{9}
  (\bibinfo{year}{2018}). \DOIprefix\doi{10.3389/fphys.2018.01002}.
\bibitem[{Frankel et~al.(2020)Frankel, Jones, and Swiler}]{frankel2020tensor}
\bibinfo{author}{A.~L. Frankel}, \bibinfo{author}{R.~E. Jones},
  \bibinfo{author}{L.~P. Swiler},
\newblock \bibinfo{title}{Tensor basis gaussian process models of hyperelastic
  materials},
\newblock \bibinfo{journal}{Journal of Machine Learning for Modeling and
  Computing} \bibinfo{volume}{1} (\bibinfo{year}{2020}).
  \DOIprefix\doi{10.1615/JMachLearnModelComput.2020033325}.
\bibitem[{Holzapfel(2000)}]{holzapfel-book}
\bibinfo{author}{G.~A. Holzapfel}, \bibinfo{title}{Nonlinear Solid Mechanics: A
  Continuum Approach for Engineering}, \bibinfo{publisher}{Wiley},
  \bibinfo{address}{New York}, \bibinfo{year}{2000}.
\bibitem[{Bonet and Wood(1997)}]{bonet1997nonlinear}
\bibinfo{author}{J.~Bonet}, \bibinfo{author}{R.~D. Wood},
  \bibinfo{title}{Nonlinear continuum mechanics for finite element analysis},
  \bibinfo{publisher}{Cambridge university press}, \bibinfo{year}{1997}.
\bibitem[{Belytschko et~al.(2013)Belytschko, Liu, Moran, and
  Elkhodary}]{belytschko2013nonlinear}
\bibinfo{author}{T.~Belytschko}, \bibinfo{author}{W.~K. Liu},
  \bibinfo{author}{B.~Moran}, \bibinfo{author}{K.~Elkhodary},
  \bibinfo{title}{Nonlinear finite elements for continua and structures},
  \bibinfo{publisher}{John wiley \& sons}, \bibinfo{year}{2013}.
\bibitem[{Holzapfel et~al.(2000)Holzapfel, Gasser, and
  Ogden}]{holzapfel2000new}
\bibinfo{author}{G.~A. Holzapfel}, \bibinfo{author}{T.~C. Gasser},
  \bibinfo{author}{R.~W. Ogden},
\newblock \bibinfo{title}{A new constitutive framework for arterial wall
  mechanics and a comparative study of material models},
\newblock \bibinfo{journal}{Journal of elasticity and the physical science of
  solids} \bibinfo{volume}{61} (\bibinfo{year}{2000}) \bibinfo{pages}{1--48}.
  \DOIprefix\doi{10.1023/A:1010835316564}.
\bibitem[{Murphy(2013)}]{MURPHY201390}
\bibinfo{author}{J.~Murphy},
\newblock \bibinfo{title}{Transversely isotropic biological, soft tissue must
  be modelled using both anisotropic invariants},
\newblock \bibinfo{journal}{European Journal of Mechanics - A/Solids}
  \bibinfo{volume}{42} (\bibinfo{year}{2013}) \bibinfo{pages}{90--96}.
  \DOIprefix\doi{10.1016/j.euromechsol.2013.04.003}.
\bibitem[{Schr{\"o}der et~al.(2005)Schr{\"o}der, Neff, and
  Balzani}]{SCHRODER20054352}
\bibinfo{author}{J.~Schr{\"o}der}, \bibinfo{author}{P.~Neff},
  \bibinfo{author}{D.~Balzani},
\newblock \bibinfo{title}{A variational approach for materially stable
  anisotropic hyperelasticity},
\newblock \bibinfo{journal}{International Journal of Solids and Structures}
  \bibinfo{volume}{42} (\bibinfo{year}{2005}) \bibinfo{pages}{4352--4371}.
  \DOIprefix\doi{10.1016/j.ijsolstr.2004.11.021}.
\bibitem[{Markert et~al.(2005)Markert, Ehlers, and
  Karajan}]{https://doi.org/10.1002/pamm.200510099}
\bibinfo{author}{B.~Markert}, \bibinfo{author}{W.~Ehlers},
  \bibinfo{author}{N.~Karajan},
\newblock \bibinfo{title}{A general polyconvex strain-energy function for
  fiber-reinforced materials},
\newblock \bibinfo{journal}{PAMM} \bibinfo{volume}{5} (\bibinfo{year}{2005})
  \bibinfo{pages}{245--246}. \DOIprefix\doi{10.1002/pamm.200510099}.
\bibitem[{Schr{\"o}der(2010)}]{schroder2010anisotropie}
\bibinfo{author}{J.~Schr{\"o}der},
\newblock \bibinfo{title}{Anisotropie polyconvex energies},
\newblock \bibinfo{journal}{Poly-, quasi-and rank-one convexity in applied
  mechanics}  (\bibinfo{year}{2010}) \bibinfo{pages}{53--105}.
\bibitem[{Dacorogna(2008)}]{dacorogna2008polyconvex}
\bibinfo{author}{B.~Dacorogna},
\newblock \bibinfo{title}{Polyconvex, quasiconvex and rank one convex
  functions},
\newblock \bibinfo{journal}{Direct methods in the calculus of variations}
  (\bibinfo{year}{2008}) \bibinfo{pages}{155--263}.
  \DOIprefix\doi{10.1007/978-0-387-55249-1\_6}.
\bibitem[{Billiar and Sacks(2000)}]{billiar2000biaxial}
\bibinfo{author}{K.~L. Billiar}, \bibinfo{author}{M.~S. Sacks},
\newblock \bibinfo{title}{Biaxial mechanical properties of the native and
  glutaraldehyde-treated aortic valve cusp: part ii--a structural constitutive
  model},
\newblock \bibinfo{journal}{Journal of biomechanical engineering}
  \bibinfo{volume}{122} (\bibinfo{year}{2000}) \bibinfo{pages}{327--335}.
  \DOIprefix\doi{10.1115/1.1287158}.
\bibitem[{Fan and Sacks(2014)}]{FAN20142043}
\bibinfo{author}{R.~Fan}, \bibinfo{author}{M.~S. Sacks},
\newblock \bibinfo{title}{Simulation of planar soft tissues using a structural
  constitutive model: Finite element implementation and validation},
\newblock \bibinfo{journal}{Journal of Biomechanics} \bibinfo{volume}{47}
  (\bibinfo{year}{2014}) \bibinfo{pages}{2043--2054}.
  \DOIprefix\doi{10.1016/j.jbiomech.2014.03.014}.
\bibitem[{Kiendl et~al.(2015)Kiendl, Hsu, Wu, and Reali}]{KIENDL2015280}
\bibinfo{author}{J.~Kiendl}, \bibinfo{author}{M.-C. Hsu},
  \bibinfo{author}{M.~C. Wu}, \bibinfo{author}{A.~Reali},
\newblock \bibinfo{title}{Isogeometric kirchhoff--love shell formulations for
  general hyperelastic materials},
\newblock \bibinfo{journal}{Computer Methods in Applied Mechanics and
  Engineering} \bibinfo{volume}{291} (\bibinfo{year}{2015})
  \bibinfo{pages}{280--303}. \DOIprefix\doi{10.1016/j.cma.2015.03.010}.
\bibitem[{Riihim{\"a}ki and Vehtari(2010)}]{riihimaki2010gaussian}
\bibinfo{author}{J.~Riihim{\"a}ki}, \bibinfo{author}{A.~Vehtari},
\newblock \bibinfo{title}{Gaussian processes with monotonicity information},
\newblock in: \bibinfo{booktitle}{Proceedings of the thirteenth international
  conference on artificial intelligence and statistics},
  \bibinfo{organization}{JMLR Workshop and Conference Proceedings},
  \bibinfo{year}{2010}, pp. \bibinfo{pages}{645--652}.
\bibitem[{Rasmussen and Williams(2005)}]{books/lib/RasmussenW06}
\bibinfo{author}{C.~E. Rasmussen}, \bibinfo{author}{C.~K.~I. Williams},
  \bibinfo{title}{Gaussian processes for machine learning.}, Adaptive
  computation and machine learning, \bibinfo{publisher}{MIT Press},
  \bibinfo{year}{2005}.
\bibitem[{Titsias et~al.(2011)Titsias, Rattray, and
  Lawrence}]{titsias_rattray_lawrence_2011}
\bibinfo{author}{M.~K. Titsias}, \bibinfo{author}{M.~Rattray},
  \bibinfo{author}{N.~D. Lawrence}, \bibinfo{title}{Markov chain Monte Carlo
  algorithms for Gaussian processes}, \bibinfo{publisher}{Cambridge University
  Press}, \bibinfo{year}{2011}, pp. \bibinfo{pages}{295--316}.
  \DOIprefix\doi{10.1017/CBO9780511984679.015}.
\bibitem[{Titsias et~al.(2019)Titsias, Schwarz, Matthews, Pascanu, and
  Teh}]{Titsias2019}
\bibinfo{author}{M.~K. Titsias}, \bibinfo{author}{J.~Schwarz},
  \bibinfo{author}{A.~G. d.~G. Matthews}, \bibinfo{author}{R.~Pascanu},
  \bibinfo{author}{Y.~W. Teh},
\newblock \bibinfo{title}{Functional regularisation for continual learning with
  {Gaussian Processes}}  (\bibinfo{year}{2019}).
  \DOIprefix\doi{10.48550/ARXIV.1901.11356}.
  \href{http://arxiv.org/abs/1901.11356}{{\tt arXiv:1901.11356}}.
\bibitem[{Hensman et~al.(2013)Hensman, Fusi, and Lawrence}]{Hensman2013}
\bibinfo{author}{J.~Hensman}, \bibinfo{author}{N.~Fusi}, \bibinfo{author}{N.~D.
  Lawrence},
\newblock \bibinfo{title}{Gaussian processes for big data},
\newblock in: \bibinfo{booktitle}{Proceedings of the Conference on Uncertainty
  in Artificial Intelligence}, \bibinfo{year}{2013}, pp.
  \bibinfo{pages}{282--290}.
\bibitem[{Gasser et~al.(2006)Gasser, Ogden, and
  Holzapfel}]{gasser2006hyperelastic}
\bibinfo{author}{T.~C. Gasser}, \bibinfo{author}{R.~W. Ogden},
  \bibinfo{author}{G.~A. Holzapfel},
\newblock \bibinfo{title}{Hyperelastic modelling of arterial layers with
  distributed collagen fibre orientations},
\newblock \bibinfo{journal}{Journal of the royal society interface}
  \bibinfo{volume}{3} (\bibinfo{year}{2006}) \bibinfo{pages}{15--35}.
  \DOIprefix\doi{10.1098/rsif.2005.0073}.
\bibitem[{Hudson et~al.(2022)Hudson, Laurence, Lau, Mullins, Doan, and
  Lee}]{HUDSON2022104907}
\bibinfo{author}{L.~T. Hudson}, \bibinfo{author}{D.~W. Laurence},
  \bibinfo{author}{H.~M. Lau}, \bibinfo{author}{B.~T. Mullins},
  \bibinfo{author}{D.~D. Doan}, \bibinfo{author}{C.-H. Lee},
\newblock \bibinfo{title}{Linking collagen fiber architecture to tissue-level
  biaxial mechanical behaviors of porcine semilunar heart valve cusps},
\newblock \bibinfo{journal}{Journal of the Mechanical Behavior of Biomedical
  Materials} \bibinfo{volume}{125} (\bibinfo{year}{2022})
  \bibinfo{pages}{104907}. \DOIprefix\doi{10.1016/j.jmbbm.2021.104907}.
\bibitem[{Lee et~al.(2014)Lee, Amini, Gorman, Gorman, and Sacks}]{LEE20142055}
\bibinfo{author}{C.-H. Lee}, \bibinfo{author}{R.~Amini}, \bibinfo{author}{R.~C.
  Gorman}, \bibinfo{author}{J.~H. Gorman}, \bibinfo{author}{M.~S. Sacks},
\newblock \bibinfo{title}{An inverse modeling approach for stress estimation in
  mitral valve anterior leaflet valvuloplasty for in-vivo valvular biomaterial
  assessment},
\newblock \bibinfo{journal}{Journal of Biomechanics} \bibinfo{volume}{47}
  (\bibinfo{year}{2014}) \bibinfo{pages}{2055--2063}.
  \DOIprefix\doi{10.1016/j.jbiomech.2013.10.058}.
\bibitem[{May-Newman and Yin(1998)}]{may1998constitutive}
\bibinfo{author}{K.~May-Newman}, \bibinfo{author}{F.~C.~P. Yin},
\newblock \bibinfo{title}{A constitutive law for mitral valve tissue},
\newblock \bibinfo{journal}{Journal of Biomechanical Engineering}
  \bibinfo{volume}{120} (\bibinfo{year}{1998}) \bibinfo{pages}{38--47}.
  \DOIprefix\doi{10.1115/1.2834305}.
\bibitem[{Holzapfel et~al.(2005)Holzapfel, Sommer, Gasser, and
  Regitnig}]{doi:10.1152/ajpheart.00934.2004}
\bibinfo{author}{G.~A. Holzapfel}, \bibinfo{author}{G.~Sommer},
  \bibinfo{author}{C.~T. Gasser}, \bibinfo{author}{P.~Regitnig},
\newblock \bibinfo{title}{Determination of layer-specific mechanical properties
  of human coronary arteries with nonatherosclerotic intimal thickening and
  related constitutive modeling},
\newblock \bibinfo{journal}{American Journal of Physiology-Heart and
  Circulatory Physiology} \bibinfo{volume}{289} (\bibinfo{year}{2005})
  \bibinfo{pages}{H2048--H2058}. \DOIprefix\doi{10.1152/ajpheart.00934.2004}.
\bibitem[{Gasser et~al.(2006)Gasser, Ogden, and
  Holzapfel}]{doi:10.1098/rsif.2005.0073}
\bibinfo{author}{T.~C. Gasser}, \bibinfo{author}{R.~W. Ogden},
  \bibinfo{author}{G.~A. Holzapfel},
\newblock \bibinfo{title}{Hyperelastic modelling of arterial layers with
  distributed collagen fibre orientations},
\newblock \bibinfo{journal}{Journal of The Royal Society Interface}
  \bibinfo{volume}{3} (\bibinfo{year}{2006}) \bibinfo{pages}{15--35}.
  \DOIprefix\doi{10.1098/rsif.2005.0073}.
\bibitem[{Humphrey and Yin(1987)}]{HUMPHREY1987563}
\bibinfo{author}{J.~Humphrey}, \bibinfo{author}{F.~Yin},
\newblock \bibinfo{title}{A new constitutive formulation for characterizing the
  mechanical behavior of soft tissues},
\newblock \bibinfo{journal}{Biophysical Journal} \bibinfo{volume}{52}
  (\bibinfo{year}{1987}) \bibinfo{pages}{563--570}.
  \DOIprefix\doi{10.1016/S0006-3495(87)83245-9}.
\bibitem[{Uhlmann(1995)}]{uhlmann1995dynamic}
\bibinfo{author}{J.~Uhlmann}, \bibinfo{title}{Dynamic map building and
  localization: new theoretical foundations.}, Ph.D. thesis, University of
  Oxford, \bibinfo{year}{1995}.
\bibitem[{Van Der~Merwe(2004)}]{van2004sigma}
\bibinfo{author}{R.~Van Der~Merwe}, \bibinfo{title}{Sigma-point Kalman filters
  for probabilistic inference in dynamic state-space models},
  \bibinfo{publisher}{Oregon Health \& Science University},
  \bibinfo{year}{2004}.
\bibitem[{Maas et~al.(2012)Maas, Ellis, Ateshian, and Weiss}]{maas2012febio}
\bibinfo{author}{S.~A. Maas}, \bibinfo{author}{B.~J. Ellis},
  \bibinfo{author}{G.~A. Ateshian}, \bibinfo{author}{J.~A. Weiss},
\newblock \bibinfo{title}{Febio: finite elements for biomechanics},
\newblock \bibinfo{journal}{Journal of biomechanical engineering}
  \bibinfo{volume}{134} (\bibinfo{year}{2012}) \bibinfo{pages}{011005}.
  \DOIprefix\doi{10.1115/1.4005694}.
\bibitem[{Betsch et~al.(1996)Betsch, Gruttmann, and Stein}]{betsch19964}
\bibinfo{author}{P.~Betsch}, \bibinfo{author}{F.~Gruttmann},
  \bibinfo{author}{E.~Stein},
\newblock \bibinfo{title}{A 4-node finite shell element for the implementation
  of general hyperelastic 3d-elasticity at finite strains},
\newblock \bibinfo{journal}{Computer Methods in Applied Mechanics and
  Engineering} \bibinfo{volume}{130} (\bibinfo{year}{1996})
  \bibinfo{pages}{57--79}. \DOIprefix\doi{10.1016/0045-7825(95)00920-5}.
\bibitem[{Aggarwal and Sacks(2016)}]{aggarwal_inverse_2015}
\bibinfo{author}{A.~Aggarwal}, \bibinfo{author}{M.~S. Sacks},
\newblock \bibinfo{title}{An inverse modeling approach for semilunar heart
  valve leaflet mechanics: exploitation of tissue structure},
\newblock \bibinfo{journal}{Biomechanics and Modeling in Mechanobiology}
  \bibinfo{volume}{15} (\bibinfo{year}{2016}) \bibinfo{pages}{909--932}.
  \DOIprefix\doi{10.1007/s10237-015-0732-7}.
\bibitem[{Aggarwal et~al.(2022)Aggarwal, Hudson, Laurence, Lee, and
  Pant}]{https://doi.org/10.48550/arxiv.2209.13038}
\bibinfo{author}{A.~Aggarwal}, \bibinfo{author}{L.~T. Hudson},
  \bibinfo{author}{D.~W. Laurence}, \bibinfo{author}{C.-H. Lee},
  \bibinfo{author}{S.~Pant}, \bibinfo{title}{A bayesian constitutive model
  selection framework for biaxial mechanical testing of planar soft tissues:
  application to porcine aortic valves}, \bibinfo{year}{2022}.
  \DOIprefix\doi{10.48550/ARXIV.2209.13038}.
\bibitem[{Hauseux et~al.(2018)Hauseux, Hale, Cotin, and Bordas}]{HAUSEUX201886}
\bibinfo{author}{P.~Hauseux}, \bibinfo{author}{J.~S. Hale},
  \bibinfo{author}{S.~Cotin}, \bibinfo{author}{S.~P. Bordas},
\newblock \bibinfo{title}{Quantifying the uncertainty in a hyperelastic soft
  tissue model with stochastic parameters},
\newblock \bibinfo{journal}{Applied Mathematical Modelling}
  \bibinfo{volume}{62} (\bibinfo{year}{2018}) \bibinfo{pages}{86--102}.
  \DOIprefix\doi{10.1016/j.apm.2018.04.021}.
\bibitem[{Hauseux et~al.(2017)Hauseux, Hale, and Bordas}]{HAUSEUX2017917}
\bibinfo{author}{P.~Hauseux}, \bibinfo{author}{J.~S. Hale},
  \bibinfo{author}{S.~P. Bordas},
\newblock \bibinfo{title}{Accelerating monte carlo estimation with derivatives
  of high-level finite element models},
\newblock \bibinfo{journal}{Computer Methods in Applied Mechanics and
  Engineering} \bibinfo{volume}{318} (\bibinfo{year}{2017})
  \bibinfo{pages}{917--936}. \DOIprefix\doi{10.1016/j.cma.2017.01.041}.
\bibitem[{Joshi et~al.(2022)Joshi, Thakolkaran, Zheng, Escande, Flaschel, {De
  Lorenzis}, and Kumar}]{JOSHI2022115225}
\bibinfo{author}{A.~Joshi}, \bibinfo{author}{P.~Thakolkaran},
  \bibinfo{author}{Y.~Zheng}, \bibinfo{author}{M.~Escande},
  \bibinfo{author}{M.~Flaschel}, \bibinfo{author}{L.~{De Lorenzis}},
  \bibinfo{author}{S.~Kumar},
\newblock \bibinfo{title}{Bayesian-euclid: Discovering hyperelastic material
  laws with uncertainties},
\newblock \bibinfo{journal}{Computer Methods in Applied Mechanics and
  Engineering} \bibinfo{volume}{398} (\bibinfo{year}{2022})
  \bibinfo{pages}{115225}. \DOIprefix\doi{10.1016/j.cma.2022.115225}.
\bibitem[{Thakolkaran et~al.(2022)Thakolkaran, Joshi, Zheng, Flaschel, {De
  Lorenzis}, and Kumar}]{THAKOLKARAN2022105076}
\bibinfo{author}{P.~Thakolkaran}, \bibinfo{author}{A.~Joshi},
  \bibinfo{author}{Y.~Zheng}, \bibinfo{author}{M.~Flaschel},
  \bibinfo{author}{L.~{De Lorenzis}}, \bibinfo{author}{S.~Kumar},
\newblock \bibinfo{title}{Nn-euclid: Deep-learning hyperelasticity without
  stress data},
\newblock \bibinfo{journal}{Journal of the Mechanics and Physics of Solids}
  \bibinfo{volume}{169} (\bibinfo{year}{2022}) \bibinfo{pages}{105076}.
  \DOIprefix\doi{10.1016/j.jmps.2022.105076}.
\bibitem[{Deshpande et~al.(2022)Deshpande, Lengiewicz, and
  Bordas}]{DESHPANDE2022115307}
\bibinfo{author}{S.~Deshpande}, \bibinfo{author}{J.~Lengiewicz},
  \bibinfo{author}{S.~P. Bordas},
\newblock \bibinfo{title}{Probabilistic deep learning for real-time large
  deformation simulations},
\newblock \bibinfo{journal}{Computer Methods in Applied Mechanics and
  Engineering} \bibinfo{volume}{398} (\bibinfo{year}{2022})
  \bibinfo{pages}{115307}. \DOIprefix\doi{10.1016/j.cma.2022.115307}.
\bibitem[{Rappel et~al.(2020)Rappel, Beex, Hale, Noels, and
  Bordas}]{rappel2020tutorial}
\bibinfo{author}{H.~Rappel}, \bibinfo{author}{L.~A. Beex},
  \bibinfo{author}{J.~S. Hale}, \bibinfo{author}{L.~Noels},
  \bibinfo{author}{S.~Bordas},
\newblock \bibinfo{title}{A tutorial on bayesian inference to identify material
  parameters in solid mechanics},
\newblock \bibinfo{journal}{Archives of Computational Methods in Engineering}
  \bibinfo{volume}{27} (\bibinfo{year}{2020}) \bibinfo{pages}{361--385}.
  \DOIprefix\doi{10.1007/s11831-018-09311-x}.
\bibitem[{Ustyuzhaninov et~al.(2019)Ustyuzhaninov, Kazlauskaite, Ek, and
  Campbell}]{https://doi.org/10.48550/arxiv.1905.12930}
\bibinfo{author}{I.~Ustyuzhaninov}, \bibinfo{author}{I.~Kazlauskaite},
  \bibinfo{author}{C.~H. Ek}, \bibinfo{author}{N.~D.~F. Campbell},
  \bibinfo{title}{Monotonic gaussian process flow}, \bibinfo{year}{2019}.
  \DOIprefix\doi{10.48550/ARXIV.1905.12930}.
\bibitem[{Andersen et~al.(2018)Andersen, Siivola, Riutort-Mayol, and
  Vehtari}]{andersen2018non}
\bibinfo{author}{M.~Andersen}, \bibinfo{author}{E.~Siivola},
  \bibinfo{author}{G.~Riutort-Mayol}, \bibinfo{author}{A.~Vehtari},
\newblock \bibinfo{title}{A non-parametric probabilistic model for monotonic
  functions.},
\newblock in: \bibinfo{booktitle}{All Of Bayesian Nonparametrics Workshop at
  NeurIPS}, \bibinfo{year}{2018}.
\bibitem[{Sutula et~al.(2020)Sutula, Elouneg, Sensale, Chouly, Chambert,
  Lejeune, Baroli, Hauseux, Bordas, and Jacquet}]{SUTULA2020103999}
\bibinfo{author}{D.~Sutula}, \bibinfo{author}{A.~Elouneg},
  \bibinfo{author}{M.~Sensale}, \bibinfo{author}{F.~Chouly},
  \bibinfo{author}{J.~Chambert}, \bibinfo{author}{A.~Lejeune},
  \bibinfo{author}{D.~Baroli}, \bibinfo{author}{P.~Hauseux},
  \bibinfo{author}{S.~Bordas}, \bibinfo{author}{E.~Jacquet},
\newblock \bibinfo{title}{An open source pipeline for design of experiments for
  hyperelastic models of the skin with applications to keloids},
\newblock \bibinfo{journal}{Journal of the Mechanical Behavior of Biomedical
  Materials} \bibinfo{volume}{112} (\bibinfo{year}{2020})
  \bibinfo{pages}{103999}. \DOIprefix\doi{10.1016/j.jmbbm.2020.103999}.
\bibitem[{Elouneg et~al.(2021)Elouneg, Sutula, Chambert, Lejeune, Bordas, and
  Jacquet}]{ELOUNEG2021106620}
\bibinfo{author}{A.~Elouneg}, \bibinfo{author}{D.~Sutula},
  \bibinfo{author}{J.~Chambert}, \bibinfo{author}{A.~Lejeune},
  \bibinfo{author}{S.~Bordas}, \bibinfo{author}{E.~Jacquet},
\newblock \bibinfo{title}{An open-source fenics-based framework for
  hyperelastic parameter estimation from noisy full-field data: Application to
  heterogeneous soft tissues},
\newblock \bibinfo{journal}{Computers \& Structures} \bibinfo{volume}{255}
  (\bibinfo{year}{2021}) \bibinfo{pages}{106620}.
  \DOIprefix\doi{10.1016/j.compstruc.2021.106620}.
\bibitem[{Clyde(2001)}]{clyde2001experimental}
\bibinfo{author}{M.~A. Clyde},
\newblock \bibinfo{title}{Experimental design: A bayesian perspective},
\newblock \bibinfo{journal}{International Encyclopia Social and Behavioral
  Sciences} \bibinfo{volume}{8} (\bibinfo{year}{2001})
  \bibinfo{pages}{5075--5081}.
\bibitem[{Ryan et~al.(2016)Ryan, Drovandi, and Pettitt}]{10.1214/15-BA977}
\bibinfo{author}{C.~M. Ryan}, \bibinfo{author}{C.~C. Drovandi},
  \bibinfo{author}{A.~N. Pettitt},
\newblock \bibinfo{title}{{Optimal Bayesian Experimental Design for Models with
  Intractable Likelihoods Using Indirect Inference Applied to Biological
  Process Models}},
\newblock \bibinfo{journal}{Bayesian Analysis} \bibinfo{volume}{11}
  (\bibinfo{year}{2016}) \bibinfo{pages}{857 -- 883}.
  \DOIprefix\doi{10.1214/15-BA977}.
\bibitem[{Dsouza et~al.(2021)Dsouza, Varghese, Ooi, Natarajan, and
  Bordas}]{DSOUZA2021538}
\bibinfo{author}{S.~M. Dsouza}, \bibinfo{author}{T.~M. Varghese},
  \bibinfo{author}{E.~T. Ooi}, \bibinfo{author}{S.~Natarajan},
  \bibinfo{author}{S.~P. Bordas},
\newblock \bibinfo{title}{Treatment of multiple input uncertainties using the
  scaled boundary finite element method},
\newblock \bibinfo{journal}{Applied Mathematical Modelling}
  \bibinfo{volume}{99} (\bibinfo{year}{2021}) \bibinfo{pages}{538--554}.
  \DOIprefix\doi{10.1016/j.apm.2021.06.021}.
\bibitem[{Hauseux et~al.(2017)Hauseux, Hale, and
  Bordas}]{10.1371/journal.pone.0189994}
\bibinfo{author}{P.~Hauseux}, \bibinfo{author}{J.~S. Hale},
  \bibinfo{author}{S.~P.~A. Bordas},
\newblock \bibinfo{title}{Calculating the malliavin derivative of some
  stochastic mechanics problems},
\newblock \bibinfo{journal}{PLOS ONE} \bibinfo{volume}{12}
  (\bibinfo{year}{2017}) \bibinfo{pages}{1--18}.
  \DOIprefix\doi{10.1371/journal.pone.0189994}.
\bibitem[{Bartels et~al.(1987)Bartels, Beatty, and
  Barsky}]{bartels1995introduction}
\bibinfo{author}{R.~H. Bartels}, \bibinfo{author}{J.~C. Beatty},
  \bibinfo{author}{B.~A. Barsky}, \bibinfo{title}{An introduction to splines
  for use in computer graphics and geometric modeling},
  \bibinfo{publisher}{Morgan Kaufmann Publishers Inc.}, \bibinfo{address}{San
  Francisco, CA, USA}, \bibinfo{year}{1987}.

\end{thebibliography}





\end{document}